\documentclass[12pt]{article}

\usepackage{amsfonts, graphicx,color,multirow, amsthm, amsmath,natbib,hypernat}
\usepackage{setspace,url}
\usepackage{bm,xr,authblk}
\usepackage{epsfig,amssymb,amsfonts,epsfig,verbatim}
\usepackage{color,graphicx,lscape,longtable,subcaption}
\newcommand{\sumi}{\ensuremath{\sum_{i=1}^{n}}}
\newcommand{\sumj}{\ensuremath{\sum_{j=1}^{n}}}
\newcommand{\sumk}{\ensuremath{\sum_{k=1}^{n}}}

\newtheorem{Th}{\underline{\bf Theorem}}

\newtheorem{Rem}{\underline{\bf Remark}}

\def\var{\hbox{var}}
\def\sd{\hbox{sd}}

\def\wh{\widehat}
\def\wt{\widetilde}
\def\eff{_{\rm eff}}

\def\mR{\mathbb{R}}
\def\n{\nonumber}

\def\argmax{\mbox{argmax}}

\def\sumi{\sum_{i=1}^n}
\def\sumj{\sum_{j=1}^n}
\def\trans{^{\rm T}}
\def\evf{\widehat{\mbox{VF}}}

\def\ba{{\boldsymbol\alpha}}

\def\bb{{\boldsymbol\beta}}
\def\bg{{\boldsymbol\gamma}}

\def\A{{\bf A}}
\def\a{{\bf a}}
\def\B{{\bf B}}

\def\V{{\bf V}}

\def\r{{\bf r}}

\def\h{{\bf h}}
\def\b{{\bf b}}

\def\U{{\bf U}}
\def\S{{\bf S}}
\def\u{{\bf u}}
\def\v{{\bf v}}

\def\w{{\bf w}}
\def\X{{\bf X}}

\def\O{{\bf O}}
\def\bo{{\bf o}}
\def\x{{\bf x}}

\def\Z{{\bf Z}}
\def\z{{\bf z}}

\def\pr{\hbox{pr}}
\def\trans{^{\rm T}}

\def\squarebox#1{\hbox to #1{\hfill\vbox to #1{\vfill}}}

\def\0{{\bf 0}}

\def\mA{\mathcal{A}}

\def\var{\hbox{var}}

\def\bse{\begin{eqnarray*}}
	\def\ese{\end{eqnarray*}}
\def\be{\begin{eqnarray}}
	\def\ee{\end{eqnarray}}
\def\bsq{\begin{equation*}}
	\def\esq{\end{equation*}}
\def\bq{\begin{equation}}
	\def\eq{\end{equation}}
\def\pr{\hbox{pr}}
\def\wh{\widehat}
\def\wt{\widetilde}

\def\trans{^{\rm T}}

\def\boxit#1{\vbox{\hrule\hbox{\vrule\kern6pt\vbox{\kern6pt#1\kern6pt}\kern6pt\vrule}\hrule}}

\begin{document}
	\thispagestyle{empty}
	\baselineskip=18pt
	\title{Flexible Inference of Optimal Individualized Treatment Strategy in Covariate
		Adjusted Randomization with Multiple Covariates}
\author[1] {Trinetri Ghosh}
\author[2]{Yanyuan Ma}
\author[3]{Rui Song}
\author[4]{Pingshou Zhong}
\affil[1]{University of Wisconsin-Madison}
\affil[2]{The Pennsylvania State University}
\affil[3]{North Carolina State University}
\affil[4]{University of Illinois at Chicago}
	\maketitle

\bigskip
\begin{abstract}
To maximize clinical benefit, clinicians routinely tailor treatment to the individual characteristics of each patient, where individualized treatment rules are needed and are of significant research interest to statisticians.  In the covariate-adjusted randomization clinical trial with many covariates, we model the treatment effect with an unspecified function of a single index of the covariates and leave the baseline response  completely arbitrary. We devise a class of estimators to consistently estimate the treatment effect function and its associated index while bypassing the estimation of the baseline response, which is subject to the curse of dimensionality.  We further develop inference tools to identify predictive covariates and isolate effective treatment region. The usefulness of the methods is demonstrated in both simulations and a clinical data example. 
\end{abstract}
	
	\noindent%
	\vfill
	
	\section{Introduction}\label{sec:intro}
	
	Precision medicine, which is defined as treatments targeted to individual patients' needs based on genetics, biomarker, phenotypic, or psychosocial characteristics that distinguish a given patient from other patients with similar clinical presentations \citep{jameson2015precision}, has generated tremendous interest in statistical research. Precision medicine based on individual's health-related metrics and environmental factors is used to discover individualized treatment regimes (ITRs); methodology for such discovery is an expanding field of statistics. 
	
	Various methods have been proposed in the statistical literature to estimate the optimal ITRs.  Q-learning
	\citep{watkins1989,watkins1992q,murphy2005experimental,zhao2009reinforcement,
		chakraborty2010inference,
		zhao2011reinforcement,qian2011performance,goldberg2012q,SWZK12}and A-learning
	\citep{murphy2003optimal,robins2004optimal,blatt2004learning,orellana2010dynamic,henderson2010regret,liang2018deep,shi2018high} are {two backward induction methods for deriving optimal dynamic treatment regimes.} Other related approaches include parametric methods \citep{8Thall00,13Thall02,14Thall07}, model-free or direct value search method by maximizing a nonparametric estimator of value function
	\citep{fan2017concordance,zhang_2012b,zhang_2012a,zhou2019restricted,KAL2020,shi2021concordance}, semiparametric methods  
	\citep{15Lunceford02,9Wahed04,16Wahed06,Moodie09,Single13,kang2018estimation,xiao2019robust}, and
	machine learning methods \citep{OWL-Learning}.  \cite{OWL-Learning} proposed outcome weighted learning, which can be viewed as a weighted classification problem using the covariate information weighted by the individual response to maximize the overall outcome.
	
	The rules based on parametric models of clinical outcomes given treatment and other covariates are simple, yet can be incorrect when parametric models are misspecified. The rules based on machine learning techniques to determine the relationship between clinical outcomes and treatment plus covariates, such as in Zhao et al. (2009, 2011), are nonparametric and flexible but are often complex and may have
	large variability. Existing semiparametric methodologies, such as \cite{murphy2003optimal} and \cite{Single13}, flexibly incorporate the relationship between the covariates and the response variables, but
	are not efficient due to the challenges in estimating the decision rules. Moreover, to the best of our knowledge, these existing semiparametric models do not allow covariate adjusted randomization, where a patient is randomized to one of the treatment arms based on the patient's covariate and a predetermined randomization scheme. Therefore we are  motivated to derive semiparametric efficient estimators for flexible models that allow covariate adjusted randomization. 
	
	Let $\X_i$ be the $p$-dimensional covariate
	vector, $Z_i$ be the treatment indicator, and $Y_i$ be the
	response, and assume $(\X_i, Z_i, Y_i)$'s are independent and identically
	distributed (iid) for $i=1, \dots, n$. Here, 
	the treatment indicator $Z_i$ is often categorical (i.e. when several treatment
	arms are considered) but can also be continuous (i.e. when a continuous of
	dosage range is considered). In the simplest case
	when $Z_i$ is binary, we denote $Z_i=1$ if
	the $i$th patient is randomized to treatment and $Z_i=0$ to placebo.  
	When the randomization is covariate adjusted, the probability
	distribution of
	$Z_i$ depends on $\X_i$.
	Our goal is to analyze the data $(\X_i, Z_i, Y_i)$ for $i=1, \dots, n$,
	so that we can develop a treatment rule that is best for each new
	individual. Without loss of generality, we 
	assume larger value of $Y_0$ indicates better outcome.
	Equivalently, we want to 
	identify a deterministic decision rule,  $d(\x)$, which takes as input
	a given value $\x$ of $\X_0$ and outputs a treatment choice of $Z_0$ that
	will maximize the potential treatment result $Y_0$. 
	Here $\X_0$ denotes the covariate of the new individual, $Z_0$ the treatment
	decision and $Y_0$ the potential treatment result given $Z_0$.  
	Let $P^d$ be the distribution of $(\X_0, Z_0, Y_0)$  and $E^d$ be the expectation
	with respect to this distribution. 
	The value function is defined as $V(d)=E^d(Y_0)$. An optimal
	individualized treatment rule $d_0$ maximizes $V(d)$ over all decision rules.
	That is, $d_0= \argmax_d V(d) = \argmax_{d(\x)} E[Y_0|Z_0=d(\x), \X_0=\x]$.
	It is important to recognize the difference between the clinical trial data 
	$(\X_i, Z_i, Y_i), i=1, \dots, n$ and the potential data associated
	with the new individual $(\X_0, Z_0, Y_0)$.
	
	In the covariate adjusted randomization, let the probability of
	assigning to the treatment or placebo arm be 
	a known function $f_{Z\mid\X}(\x,z)$. Here and throughout, we omit the
	subindex $_i$ when describing the data from a randomized clinical
	trial as long as it does not cause confusion. 
	We model the treatment effect using a single index model of the
	covariates, while leaving the baseline response unspecified. 
	This yields the model
	\be\label{eq:modelbin}
	Y=f(\X)+Zg(\bb\trans\X)+\epsilon.
	\ee
	We assume the regression error  $\epsilon$ is independent of the covariates $\X$ and the
	treatment indicator $Z$, and $\epsilon\sim N(0,\sigma^2)$. We do not
	assume $Z$ and $\X$ to be independent, hence allow the covariate
	adjusted randomization procedure.
	Naturally from (\ref{eq:modelbin}), $g(\bb\trans\x)$ is the treatment
	effect as a function of the covariate value $\x$. We assume $g$ to be a
	smooth function. To retain the identifiability of $g(\cdot)$ and $\bb$, we fix
	the first component of $\bb$ to be 1.
	Our goal is to estimate $\bb$
	and $g(\cdot)$, hence to estimate the treatment effect given
	any covariate $\x$.

	The contribution of our proposed method beyond existing literature can
	be summarized in the following four folds. First, we propose an
	estimator and show that it is locally semiparametric efficient. 
	Second, our method allows  randomization to depend on patients'
	covariates, which significantly generalizes the practicality of the
	semiparametric methodology in real studies. Third, although
	illustrated for binary treatment selection, the proposed method does
	not restrict to the binary treatment case and is directly applicable
	even when $Z$ is categorical or 
	continuous. Thus, the method applies to slope estimation while the
	intercept, i.e. $f(\X)$ in (\ref{eq:modelbin}), is unspecified and inestimable
	due to the curse of dimensionality caused by the multi-dimension of $\X$.  
	Fourth, different from 
	\cite{Single13} where B-spline expansion were used, we employ kernel
	smoothing techniques into the estimation and inference for optimal treatment
	regimes.  Fifth, our method is more flexible in terms of model 
	assumptions compared to \cite{KAL2020}.
	
	The rest of the article is organized as the following. In Section
	\ref{sec:method}, we devise a class of estimators for $\bb$ and
	$g(\cdot)$ while bypassing the estimation of $f(\cdot)$. We study the
	large sample properties of the estimators and carry out inference to
	detect effective treatment region in Section
	\ref{sec:theory}. Simulation experiments are carried out in Section
	\ref{sec:simulation}, and the method is implemented on a clinical trial
	data in Section \ref{sec:example}. We conclude the article with a
	discussion in Section \ref{sec:discussion} and collect all the
	technical derivations and proofs in the Supplementary Material. 
	
	\section{Estimation of $\bb$ and $g(\cdot)$}\label{sec:method}
	
	\subsection{Basic estimation}\label{sec:basic}
	The estimation of $\bb$ and $g(\cdot)$ is complicated by the presence
	of the intercept term $f(\X)$. When $\X$ is of high or even moderate
	dimension, $f(\X)$ is challenging to estimate due to the well known
	curse of dimensionality. Thus, a simple treatment is to eliminate the
	effect of $f(\X)$. Following this approach, 
	multiplying $Z$ and $E(Z\mid\X)$ on (\ref{eq:modelbin}), we have
	\bse
	ZY&=&Z^2g(\bb\trans\X)+Zf(\X)+Z\epsilon, \\
	\mbox{ and}\; E(Z\mid\X)Y&=&ZE(Z\mid\X)g(\bb\trans\X)+E(Z\mid\X)f(\X)+E(Z\mid\X)\epsilon.
	\ese
	Taking difference of the above two equations yields $\{Z-E(Z\mid\X)\}Y=\{Z^2-ZE(Z\mid\X)\}g(\bb\trans\X)
	+\{Z-E(Z\mid\X)\}\{f(\X)+\epsilon\}$.
	Note that $E[\{Z-E(Z\mid\X)\}\{f(\X)+\epsilon\}\mid\X]=E\{Z-E(Z\mid\X)\mid\X\}f(\X)+0=0$
	and $E [\{Z^2-ZE(Z\mid\X)\}g(\bb\trans\X)\mid\X]=\var(Z\mid\X) g(\bb\trans\X)$.
	So we obtain $E[\{Z-E(Z\mid\X)\}Y\mid\X]=\var(Z\mid\X) g(\bb\trans\X)$.
	This indicates that if we write $\wt Y\equiv\{Z-E(Z\mid\X)\}Y\{\var(Z\mid\X)\}^{-1}$, 
	then $E(\wt Y\mid\X)=g(\bb\trans\X)$. Viewing $\X$ and $\wt Y$ as the
	covariate and response respectively, this is a classical single index
	model, and we can estimate $\bb$ and $g$ using standard methods
	(Ichimura 1993). Considering that $\var(Z\mid\X)$ might be close to
	zero, one may want to avoid taking its inverse. In this case, a more
	stable estimator would be to minimize
	\vspace*{-0.1in}
	\bse
	\sumj\sumi w_{ij}[\{z_i-E(Z_i\mid\x_i)\} y_i-\var(Z_i\mid\x_i)\{a_j+b_j\bb\trans(\x_i-\x_j)\}]^2
	\ese
	\vspace*{-0.1in}
	with respect to $a_1,\cdots, a_n, b_1,\cdots, b_n$ and parameters in $\bb$, where
	\bse
	w_{ij}=\frac{K_h\{\bb\trans(\x_i-\x_j)\}}{\sumi K_h\{\bb\trans(\x_i-\x_j)\}},
	\ese
	$K_h(\cdot)=h^{-1}K(\cdot/h)$, $h$ is a bandwidth and $K(\cdot)$
	is a kernel function. 
	At a fixed $\bb$ value, the minimization with respect to $(a_j, b_j)$
	does not rely on other $(a_k, b_k)$ values for $k\neq j$, hence can be
	done separately, while the optimization with respect to $\bb$ involves
	all the terms.  
	The resulting $\wh\bb$ estimates $\bb$ and the resulting $\wh a_j$
	estimates $g(\bb\trans\x_j)$. 
	We can further estimate $g(\bb\trans\x_0)$ by $\wh a_0$ which is obtained by
	\vskip 3mm
	$
	(\wh a_0, \wh b_0)=\arg\min_{a_0, b_0} \sumi w_{i0}[\{z_i-E(Z_i\mid\x_i)\} y_i-\var(Z_i\mid\x_i)\{a_0+b_0\wh\bb\trans(\x_i-\x_0)\}]^2.
	$
	
	\subsection{Proposed estimator}\label{sec:best}
	
	Although the above procedure provides one way of estimating $\bb$ and
	$g(\cdot)$, it is somewhat ad hoc, and it is unclear if other estimators
	exist to achieve similar or better estimation.  To investigate the
	problem more thoroughly and systematically, we start with the
	likelihood function of a single observation $(\x, z, y)$, 
	\be\label{eq:pdfbin}
	f_{\X,Z,Y}(\x,z,y,\bb,\sigma,g,f,\eta)=\eta(\x)f_{Z\mid\X}(\x,z)\sigma^{-1}\phi[\sigma^{-1}\{y-f(\x)-zg(\bb\trans\x)\}].
	\ee
	Here $\eta(\x)$ is the marginal probability density or mass function
	(pdf or pmf) of $\X$, and $\phi(\cdot)$ is the standard normal pdf. We
	first focus our attention on estimating $\bb$ alone, thus we
	view $g$ together with $f$, $\eta$ and $\sigma$ as nuisance parameters. In this case, (\ref{eq:pdfbin}) is a
	semiparametric model, thus we derive the estimators for $\bb$ through
	deriving its influence functions and constructing estimating
	equations. It is easy to obtain the score function with respect
	to $\bb$ through taking partial derivative of the loglikelihood
	function with respect to the parameter. In
	Supplementary Material \ref{sec:nuisanceg}, we further project the score functions
	$\S_\beta(\x,z,y,\bb,g,f,\sigma)$  onto the
	nuisance tangent space, a space spanned by the nuisance score
	functions, and obtain the efficient score function 
	$\S\eff(\x,z,y,\bb,g,f,\sigma)=
	\sigma^{-2}\epsilon\{z-
	E(Z\mid\x)\}g'(\bb\trans\x)\{\x_L-\u(\bb\trans\x)\},
	$ where $g'(\cdot)$ is the first derivative of $g(\cdot)$, $\x_L$ is the lower $(p-1)$-dimensional sub-vector of $\x$,
	$\epsilon\equiv y-f(\x)-zg(\bb\trans\x)$ and 
	$
	\u(\bb\trans\x)\equiv E\left\{\var(Z\mid\X)\X_L\mid\bb\trans\x\right\}/
	E\left\{\var(Z\mid\X)\mid\bb\trans\x\right\}.
	$
	Inspired by the form of the efficient score function, we propose a general class
	of  consistent estimating equations for $\bb$ as 
	\bse
	\sumi \{y_i-f^*(\x_i)-z_i g^*(\bb\trans\x_i)\}\{z_i-E(Z_i\mid\x_i)\}
	h^*(\bb\trans\x_i)\{\x_{Li}-\u^*(\bb\trans\x_i)\}=\0,
	\ese
	where $f^*$ is an arbitrary function of $\x$ and $h^*$ is an arbitrary 
	function of $\bb\trans\x$. 
	Regarding $g^*$ and $\u^*$, we have the freedom of estimating one of
	the two functions and replacing the other with an arbitrary function
	of $\bb\trans\x$, or estimating both. 
	
	To explore the various flexibilities suggested above, let $f^*(\x)$ be an arbitrary predecided
	function. For example, $f^*(\x)=0$.  We first examine the choice of
	approximating both $\u(\cdot)$ and $g(\cdot)$. As a by-product of
	local linear estimation, we also approximate $g'(\cdot)$.  
	Let $\wh\u(\bb\trans\x)$ be a nonparametric estimation of
	$\u(\bb\trans\x)$. Note that $\wh\u(\bb\trans\x)$ involves only
	univariate nonparametric regression. Specifically, using kernel
	method, 
	\be\label{eq:u}
	\wh\u(\bb\trans\x)=
	\frac{\sumi K_h\{\bb\trans(\x_i-\x)\}\var(Z\mid\x_i)\x_{Li}}
	{\sumi K_h\{\bb\trans(\x_i-\x)\}\var(Z\mid\x_i)}.
	\ee
	Let $\ba=(\alpha_c, \alpha_1)\trans$ and for $j=1, \dots, n$, let $\wh\ba(\bb,j)$ solve
	the estimating equation  w.r.t. $\alpha_c$ and $\alpha_1$ for given $\bb$ and $\x_j$
	\be\label{eq:g}
	\sumi w_{ij}
	[y_i-f^*(\x_i)-z_i\{\alpha_c+\alpha_1\bb\trans(\x_i-\x_j)\}
	]\{z_i-E(Z\mid\x_i)\}
	\left\{
	1,\bb\trans(\x_i-\x_j)
	\right\}\trans=\0.
	\ee
	Specifically, let $v_{0j}=\sumi w_{ij}z_i\{z_i-E(Z\mid\x_i)\}$, 
	$v_{1j}=\sumi w_{ij}z_i\{z_i-E(Z\mid\x_i)\}\bb\trans(\x_i-\x_j)$, 
	$v_{2j}=\sumi w_{ij}z_i\{z_i-E(Z\mid\x_i)\}\{\bb\trans(\x_i-\x_j)\}^2$, 
	then
	\bse
	\wh\ba(\bb,j)
	=
	\left[\begin{array}{c}
		\left(v_{0j}v_{2j}-v_{1j}^2\right)^{-1} \sumi w_{ij}\{y_i-f^*(\x_i)\}\{z_i-E(Z\mid\x_i)\}\{v_{2j}
		-v_{1j}\bb\trans(\x_i-\x_j)\}\\
		\left(v_{0j}v_{2j}-v_{1j}^2\right)^{-1}\sumi w_{ij}\{y_i-f^*(\x_i)\}\{z_i-E(Z\mid\x_i)\}\{v_{0j}\bb\trans(\x_i-\x_j)-v_{1j}\}
	\end{array}\right].
	\ese
	In (\ref{eq:g}), we can replace $w_{ij}$ with
	$K_h\{\bb\trans(\x_i-\x_j)\}$ and the resulting estimating equation is identical. 
	Note that the above procedure enables us to obtain $\wh\alpha_c(\bb,
	j)$ as an approximation of $g(\bb\trans\x_j)$ and $\wh\alpha_1(\bb,j)$ 
	as an approximation of $g'(\bb\trans\x_j)$.
	We then plug in the estimated $\wh\u, \wh g$ and $\wh g'$ and solve
	\be\label{eq:best}
	\sumi\{y_i-f^*(\x_i)-z_i\wh\alpha_c(\bb,i)\}
	\{z_i-E(Z\mid\x_i)\}\wh\alpha_1(\bb,i)\{\x_{Li}-\wh\u(\bb\trans\x_i)\}
	=\0
	\ee
	to obtain $\wh\bb$. It is easily seen that the procedure we described
	above is a type of profiling estimator, where we estimate $g(\cdot),
	g'(\cdot)$, $\u(\cdot)$ as functions of  a given $\bb$, and then estimate
	$\bb$. The idea behind the construction of (\ref{eq:g}) is similar to the
	consideration of the efficient score function, and we describe the
	detailed derivation in Supplementary Material \ref{sec:nuisancenog}. The estimator
	based on solving (\ref{eq:best}) is guaranteed to be consistent, and
	has the potential to be efficient. In fact, it will be efficient if
	$f^*(\cdot)$ happens to be the true $f(\cdot)$ function. Of course, 
	this is not very likely to happen in practice. However, this is still the
	estimator we propose, because $f(\cdot)$ is formidable to estimate for
	large or even moderate $p$.
	Compared to all other estimators that rely on the same
	$f^*(\cdot)$, the estimator proposed here yields the smallest
	estimation variability and is the most stable, as we will demonstrate in Section
	\ref{sec:theory}. 
	
	The explicit algorithm of estimating $\bb$ and $g$ is described below.
	\begin{enumerate}
		\item
		Select a candidate function $f^*(\x)$.
		\item
		Solve the estimating equation (\ref{eq:best}) to obtain $\wh\bb$, where  $\wh\alpha_c(\bb,i)$
		and $\wh\alpha_1(\bb,i)$ are solutions  given right below (\ref{eq:g})
		and $\wh\u(\bb\trans\x_i)$ is given in (\ref{eq:u}).
		\item
		Solve (\ref{eq:g}) once more at $\x_j=\x_0$  while fix $\bb$ at $\wh\bb$ 
		to obtain $\wh g(\bb\trans\x_0)=\wh\alpha_c(\wh\bb,0)$.
	\end{enumerate}
	
	\begin{Rem}
		We left out the details on how to select the bandwidths in the
		algorithm. In Section \ref{sec:theory}, we will show that a large
		range of bandwidth will yield the same asymptotic results and no under-smoothing is required. In other
		words, when the sample size is sufficiently large, the estimator of $\bb$ is very
		insensitive to the bandwidth. However, for small or moderate sample
		size and when the estimation of the functional form $g$ is of
		interest, then different bandwidths may yield different results. 
		In this case, a careful bandwidth selection procedure such as
		leave-one-out 
		cross-validation needs to be implemented.  
	\end{Rem}
	
	\subsection{Alternative simpler estimators}\label{sec:alter}
	
	To estimate $\bb$, 
	instead of estimating both $\u(\cdot)$ and $g(\cdot)$, we can estimate 
	only $\u(\cdot)$, as we now investigate. The basic idea is to replace
	$g$ and $g'$ using arbitrary functions, say $g^*$ and $h^*$
	respectively in the efficient score function
	construction, and solve
	\be\label{eq:simple1}
	\sumi
	\{y_i-f^*(\x_i)-z_ig^*(\bb\trans\x_i)\}
	\{z_i-E(Z\mid\x_i)\}h^*(\bb\trans\x_i)\{\x_{Li}-\wh\u(\bb\trans\x_i)\}
	=\0
	\ee
	to obtain an estimator for $\bb$. The simplest choice will be to set
	$f^*=g^*=0, h^*=1$.  Denote the estimator $\wt\bb$. To further estimate the
	function $g(\cdot)$, we can use a simple local constant estimator via solving
	\bse
	\sumi K_h\{\wt\bb\trans(\x_i-\x_0)\}
	\{y_i-f^*(\x_i)-z_i\alpha_c\}
	\{z_i-E(Z\mid\x_i)\}=0.
	\ese
	The resulting solution $\wh\alpha_c(\wt\bb,0)$ is then our estimate of $g$ at
	$\bb\trans\x_0$, i.e. $\wh
	g(\bb\trans\x_0)=\wh\alpha_c(\wt\bb,0)$.

	Likewise, we can also opt to estimate $g(\cdot)$ instead of
	$\u(\cdot)$. We can choose either to estimate $g'(\cdot)$ or to 
	make a subjective choice for it. When we estimate $g'(\cdot)$ along
	with $g(\cdot)$, 
	the procedure is the following. Solve (\ref{eq:g})
	to obtain $\wh\ba(\bb,j)$
	for $j=1, \dots, n$.
	Then solve
	\be\label{eq:simple2}
	\sumi\{y_i-f^*(\x_i)-z_i\wh\alpha_c(\bb,i)\}
	\{z_i-E(Z\mid\x_i)\}\wh\alpha_1(\bb,i)\{\x_{Li}-\u^*(\bb\trans\x_i)\}
	=\0
	\ee
	to obtain $\wh\bb$. Here 
	$\u^*$ is an arbitrarily chosen function, for example, the simplest is to set
	$\u^*=\0$.
	
	Because the procedures involved in these two simpler estimators are
	similar and are both simpler compared with the estimator described in Section
	\ref{sec:best}, we omit the details on the computational algorithms. 
	
	\section{Large sample property and inference}\label{sec:theory}
	
	\begin{Th}\label{th:best}
		Assume the regularity conditions listed in Supplementary Material
		\ref{sec:conditions} hold. 
		When $n\to\infty$, the estimator described in Section \ref{sec:best}
		satisfies the property that
		$
		\sqrt{n}(\wh\bb-\bb)\to N(\0,\A^{-1}\B{\A^{-1}}\trans)
		$ in distribution, where
		\bse
		\A&=&E\left[\var(Z\mid\X) g'^2(\bb\trans\X)\{\X_L-\u(\bb\trans\X)\}^{\otimes2}\right],\\
		\B&=&\sigma^2E[\var(Z\mid\X)g'^2(\bb\trans\X)\{\X_L-\u(\bb\trans\X)\}^{\otimes2}]\\
		&&+E[\{f(\X)-f^*(\X)\}^2\var(Z\mid\X)g'^2(\bb\trans\X)\{\X_L-\u(\bb\trans\X)\}^{\otimes2}].
		\ese
		Here and throughout the text, $\a^{\otimes2}\equiv\a\a\trans$ for a generic
		vector or matrix $\a$.
		When $f^*(\X)=f(\X)$, the estimator obtains the optimal efficiency
		bound.
	\end{Th}
	We also have similar large sample properties for the two alternative
	estimators given in Section \ref{sec:alter}, stated in the following
	Theorems \ref{th:simple1} and \ref{th:simple2}. 
	The proofs of Theorems \ref{th:best}, \ref{th:simple1} and \ref{th:simple2}  are given in the Supplementary Material.

	\begin{Th}\label{th:simple1}
		Assume the regularity conditions listed in Supplementary Material
		\ref{sec:conditions} hold. When $n\to\infty$, the estimator obtained
		from solving (\ref{eq:simple1})
		satisfies the property that $ \sqrt{n}(\wh\bb-\bb)\to N(\0,\A^{-1}\B{\A^{-1}}\trans)
		$ in distribution, 
		where 
		\be\label{eq:B1}
		\A&=&
		E \left[\var(Z\mid\X)h^{*}(\bb\trans\X) g'^{*}(\bb\trans\X)\{\X_L-\u(\bb\trans\X)\}^{\otimes2}\right] \n \\
		&&-E \left[ \var(Z\mid\X)\{g(\bb\trans\X)-g^*(\bb\trans\X)\}\frac{\partial
			h^*(\bb\trans\X)\{\X_L-\u(\bb\trans\X)\}}{\partial\bb_L\trans}\right], \n\\
		\B&=&
		\sigma^2E\left[ \var(Z\mid\X) h^{*2}(\bb\trans\X)
		\left\{\X_{L}-\u(\bb\trans\X)\right\}^{\otimes2}\right]\nonumber\\
		&&+E\left[\{f(\X)-f^*(\X)\}^2
		\var(Z\mid\X)h^{*2}(\bb\trans\X)\left\{\X_{L}-\u(\bb\trans\X)\right\}^{\otimes2}\right]\nonumber\\
		&&+E\left[Z^2\{Z-E(Z\mid\X)\}^2\{g(\bb\trans\X)-g^*(\bb\trans\X)\}^2
		h^{*2}(\bb\trans\X)\left\{\X_{L}-\u(\bb\trans\X)\right\}^{\otimes2}\right]\nonumber\\
		&&+E\left(\frac{
			\{\var(Z\mid\X)\}^2
			\{\X_{L}-\u(\bb\trans\X)\}^{\otimes2}}{\left[E\left\{\var(Z\mid\X)\mid\bb\trans\X\right\}\right]^2}\right)\nonumber\\
		&&+2E\left[Z\{Z-E(Z\mid\X)\}^2
		\{f(\X)-f^*(\X)\}\{g(\bb\trans\X)-g^*(\bb\trans\X)\}
		h^{*2}(\bb\trans\X)\left\{\X_{L}-\u(\bb\trans\X)\right\}^{\otimes2}
		\right]\nonumber\\
		&&-2E\left[
		\frac{
			\{\var(Z\mid\X)\}^2
			\{g(\bb\trans\X)-g^*(\bb\trans\X)\}
			h^*(\bb\trans\X)\left\{\X_{L}-\u(\bb\trans\X)\right\}^{\otimes2}
		}{E\left\{\var(Z\mid\X)\mid\bb\trans\X\right\}}\right].
		\ee
		Here, $g'^*(\cdot)$ is the first derivative of $g^*(\cdot)$. 
	\end{Th}
	Interestingly, even when $f^*(\X)=f(\X)$,
	$g^*(\bb\trans\X)=g(\bb\trans\X)$ and
	$h^*(\bb\trans\X)=g'(\bb\trans\X)$, the estimator still does not
	achieve the optimal efficiency bound. This is in stark contrast with
	the proposed estimator, which is optimal as long as $f^*(\X)=f(\X)$.

	\begin{Th}\label{th:simple2}
		Assume the regularity conditions listed in Supplementary Material
		\ref{sec:conditions} hold.
		When $n\to\infty$, the estimator obtained
		from solving (\ref{eq:simple2})
		satisfies the property that $
		\sqrt{n}(\wh\bb-\bb)\to N(\0,\A^{-1}\B{\A^{-1}}\trans)
		$ in distribution, 
		where 
		\bse
		\A&=&
		E\left[\var(Z\mid\X)
		g'^2(\bb\trans\X)\{\X_{L}-\u^*(\bb\trans\X)\}\X_L\trans
		\right],  
		\ese
		\be\label{eq:B2}
		\B&=&\sigma^2
		E\left[
		\var(Z\mid\X)g'^2(\bb\trans\X)\{\X_{L}-\u(\bb\trans\X)\}^{\otimes2}\right]\nonumber\\
		&&+
		E\left[\{f(\X)-f^*(\X)\}^2
		\var(Z\mid\X)g'^2(\bb\trans\X)\{\X_{L}-\u(\bb\trans\X)\}^{\otimes2}\right].
		\ee
		When $f^*(\X)=f(\X)$ and $\u^*(\bb\trans\X)=\u(\bb\trans\X)$, the estimator is efficient. 
	\end{Th}
	
	\section{Estimation and Inference when $Z$ is continuous}\label{sec:cont}
	When $Z$ is continuous, typically representing the dosage, the treatment effect of the form
	$Zg(\bb\trans\X)$ may not be adequate. The only conclusion that
	can be drawn from such a model is that if $g(\bb\trans\X)$ is
	positive, the largest dosage $Z$ should be selected and when it is
	negative, the smallest dosage should be taken. More useful models in
	this case include a quadratic treatment effect
	$Zg_1(\bb\trans\X)+Z^2g_2(\bb\trans\X)$, a cubic model
	$Zg_1(\bb\trans\X)+Z^2g_2(\bb\trans\X)+Z^3g_3(\bb\trans\X)$, or some
	other nonlinear model of $Z$. For example, if the quadratic model is used, 
	a new patient with covariate $X_0$ and $g_2(\bb\trans\X_0)<0$ would
	have the best treatment result  
	if $Z_0=-g_1(\bb\trans\X_0)/\{2g_2(\bb\trans\X_0)\}$. We thus consider
	a general polynomial model of the form  
	\be\label{eq:modelcont}
	Y=f(X)+\sum_{k=1}^KZ^kg_k(\bb\trans\X)+\epsilon,
	\ee
	while keep all other aspects of the model assumption identical to that
	of (\ref{eq:modelbin}). Similar to (\ref{eq:pdfbin}), the pdf in this
	case is
	\be\label{eq:pdfcont}
	f_{\X,Z,Y}(\x,z,y,\bb,\sigma,g,f,\eta)=\eta(\x)f_{Z\mid\X}(\x,z)\sigma^{-1}\phi[\sigma^{-1}\{y-f(\x)-\sum_{k=1}^Kz^kg_k(\bb\trans\x)\}],
	\ee
	and it has efficient score
	$
	\S\eff(\x,z,y,\bb,g,f,\sigma)
	=\sigma^{-2}\epsilon 
	\left\{\u
	-\sum_{r=1}^K W_r 
	E\left(W_r\U
	\mid\bb\trans\x\right)\right\},
	$ where $
	\U=\sum_{k=1}^Kg_{k1}(\bb\trans\x) 
	\left\{ Z^k-E(Z^k\mid\X)\right\}\x_L$. Here $g_{k1}(\cdot)$ is the first derivative of $g_k(\cdot)$ for
	$k=1, \dots, K$, 
	$\epsilon\equiv y-f(\x)-\sum_{k=1}^Kz^kg_k(\bb\trans\x)$. Further, 
	$W_1, \dots, W_K$ are linear transformations of $Z-E(Z\mid\X), \dots
	Z^K-E(Z^K\mid\X)$, i.e. $\A(\bb\trans\X)(W_1, \dots, W_K)\trans=
	\{Z-E(Z\mid\X), \dots
	Z^K-E(Z^K\mid\X)\}\trans$, such that
	$E(W_jW_k\mid\bb\trans\X)=I(j=k)$, i.e. they form a set of orthonormal bases. Supplementary Material \ref{sec:conteff} 
	provides the detailed construction of $W_k$'s and the derivation of the
	efficient score function. 
	Naturally, based on the form of the efficient score function and our
	experience in the binary $Z$ case, we propose a general class
	of  consistent estimating equations for $\bb$ as
	\be\label{eq:contbeta}
	\sumi \{y_i-f^*(\x_i)-\sum_{k=1}^Kz_i^k \wh g_k(\bb\trans\x_i)\}
	\{\wh\u_i
	-\sum_{r=1}^K w_{ri} 
	\wh E(W_r\wh\U
	\mid\bb\trans\x_i)\}
	=\0,
	\ee
	where $\wh\U=\sum_{k=1}^K\wh g_{k1}(\bb\trans\x) 
	\{ Z^k-E(Z^k\mid\X)\}\x_L$, $f^*$ is an arbitrary function of $\x$. Here at any $\x_0=\x_1, \dots,
	\x_n$ and any $\bb$, $\wh g_k(\bb\trans\x_0,\bb), \wh g_{k1}(\bb\trans\x_0,\bb)$ are
	nonparametric profiling estimators obtained from
	\be\label{eq:contalpha}
	\sumi w_{i0}[y_i-f^*(\x_i)-\sum_{k=1}^Kz_i^k \{\alpha_{kc}+\alpha_{k1}\bb\trans(\x_i-\x_0)\}]
	\v_i
	=\0,
	\ee
	where 
	$
	\V=\left[\left\{ Z-E(Z\mid\X)\right\}(1, \bb\trans\X-\bb\trans\x_0), \dots, 
	\left\{ Z^K-E(Z^K\mid\X)\right\}(1, \bb\trans\X-\bb\trans\x_0)\right]\trans$.
	
	Similar to the binary $Z$ case, we also have the standard root-$n$
	consistency and local efficiency of $\wh\bb$ for continuous $Z$. We
	state the results in Theorem \ref{th:contbest} while skipping the proof
	because it is almost identical to that of Theorem \ref{th:best}. 
	
	\begin{Th}\label{th:contbest}
		Assume the regularity conditions listed in Supplementary Material
		\ref{sec:conditions} hold. 
		When $n\to\infty$, the estimator described in (\ref{eq:contbeta}) and (\ref{eq:contalpha})
		satisfies the property that $\sqrt{n}(\wh\bb-\bb)\to N(\0,\A^{-1}\B\A^{-1})$ in distribution, 
		where
		\bse
		\A&=&\sigma^2 E\left[
		\{\U
		-\sum_{r=1}^K W_r 
		E(W_r\U
		\mid\bb\trans\X)\}^{\otimes2}\right]
		\ese
		\bse
		\B&=&E\left([\sigma^2+\{f(\X)-f^*(\X)\}^2]
		\{\U
		-\sum_{r=1}^K W_r 
		E(W_r\U
		\mid\bb\trans\X)\}^{\otimes2}\right).
		\ese
		Additionally, when $f^*(\X)=f(\X)$, the estimator obtains the optimal efficiency
		bound.
	\end{Th}

	\section{Estimation and Inference when $Z$ is categorical}\label{sec:cat}
	
	To make the analysis complete, we now consider the case where $Z$ is
	categorical. This arises naturally in practice when several different
	treatment arms are compared. Assume we consider $K$ treatment arms
	in addition to the control arm. Thus, we have $K$ binary
	variables, $Z_1, \dots, Z_K$, each takes value 0 or 1. Note that at
	most one of the $Z_k, k=1, \dots, K$ values can be 1 in each observation.
	A sensible model in this scenario is
	\be\label{eq:modelcat}
	Y=f(X)+\sum_{k=1}^KZ_kg_k(\bb\trans\X)+\epsilon,
	\ee 
	and the pdf is
	\be\label{eq:pdfcat}
	f_{\X,\Z,Y}(\x,\z,y,\bb,\sigma,g,f,\eta)=\eta(\x)f_{\Z\mid\X}(\x,\z)\sigma^{-1}\phi[\sigma^{-1}\{y-f(\x)-\sum_{k=1}^Kz_kg_k(\bb\trans\x)\}],
	\ee
	where $\Z=(Z_1, \dots, Z_k)\trans$.
	The corresponding efficient score is
	$
	\S\eff(\x,\z,y,\bb,g,f,\sigma)
	=\sigma^{-2}$\\$\epsilon 
	\left\{\u
	-\sum_{r=1}^K W_r 
	E\left(W_r\U
	\mid\bb\trans\x\right)\right\},
	$ where $
	\U=\sum_{k=1}^Kg_{k1}(\bb\trans\x) 
	\left\{ Z_k-E(Z_k\mid\X)\right\}\x_L$.
	Here $g_{k1}(\cdot)$ is the first derivative of $g_k(\cdot)$ for
	$k=1, \dots, K$, 
	$\epsilon\equiv y-f(\x)-\sum_{k=1}^Kz_kg_k(\bb\trans\x)$. Further, 
	$W_1, \dots, W_K$ are linear transformations of $Z_1-E(Z_1\mid\X), \dots
	Z_K-E(Z_K\mid\X)$, i.e. $\A(\bb\trans\X)(W_1, \dots, W_K)\trans=
	\{Z-E(Z\mid\X), \dots
	Z_K-E(Z_K\mid\X)\}\trans$, such that
	$E(W_jW_k\mid\bb\trans\X)=I(j=k)$, i.e. they form a set of orthonormal
	bases. We can see that the efficient score has much resemblance with
	the case for the continuous $Z$ case, except that we are now treating
	a multivariate $\Z$. We show the derivation of the efficient score in  
	Supplementary Material \ref{sec:cateff}, where the detailed construction of $W_k$'s
	are also given.  Similar to the continuous univariate $Z$ case, we propose a general class
	of  consistent estimating equations for $\bb$ as
	\be\label{eq:catbeta}
	\sumi \{y_i-f^*(\x_i)-\sum_{k=1}^Kz_{ki} \wh g_k(\bb\trans\x_i)\}
	\{\wh\u_i
	-\sum_{r=1}^K w_{ri} 
	\wh E(W_r\wh\U
	\mid\bb\trans\x_i)\}
	=\0,
	\ee
	where $
	\wh\U=\sum_{k=1}^K\wh g_{k1}(\bb\trans\x) 
	\{ Z_k-E(Z_k\mid\X)\}\x_L$, $f^*$ is an arbitrary function of $\x$. Here at any $\x_0=\x_1, \dots,
	\x_n$ and any $\bb$, $\wh g_k(\bb\trans\x_0,\bb), \wh g_{k1}(\bb\trans\x_0,\bb)$ are
	nonparametric profiling estimators obtained from
	\be\label{eq:catalpha}
	\sumi w_{i0}[y_i-f^*(\x_i)-\sum_{k=1}^Kz_{ki} \{\alpha_{kc}+\alpha_{k1}\bb\trans(\x_i-\x_0)\}]
	\v_i
	=\0,
	\ee
	where 
	$\V=\left[\left\{ Z_1-E(Z_1\mid\X)\right\}(1, \bb\trans\X-\bb\trans\x_0), \dots, 
	\left\{ Z_K-E(Z_K\mid\X)\right\}(1, \bb\trans\X-\bb\trans\x_0)\right]\trans$.
	Similarly, we summarize th asymptotic properties of $\wh\bb$ in
	Theorem \ref{th:catbest} 
	and skip the proof
	because of its similarity to that of Theorem \ref{th:best}. 
	
	\begin{Th}\label{th:catbest}
		Assume the regularity conditions listed in Supplementary Material
		\ref{sec:conditions} hold. 
		When $n\to\infty$, the estimator described in (\ref{eq:catbeta}) and (\ref{eq:catalpha})
		satisfies the property that $ \sqrt{n}(\wh\bb-\bb)\to N(\0,\A^{-1}\B{\A^{-1}}\trans)$ in distribution, 
		where 
		\bse
		\A&=&\sigma^2 E\left[
		\{\U
		-\sum_{r=1}^K W_r 
		E(W_r\U
		\mid\bb\trans\X)\}^{\otimes2}\right]\\
		\B&=&E\left([\sigma^2+\{f(\X)-f^*(\X)\}^2]
		\{\U
		-\sum_{r=1}^K W_r 
		E(W_r\U
		\mid\bb\trans\X)\}^{\otimes2}\right).
		\ese
		Additionally, when $f^*(\X)=f(\X)$, the estimator obtains the optimal efficiency
		bound.
	\end{Th}
	
	\section{Simulation studies}\label{sec:simulation}
	
	We conduct six sets of simulation experiments, covering all the
	scenarios discussed so far, to investigate the finite sample
	performance of the methods proposed in Section \ref{sec:method}. The
	results under various scenarios reflect the superior performance of
	Method I and show reasonably accurate inference results as
	well. We also compute the percentage of making correct
	decisions based on $g(\cdot)$ functions. In most of the scenarios,
	we are able to make correct decisions more than 90\% of the times. In each
	experiment, we simulate 500 data sets. 
	
	\subsection{Simulation 1}\label{sec:simu1}
	
	In the first experiment, we consider a relatively simple setting. 
	Specifically, we let the covariate $\X_i=(X_{i1},X_{i2})\trans$, where
	$X_{i1}$ and $X_{i2}$ are independently generated from a uniform
	distribution on $(0,1)$. We set $f(\X_i)=0.05(X_{i1}+X_{i2})$ and
	$g(\bb\trans\X_i)=\exp(1.5\bb\trans\X_i)-1$,  where
	$\bb=(1,-1)\trans$. Note that the function $g$ is monotone. We
	generated the model error $\epsilon_i$ from a centered normal
	distribution with standard error 0.3, and
	generated the treatment indicator $Z_i$ from a Bernoulli
	distribution with probability 0.5. 
	
	We implemented four methods for estimating the unknown parameter $\bb$
	for comparison. Method I is the one given in (\ref{eq:best}). Method II is the 
	alternative method proposed in (\ref{eq:simple1}),  where we let
	$f^*=g^*=0, h^*=1$. Method III corresponds to (\ref{eq:simple2}) and
	Method IV refers to the  
	method proposed in Section \ref{sec:basic}. To ensure identifiability, we fix
	the first component of $\bb$ to be 1. 
	
	From the results reported in Table \ref{table1}, we can observe that
	while all the estimation methods have very small biases and standard
	errors, Method I performed significantly better than all other methods
	in terms of both estimation bias and the standard errors. This is
	within our expectation because Method I is a locally efficient
	estimator. In addition, in our experience, the computation time for
	all four methods are also comparable. Thus we recommend using Method I
	due to its superior performance. Consequently, we focus on Method I to
	proceed with the inference of $\bb$ and the estimation and inference
	of $g(\cdot)$.  
	
	Although Theorem \ref{th:best} provides an explicit form of the
	asymptotic variance, the resulting form contains $f(\X)$, whose estimation is
	subject to the curse of dimensionality and we have successfully
	avoided so far. Thus, to proceed with inference on $\bb$, we evaluate the
	coverage probabilities of the 95\% confidence regions for $\bb$ resorting
	to the bootstrap method. The coverage probabilities for $\beta_2$ are
	reported in the upper part of Table \ref{table_SIM1} under Method
	I. All results are reasonably close to the nominal level. To evaluate the accuracy of the subsequent 
	treatment assignment rule based on $\mbox{sign}\{\hat
	g(\widehat{\bb}\trans\X)\}$, we calculate the percentage of making
	correct decisions, defined as 
	${\rm PCD}=1-n^{-1}\sumi|I\{\wh g(\wh\bb\trans\X_i)>0\}-I\{g(\bb\trans\X_i)>0\}|$.
	We also evaluated the value functions via
	\bse
	{\rm VF}=\frac{1}{n}\sumi 
	\frac{Y_iI[Z_i=I\{\wh g(\wh\bb\trans\X_i)>0\}]}{\pr[Z_i=I\{\wh
		g(\wh\bb\trans\X_i)>0\}|\X_i]}. 
	\ese
	In this particular simulation, due to covariate-independent randomization,
	${\rm VF}=(2/n)$ \\$\sumi Y_i I[I\{\wh g(\wh\bb\trans\X_i)>0\}]$. The values of PCD and VF and their standard errors are reported in
	the upper part of Table \ref{table_SIM1} under Method I. 
	
	For comparison, we consider the semiparametric 
	single index model (SIM) approach by \cite{Single13} to estimate the
	optimal individualized treatment strategy. Results are provided in
	Table \ref{table_SIM1}. We can clearly see
	that our proposed Method I performs much better in that it 
	yields smaller bias in estimating 
	parameter $\beta_2$, PCD and value function, compared to the SIM
	approach. The coverage probability of estimating $\beta_2$ using SIM
	approach does not achieve the nominal level of 95\%. The estimation
	variability of our Method I is also less than SIM method.
	
	\begin{table}[!ht]
		\begin{center}
			\caption{Bias and standard deviations (SD) for the estimation of $\beta_2$ in Simulation 1.}
			\vspace{-0.2in}
			\begin{tabular}{lcc|cc}
				\\\hline
				Methods  &\multicolumn{2}{c}{I}  & \multicolumn{2}{c}{II} \\ \cline{2-5}
				$n$       &  Bias  & SD  & Bias  & SD     \\
				$600$      & -0.0005  & 0.0737  &  -0.0363 & 0.1586  \\
				$800$      & -0.0004  & 0.0642  &  -0.0259 & 0.1195  \\
				$1000$    & -0.0003  & 0.0543  &  -0.0229 & 0.0736   \\ \hline
				Methods  &\multicolumn{2}{c}{III}  &\multicolumn{2}{c}{IV} \\ \cline{2-5}
				$600$      & -0.1028 & 0.1611  & 0.0506 & 0.1518 \\
				$800$      & -0.0850 & 0.1357  & 0.0571 & 0.1153 \\
				$1000$    & -0.0667 & 0.1049  & 0.0526 & 0.1004  \\ \hline
			\end{tabular}
			\label{table1}
		\end{center}
	\end{table}
	
	\begin{table}[!ht]
		\begin{center}
			\caption{Estimation and inference results for $\beta_2$ in Simulation 1 using Method I and SIM approach \citep{Single13}. We report mean of estimated single index coefficient biases, standard errors (SE) of estimated single index coefficients, estimation results for PCD and Value function over 500 replications with their empirical standard deviations (SD). True value function is 0.4314.}
			\begin{tabular}{lccccccccc}
				\hline
				$n$ & Bias($\beta_2$) & SD($\beta_2$)& SE($\beta_2$) & CP($\beta_2$)  & PCD & SD(PCD) & $\evf$ & SD($\evf$) \\ \hline
				\multicolumn{9}{c}{Method I} \\ \hline
				$600$ & -0.0005 & 0.0738 & 0.1180 &  0.978& 0.9011& 0.0191 & 0.4257 &  0.0435\\
				$800$ & -0.0004 & 0.0643  & 0.0953 &  0.950 & 0.9013 & 0.0153 & 0.4261 & 0.0371\\
				$1000$&-0.0003&  0.0544 & 0.0614 & 0.946 & 0.9052&0.0150 & 0.4251&0.0372\\\hline
				\multicolumn{9}{c}{SIM Method} \\ \hline
				$600$ & -0.0152  & 0.1555 & 0.0527 & 0.558  & 0.9489 & 0.0404& 0.2837 & 0.0295 \\
				$800$ & 0.0052 & 0.1150 & 0.0441 & 0.528  & 0.9571 & 0.0312 & 0.2855 & 0.0245 \\
				$1000$ & -0.0078 & 0.1119 & 0.0400 & 0.584  & 0.9618 & 0.0311 & 0.2842 & 0.0206\\\hline
			\end{tabular}
			\label{table_SIM1}
		\end{center}
	\end{table}
	\vspace{-0.2in}
	\subsection{Simulation 2}\label{sec:simu2}
	In the second experiment, 
	the covariate $\X_i=(X_{i1},X_{i2}, X_{i3})\trans$,
	where $X_{i1}, X_{i2}, X_{i3}$ are also independently generated from a
	uniform(0,1) distribution. We let
	$f(\X_i)=0.05(X_{i1}+X_{i2}+X_{i3})$,
	$g(\bb\trans\X_i)=\sin(\pi\bb\trans\X_i)$ with
	$\bb\trans=(1,-1,2)$. Thus, the $g$ function here is both nonlinear
	and non-monotone. The regression errors  $\epsilon_i$ and the
	treatment indicator $Z_i$ are generated identically as in Simulation
	1.
	
\begin{figure}[!ht]
		\begin{center}
			\includegraphics[width=0.32\textwidth,height=0.35\textheight]{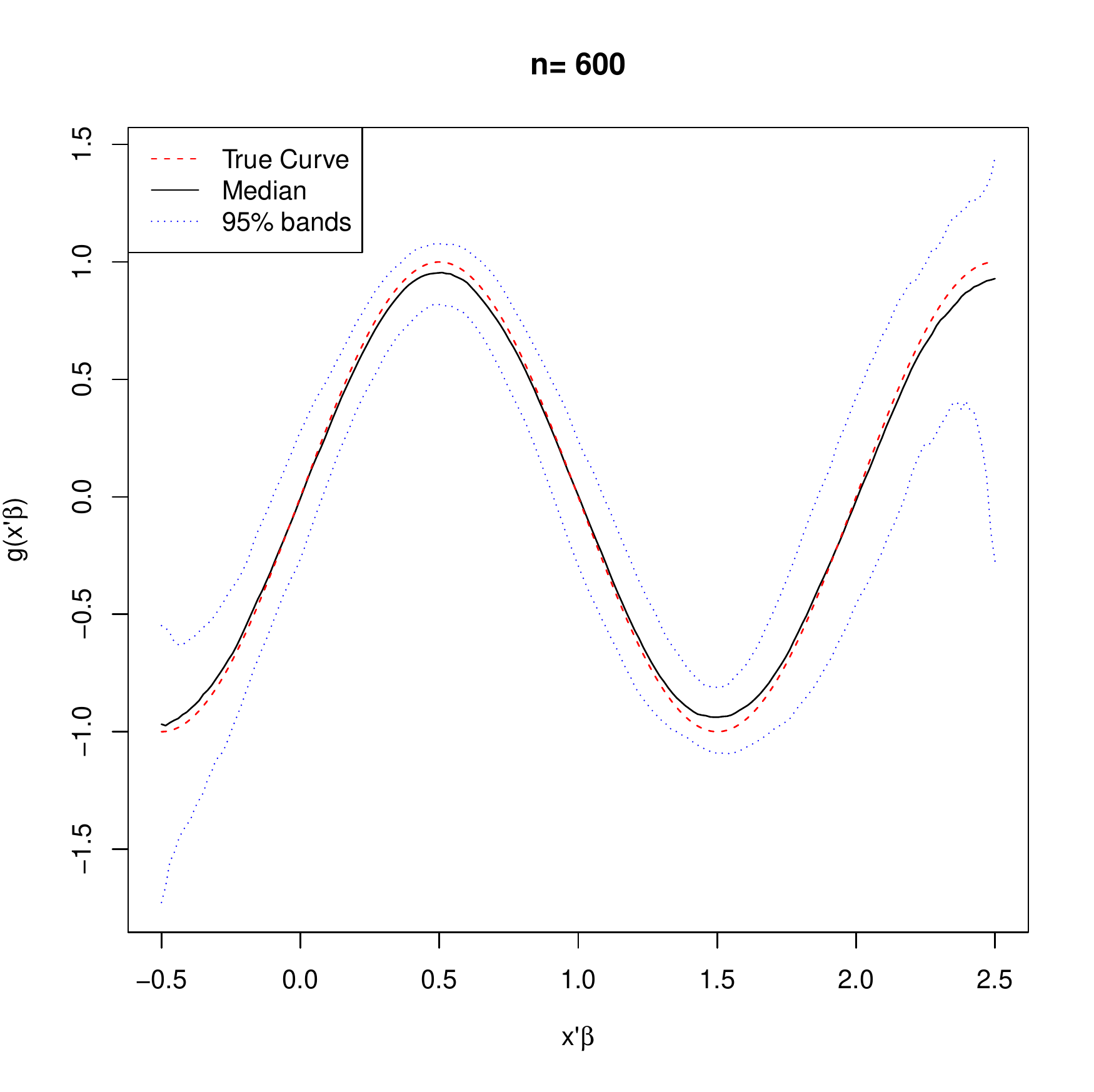}
			\includegraphics[width=0.32\textwidth,height=0.35\textheight]{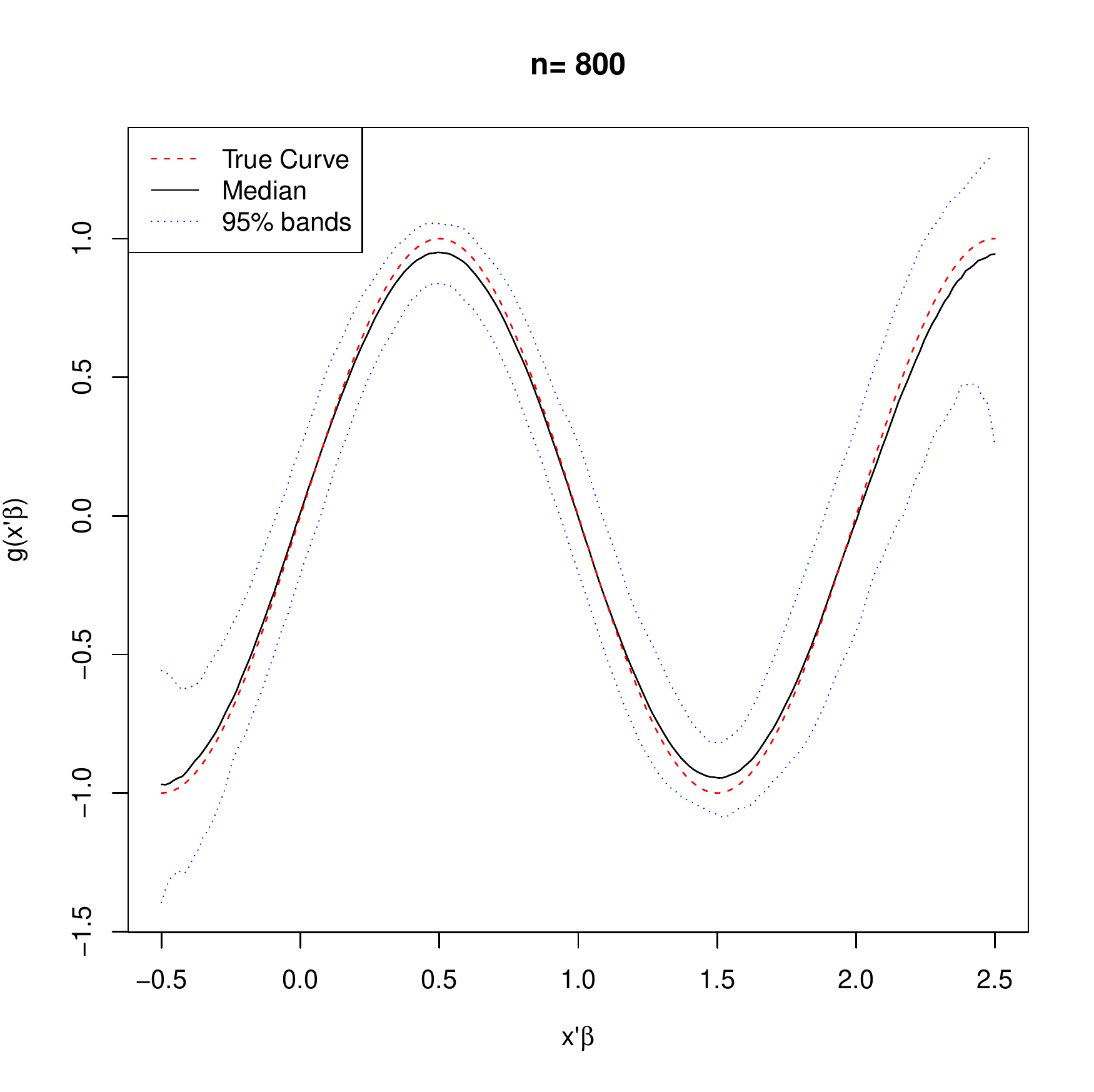}
			\includegraphics[width=0.32\textwidth,height=0.35\textheight]{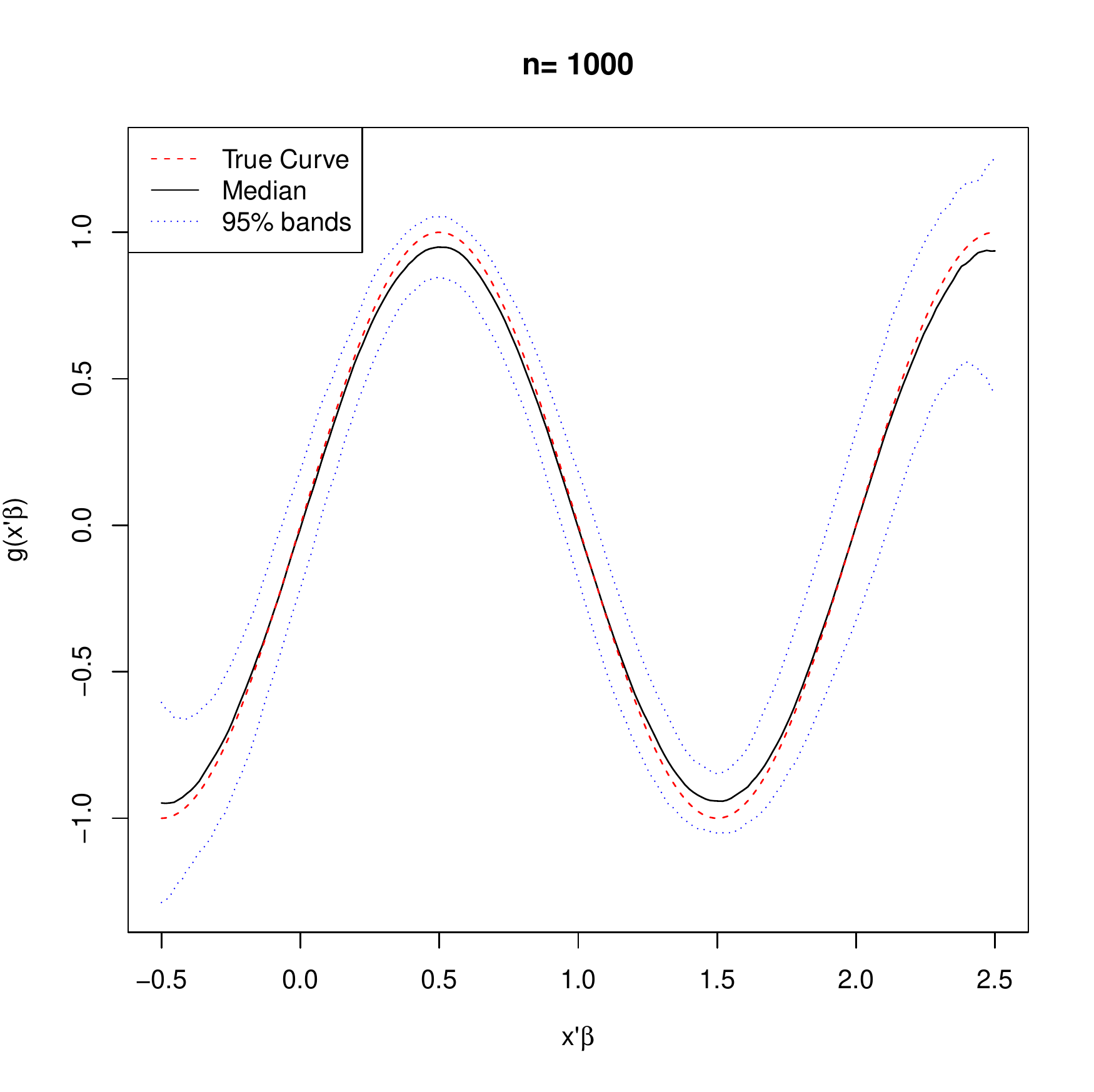}
			\caption{\footnotesize{Estimation of $g(\cdot)$, $g(\x'\bb)=\sin(\pi \x'\bb)$ in Simulation 2.}}
			\label{figure1}
		\end{center}
	\end{figure}

	Similar to Simulation 1, we implemented the four methods to compare
	the estimation of unknown parameter $\bb$, while fixing the first
	component of $\bb$ at 1. The estimation bias and standard errors are
	reported in Table \ref{table2}. From the results, we can again
	conclude that Method I performed significantly better than all other
	methods, even though all methods have acceptable bias and standard
	errors. We computed the coverage probabilities of the 95\% bootstrapped
	confidence regions for $\bb$ for Method I based on the depth function
	\citep{Liu1999depth} to accommodate the two dimensional parameter. The
	results are reported in Table \ref{table_SIM2}. All results are reasonably
	close to the nominal level. 
	We also reported the mean and standard
	deviations of VF and PCD in the lower part of Table \ref{table_SIM2}
	under Method I. To further check the performance, we plot the median,
	95\% confidence bands for the
	estimation of $g(\cdot)$ in Figure \ref{figure1} and the performance is satisfactory.
	We also reported the results obtained by the SIM
	approach \citep{Single13} in Table \ref{table_SIM2}. As expected,
	our proposed Method I performs better compared to SIM method, even
	when the treatment effect function is not monotone. Our proposed
	Method I has smaller bias and less estimation variability.
	
	\begin{table}[!ht]
		\begin{center}
			\caption{Bias and standard deviation (SD) for the estimation of $(\beta_2,\beta_3)$ in Simulation 2.}
			\vspace{-0.2in}
			\begin{tabular}{lcccc}
				\\\hline
				Methods  &\multicolumn{4}{c}{I} \\ \cline{2-5}
				$n$     &  Bias  & SD & Bias & SD \\
				$600$      &-0.0048  &0.0604 &  0.0059  &  0.0979 \\
				$800$      &-0.004  & 0.0603 &  0.0169  & 0.0816 \\
				$1000$    &-0.0017  & 0.0475 & 0.0028 & 0.0754 \\ \hline
				Methods  &\multicolumn{4}{c}{II} \\ \cline{2-5}
				$600$     &  -0.0276 &  0.1831 &  0.0796 &0.2685 \\
				$800$     &  -0.0205 &  0.1547 &  0.0334 &0.2065 \\
				$1000$   &  -0.0351 & 0.1414 &  0.0547 &0.2053\\ \hline
				Methods  & \multicolumn{4}{c}{III}  \\ \cline{2-5}
				$600$      &-0.0413 &0.0835 &  0.0886  &0.1438 \\
				$800$      &-0.0413 &0.0642 &  0.0790 & 0.0985 \\
				$1000$    &-0.0276 &0.0551 &  0.0687  &0.0919 \\ \hline
				Methods  &\multicolumn{4}{c}{IV} \\ \cline{2-5}
				$600$     &   -0.0512 &0.0906 &  0.0930 & 0.1442 \\
				$800$     &   -0.0432 &0.0788 &  0.0708  &0.1215\\
				$1000$   &   -0.0287 &0.0658 &  0.0551  &0.1071\\ \hline
			\end{tabular}
			\label{table2}
		\end{center}
	\end{table}
	\begin{table}[!ht]
		\begin{center}
			\caption{Estimation and inference results for $\beta_2$ in Simulation 2 using SIM approach \citep{Single13}. Other caption is same as Table \ref{table_SIM1}. True value function is 0.3937.}
			\begin{tabular}{lcccccc}
				\hline
				$n$ & Bias($\beta_2$) &  SD($\beta_2$) & SE($\beta_2$) & Bias($\beta_3$) & SD($\beta_3$) & SE($\beta_3$) \\ \hline
				&\multicolumn{6}{c}{Method I}\\\cline{1-7}
				$600$ & -0.0048 & 0.0604  & 0.0971 & 0.0059 & 0.0979 &  0.1456 \\ 
				$800$ & -0.0040 & 0.0603  & 0.0711 & 0.0169 & 0.0816 &  0.1120 \\
				$1000$& -0.0017& 0.0475  & 0.0567 & 0.0028 & 0.0754 &  0.0961 \\\hline
				&\multicolumn{6}{c}{SIM Method}\\\cline{1-7}
				$600$ & -0.0016 &0.2754& 0.1565 & 0.0674 & 0.5060& 0.4961\\
				$800$ & -0.0134 &0.2344& 0.1281 & 0.0523 &0.3844& 0.4296 \\
				$1000$ & -0.0099 &0.2076&0.1123 & 0.0446&0.3490 & 0.3836 \\\hline
				$n$ & CP($\beta_2$) & CP($\beta_3$) & PCD & SD(PCD) & $\evf$ & SD($\evf$)  \\ \hline
				&\multicolumn{6}{c}{Method I}\\\cline{1-7}
				$600$ & 0.972 & 0.976 &0.975 & 0.010 & 0.3964 & 0.0328\\
				$800$ & 0.974 & 0.982 & 0.978& 0.009 & 0.3961& 0.0281\\
				$1000$ & 0.984  & 0.982 & 0.980 & 0.007 & 0.3942& 0.0243\\ \hline
				&\multicolumn{6}{c}{SIM Method}\\\cline{1-7}
				$600$& 0.778 & 0.934 & 0.9554 & 0.0223& 0.7215 & 0.0270 \\
				$800$& 0.75 & 0.932 &0.9601 & 0.0203 & 0.7188 & 0.0242 \\
				$1000$&0.716 & 0.95 & 0.9608 & 0.0192 & 0.7201 & 0.0213 \\ \hline
			\end{tabular}
			\label{table_SIM2}
		\end{center}
	\end{table}
	\subsection{Simulation 3}\label{simu3}
	In the third simulation, the covariate
	$\X_i=(X_{i1},X_{i2})\trans$ is generated from bivariate normal
	distribution with zero mean and identity covariance matrix. Here,
	$g(\bb\trans\X_i)=2\bb\trans\X_i$ with $\bb=(1,-1)\trans$. The 
	regression errors  $\epsilon_i$, the baseline model $f(\X_i)$ and the
	treatment indicator $Z_i$ are generated similarly as in Simulation 1. 
	
\begin{figure}[!ht]
		\begin{center}
			\includegraphics[width=0.32\textwidth,height=0.35\textheight]{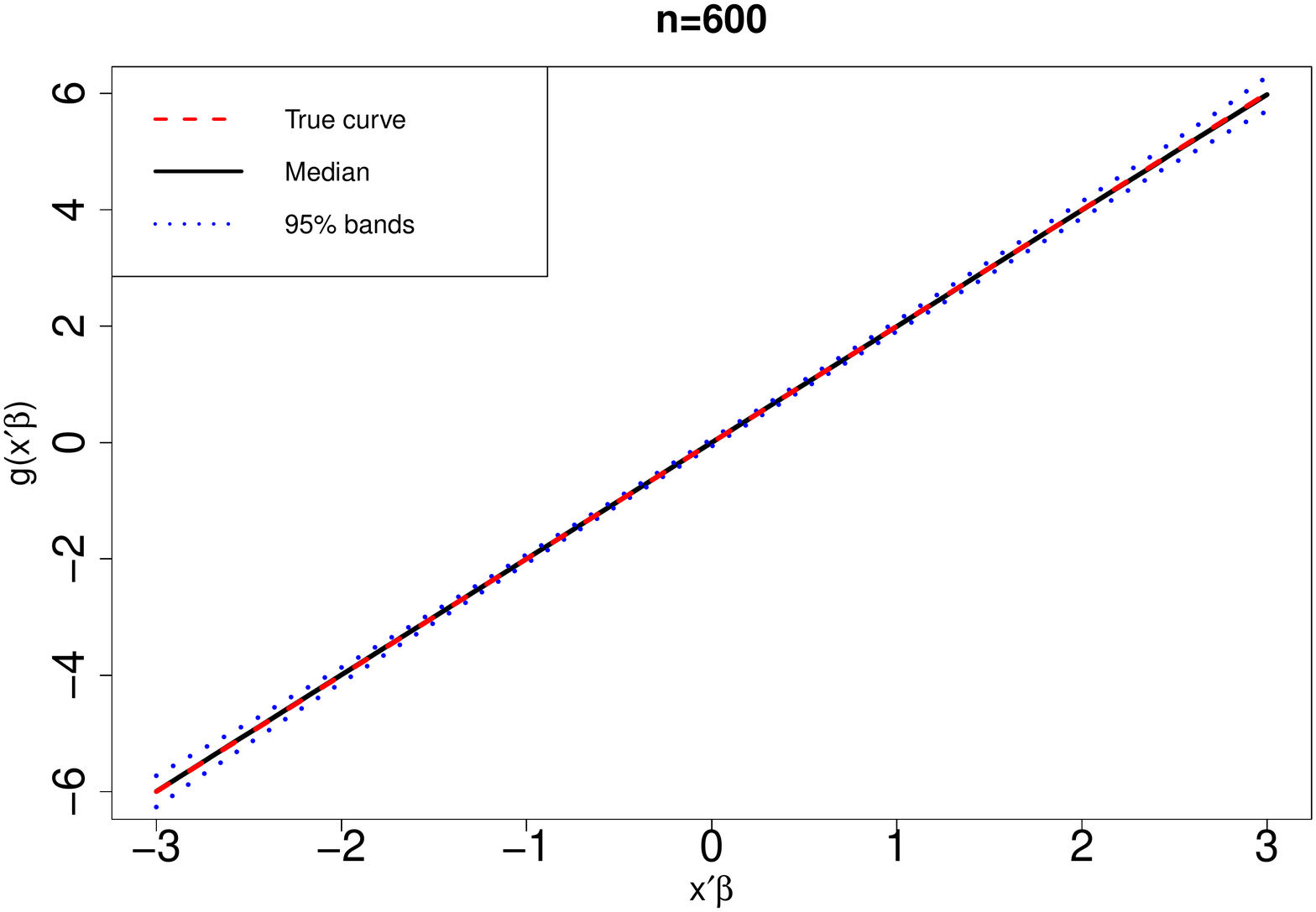}
			\includegraphics[width=0.32\textwidth,height=0.35\textheight]{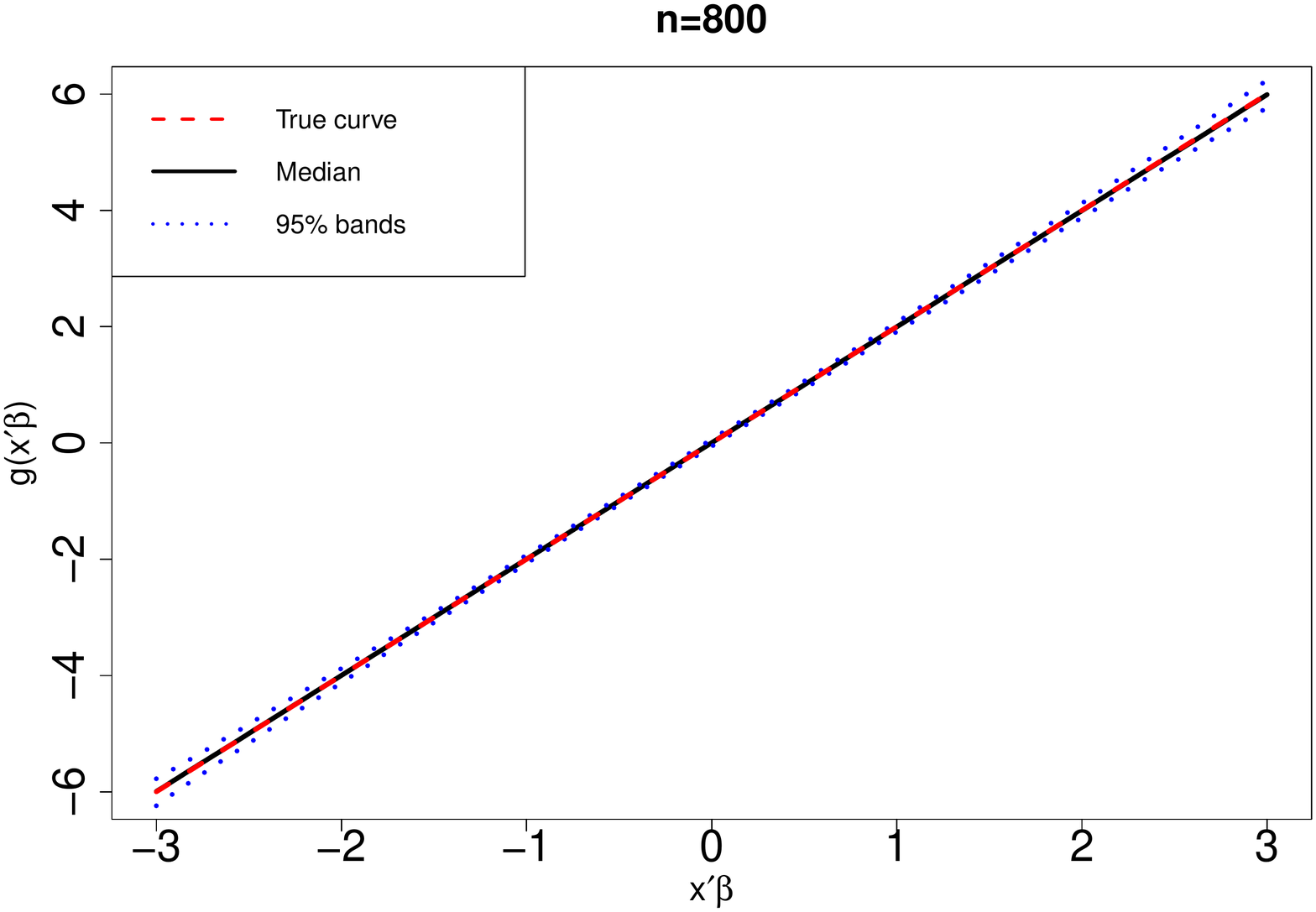}
			\includegraphics[width=0.32\textwidth,height=0.35\textheight]{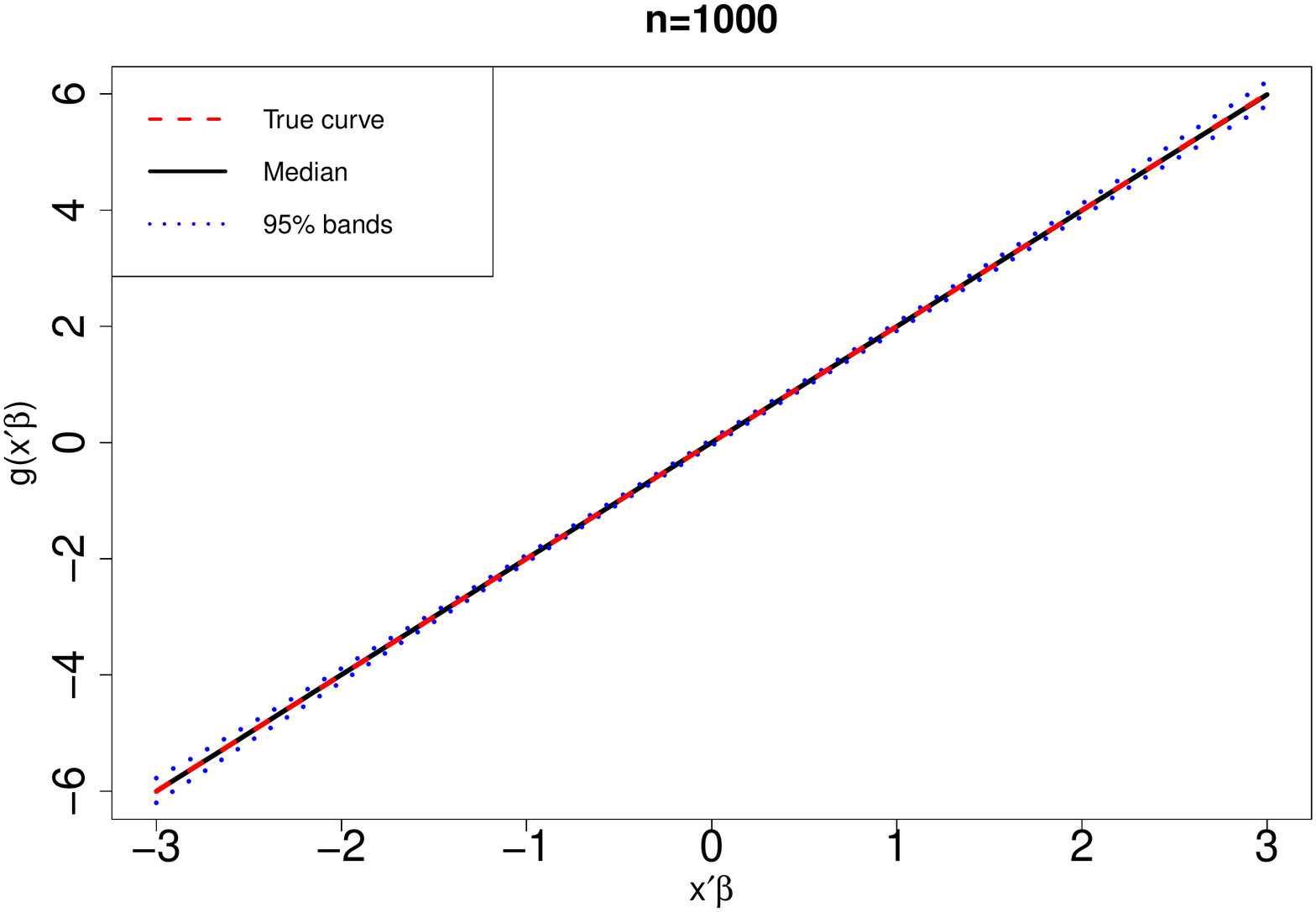}
			\caption{\footnotesize{Estimation of $g(\cdot)$, when $g(\x'\beta)=\x'\bb$ in Simulation 3.}}
			\label{figure_lin}
		\end{center}
	\end{figure}

	We summarized the estimation bias and standard errors of unknown
	parameter $\bb$ in Table \ref{table_lin_1}, obtained from four
	methods. As usual, Method I performed significantly better compared
	to other methods. In the upper part of Table \ref{table_SIM3}, we reported the
	coverage probability of the 95\% bootstrap confidence intervals of $\bb$,
	mean and standard deviations for VF and PCD
	for Method I. We also computed the 95\% confidence band and median for
	$g(\cdot)$ in Figure \ref{figure_lin}. For comparison,
	we summarized the results via SIM 
	\citep{Single13} in the lower part of Table \ref{table_SIM3}. We
	observe that our proposed Method I yields smaller bias and
	estimation variability.
	
	\begin{table}[!ht]
		\begin{center}
			\caption{Bias and standard deviation (SD) for the estimation of $\beta_2$ in Simulation 3.}
			\vspace{-0.2in}
			\begin{tabular}{lcc|cc}
				\\\hline
				Methods  &\multicolumn{2}{c}{I}  & \multicolumn{2}{c}{II} \\ \cline{2-5}
				$n$     &  Bias &SD &  Bias &SD\\
				$600$      &-0.0015&0.0177&-0.0056&0.1046\\
				$800$      &-0.0013&0.0153&-0.0056&0.0921\\
				$1000$    &-0.0004&0.0143&-0.0036&0.0824\\ \hline
				Methods  & \multicolumn{2}{c}{III}  &\multicolumn{2}{c}{IV} \\ \cline{2-5}
				$600$      &0.0028&0.1340&0.0214&0.1415\\
				$800$      &-0.0058&0.1147&0.0205&0.1317\\
				$1000$    &0.0022&0.1038&0.0145&0.1136\\ \hline
			\end{tabular}
			\label{table_lin_1}
		\end{center}
	\end{table}
	
	\begin{table}[!ht]
		\begin{center}
			\caption{Estimation and inference results for $\beta_2$ in Simulation 3 using Method I and SIM approach. Other caption is same as Table \ref{table_SIM1}.  True value function is 1.1306.}
			\begin{tabular}{lcccccccc}
				\hline
				n & Bias($\beta_2$) & SD($\beta_2$)& SE($\beta_2$) & CP($\beta_2$)  & PCD & SD(PCD) & $\evf$ & SD($\evf$) \\ \hline
				\multicolumn{9}{c}{Proposed Method I} \\ \hline
				600 & -0.0015 & 0.0177 & 0.0194 & 0.972 & 0.9953 & 0.0040 & 1.1264 & 0.1071 \\
				800 & -0.0013 & 0.0153 & 0.0168 & 0.96 & 0.9958 & 0.0033 & 1.1279 & 0.0929\\
				1000& 0.0004 & 0.0143 & 0.0147 & 0.956 & 0.9960 & 0.0030 & 1.1256 & 0.0827 \\ \hline
				\multicolumn{9}{c}{SIM Method} \\ \hline
				600 & -0.0046 & 0.0940 &  0.0867 & 0.93 & 0.9881 & 0.0108& 0.7978 & 0.0665 \\
				800 & -0.0025 & 0.0779 &  0.0745 & 0.948  & 0.9899 & 0.0083 & 0.7987 & 0.0593\\
				1000 & -0.0015 & 0.0696 & 0.0675 & 0.95 & 0.9908 & 0.0078 & 0.7968 & 0.0550 \\\hline
			\end{tabular}
			\label{table_SIM3}
		\end{center}
	\end{table}
	\vspace{-0.2in}
	\subsection{Simulation 4}\label{simu4}
	In the fourth simulation, the covariate $\X_i=(X_{i1},X_{i2})\trans$
	is generated the same as in
	Simulation 3. We set
	$g(\bb\trans\X_i)=2\bb\trans\X_i+\sin(2\bb\trans\X_i)$ with $\bb=(1,-1)\trans$. The
	regression errors  $\epsilon_i$ and the baseline response model
	$f(\X_i)$ are generated identically as in Simulation 1. We allow the
	distribution of treatment indicator $Z_i$ to depend on the covariates.
	Specifically, $Z_i$ is generated from a Bernoulli distribution
	with probability of success
	$\exp(\bg\trans\X_i)/\{1+\exp(\bg\trans\X_i)\}$, where
	$\bg=(0.3,-0.2)\trans$. Note that here we considered covariate
	adjusted randomization.  
	
\begin{figure}[!ht]
		\begin{center}
			\includegraphics[width=0.32\textwidth,height=0.35\textheight]{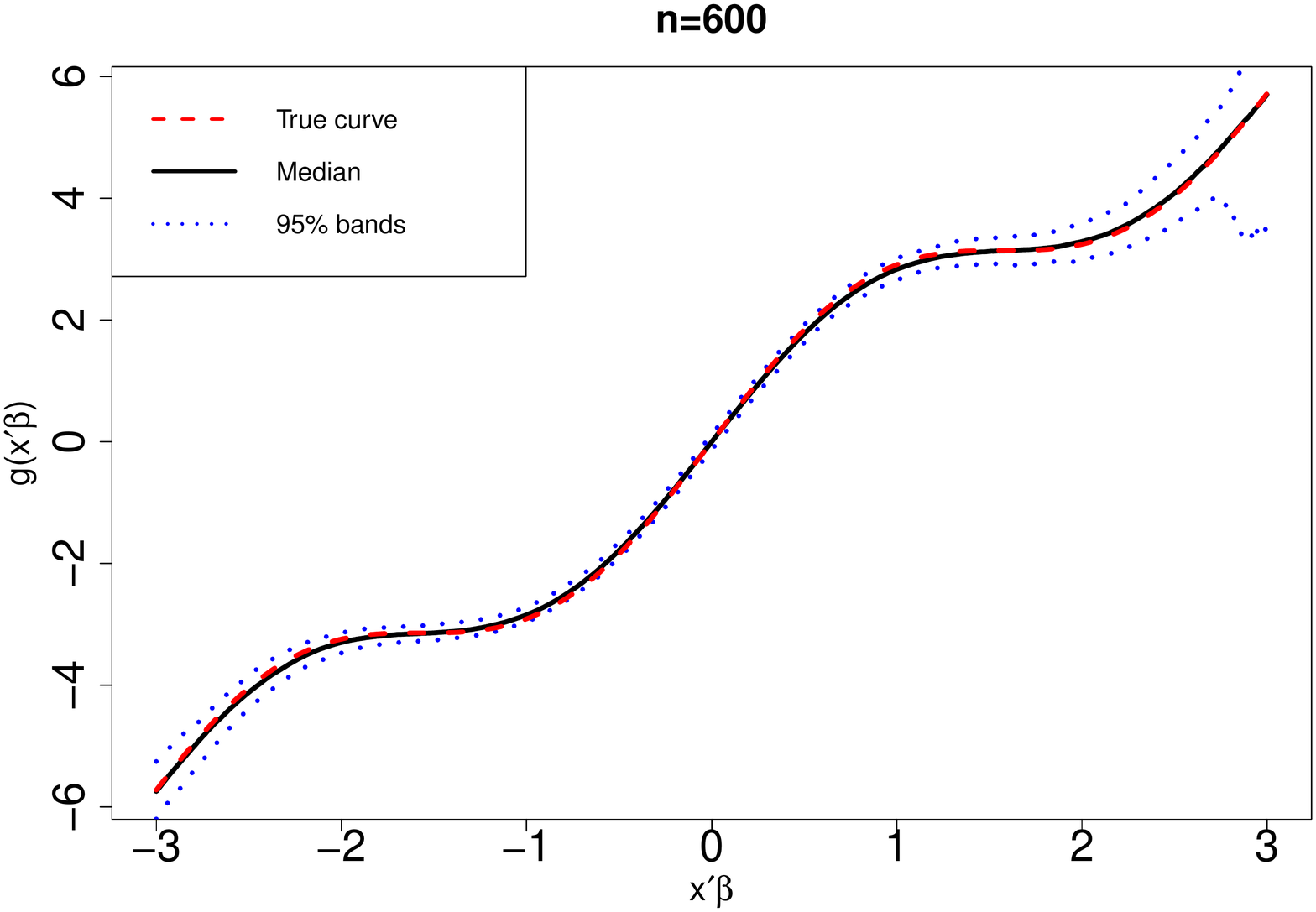}
			\includegraphics[width=0.32\textwidth,height=0.35\textheight]{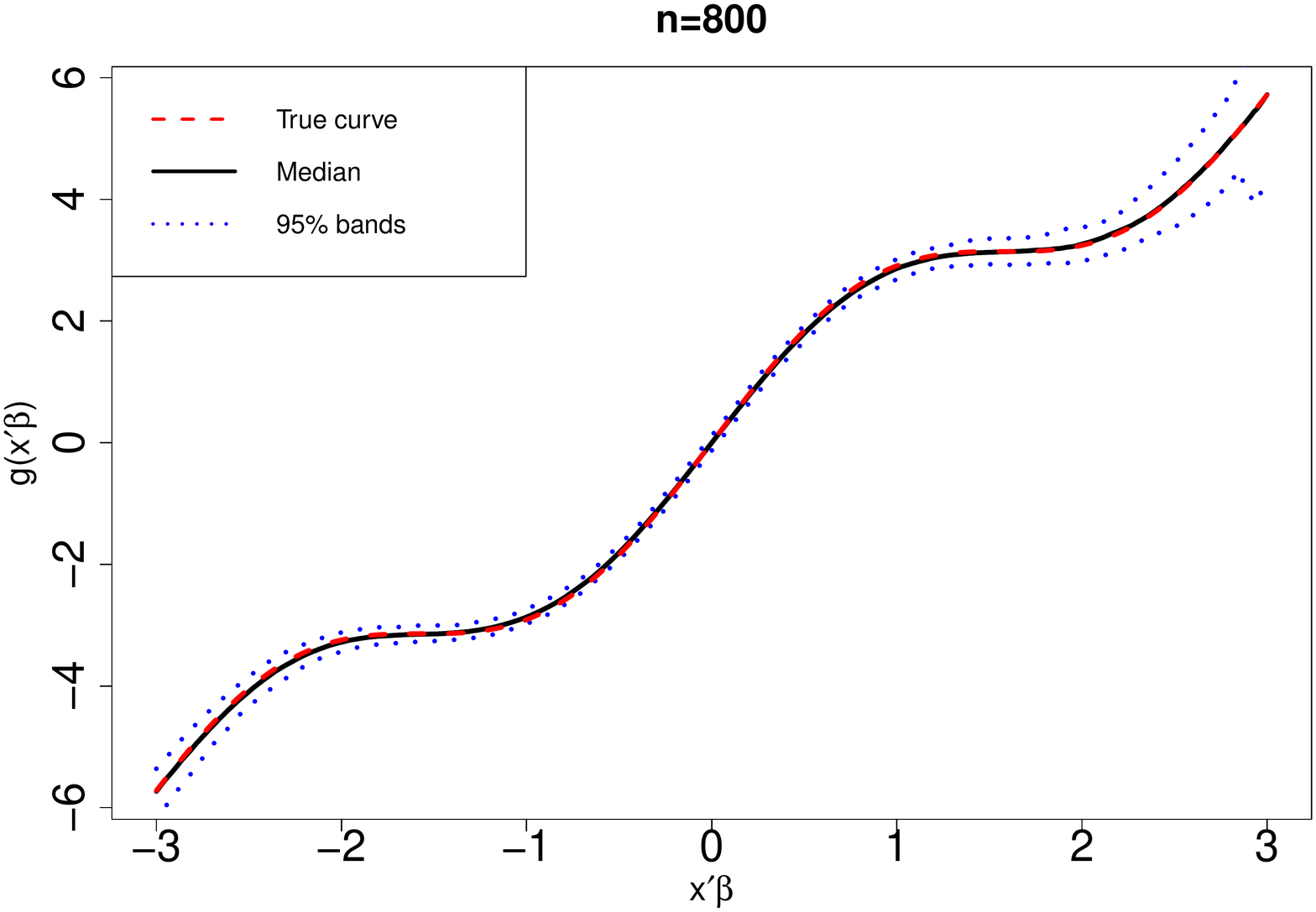}
			\includegraphics[width=0.32\textwidth,height=0.35\textheight]{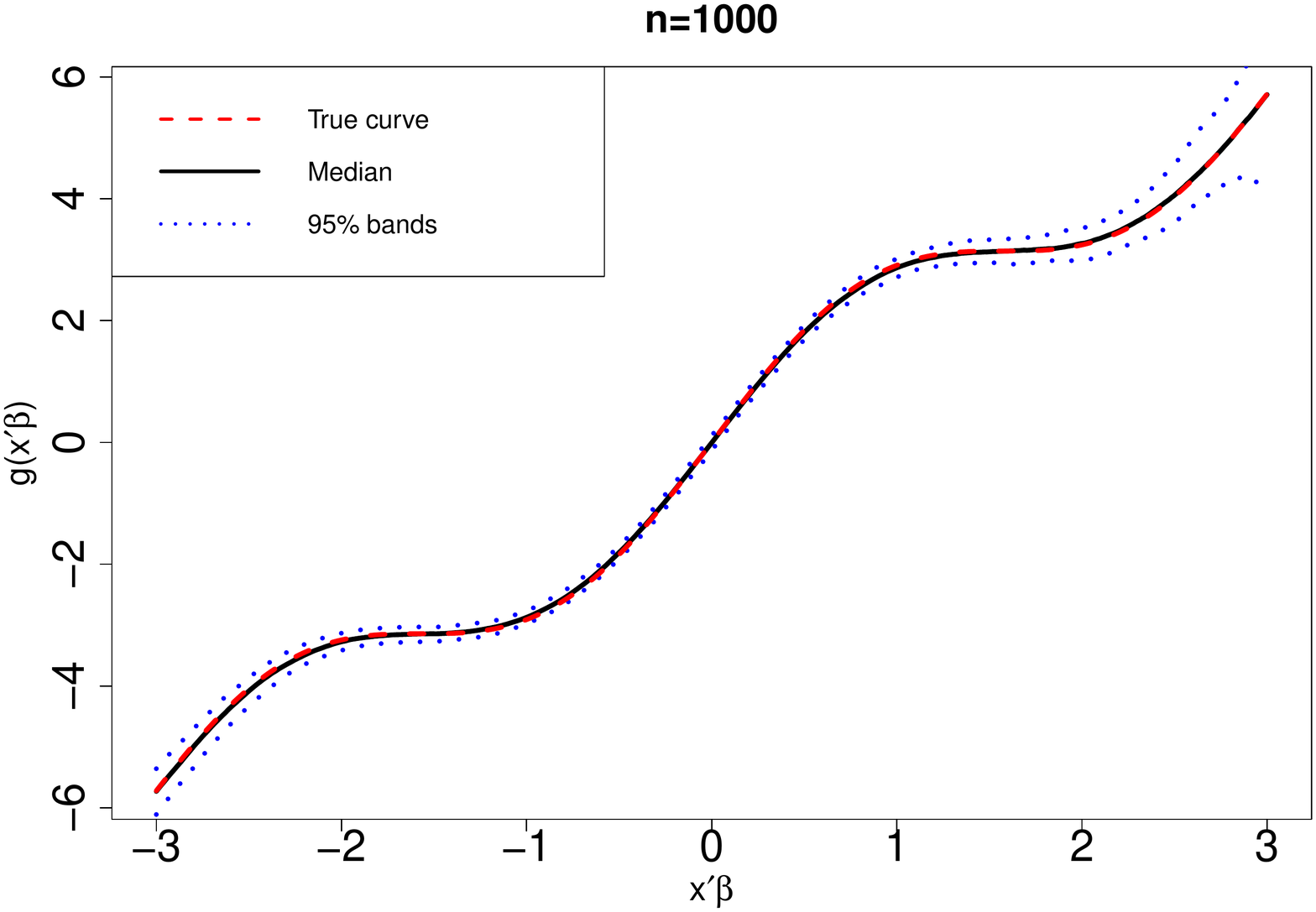}
			\caption{\footnotesize{Estimation of $g(\cdot)$, when $g(\x'\beta)=2(\x'\bb)+\sin\{2(\x'\bb)\}$ in Simulation 4.}}
			\label{figure_sin}
		\end{center}
	\end{figure}

	From the results in Table \ref{table_sin_1}, we can observe that all
	estimation methods have acceptable biases and standard errors even
	in covariate adjusted randomization setup. Method I performed
	significantly better compared to other methods, as expected. In the upper part of
	Table \ref{table_SIM4}, we summarized the coverage probability of the
	95\% bootstrap confidence intervals of $\bb$, mean and standard
	deviations for estimating VF and PCD for Method I. We also 
	plotted the 95\% confidence band and median for $g(\cdot)$ in Figure
	\ref{figure_sin}. From the results provided in Table
	\ref{table_SIM4}, We can clearly see that our proposed Method I
	performs better in estimating  
	parameter, $\beta_2$, PCD and value function, compared to the SIM
	approach. The estimation variability of our Method I is also less than
	SIM. 
	
	\begin{table}[!ht]
		\begin{center}
			\caption{Bias and standard deviation (SD) for the estimation of $\beta_2$ in Simulation 4.}
			\vspace{-0.2in}
			\begin{tabular}{lcc|cc}
				\\\hline
				Methods  &\multicolumn{2}{c}{I}  & \multicolumn{2}{c}{II} \\ \cline{2-5}
				$n$     &  Bias &SD&Bias &SD \\
				$600$      &-0.0012&0.0294&-0.0044&0.1365\\
				$800$      &-0.0007&0.0257&-0.0032&0.1217\\
				$1000$    &0.00004&0.0228&0.0044&0.1053\\ \hline
				Methods  & \multicolumn{2}{c}{III}  &\multicolumn{2}{c}{IV} \\ \cline{2-5}
				$600$      &0.0038&0.0981&0.0153&0.1069\\
				$800$      &0.0035&0.0843&0.0093&0.0916\\
				$1000$    &-0.0022&0.0740&0.0037&0.0846\\ \hline
			\end{tabular}
			\label{table_sin_1}
		\end{center}
	\end{table}
	
	\begin{table}[!ht]
		\begin{center}
			\caption{Estimation and inference results for $\beta_2$ in Simulation 4 using Method I and SIM approach. Other caption is same as Table \ref{table_SIM1}. True value function is 0.9027.}
			\begin{tabular}{lcccccccc}
				\hline
				$n$ & Bias($\beta_2$) & SD($\beta_2$)& SE($\beta_2$) & CP($\beta_2$)  & PCD & SD(PCD) & $\evf$ & SD($\evf$) \\ \hline
				\multicolumn{9}{c}{Method I} \\ \hline
				$600$ & -0.0012 & 0.0294 & 0.0296 & 0.946 & 0.9900 & 0.0064 & 0.8977 & 0.0775  \\
				$800$ & -0.0007 & 0.0257 & 0.0256 & 0.956 & 0.9913 & 0.0054 & 0.8990 & 0.0709 \\
				$1000$& 0.00004 & 0.0228 & 0.0224 & 0.96 & 0.9920 & 0.0044 & 0.8967 & 0.0664\\ \hline
				\multicolumn{9}{c}{SIM Method} \\ \hline
				$600$ & 0.0358 &0.0997 & 0.0823 & 0.834 & 0.9857 & 0.0114& 0.8555 & 0.0766\\
				$800$ & 0.0352 & 0.0896 & 0.0715 & 0.804 & 0.9867 & 0.0095 & 0.8571 & 0.0693\\
				$1000$ & 0.0364 & 0.0755& 0.0644 & 0.82 & 0.9888 & 0.0085 & 0.8544 & 0.0653\\\hline
			\end{tabular}
			\label{table_SIM4}
		\end{center}
	\end{table}
	\vspace{-0.2in}
	\subsection{Simulation 5}\label{simu5}
	In this simulation setting, we consider the categorical treatment
	indicator. Our covariate $\X_i=(X_{i1},X_{i2},X_{i3})\trans$ is
	generated from multivariate standard normal distribution.
	We consider three levels for treatment
	indicator $Z_i$, $0$, $1$ and $2$ with probability $0.4$, $0.4$ and
	$0.2$, respectively. Here, relative to the base line at $Z=0$, 
	$g_1(\bb\trans\X_i)=0.5(\bb\trans\X_i)^2-1$
	is the treatment effect, when $Z_i=1$ and
	$g_2(\bb\trans\X_i)=(\bb\trans\X_i)\sin(\bb\trans\X_i)-1$ is the
	treatment effect for $Z_i=2$ with $\bb=(1,-1,1)\trans$. Here,
	$f(\X_i)=0.5(X_{i1}+X_{i2}+X_{i3})$ and the regression error
	$\epsilon_i$ is generated identically as in Simulation 2.  
	
\begin{figure}[!ht]
		\begin{center}
			\includegraphics[width=0.32\textwidth,height=0.35\textheight]{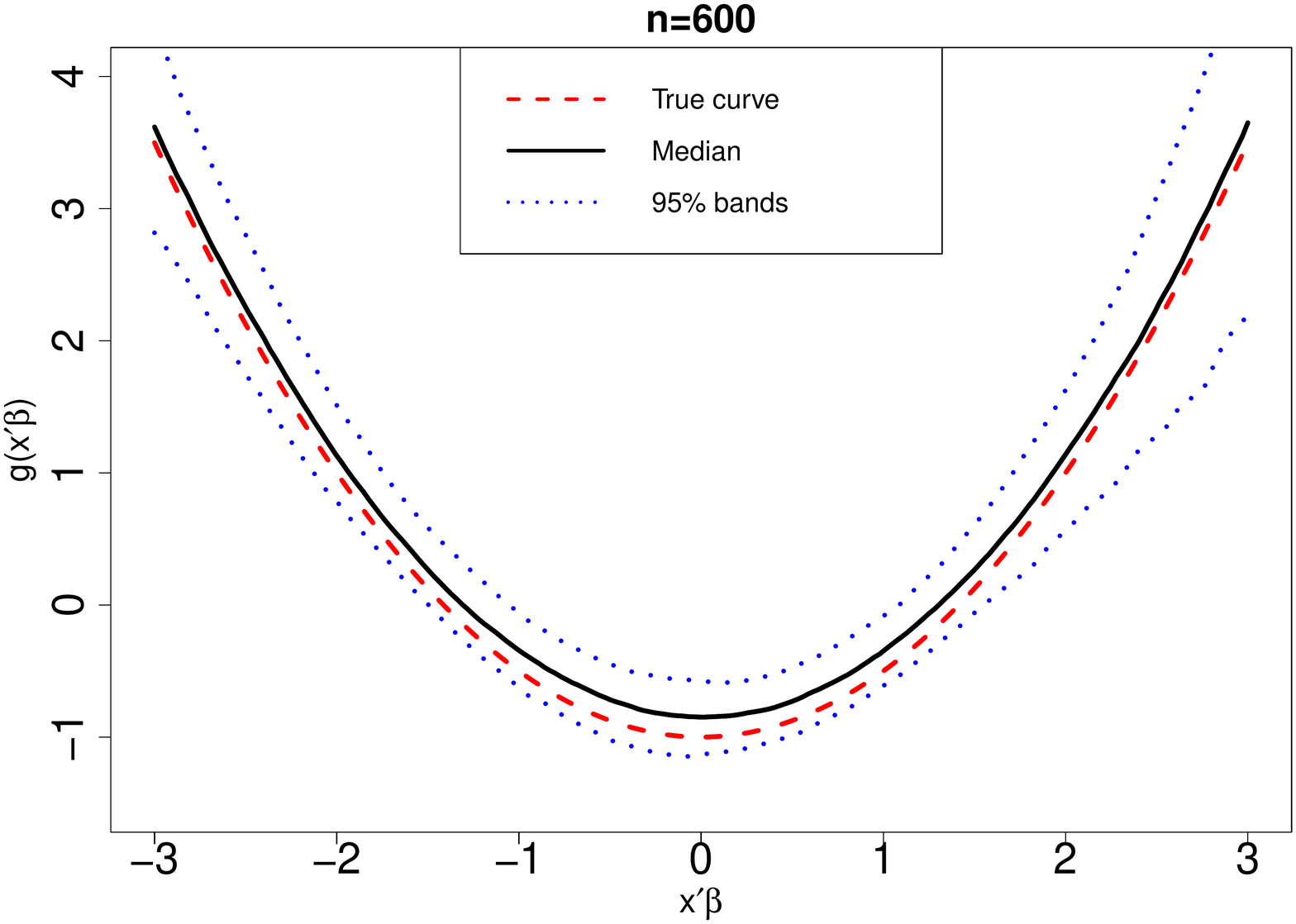}
			\includegraphics[width=0.32\textwidth,height=0.35\textheight]{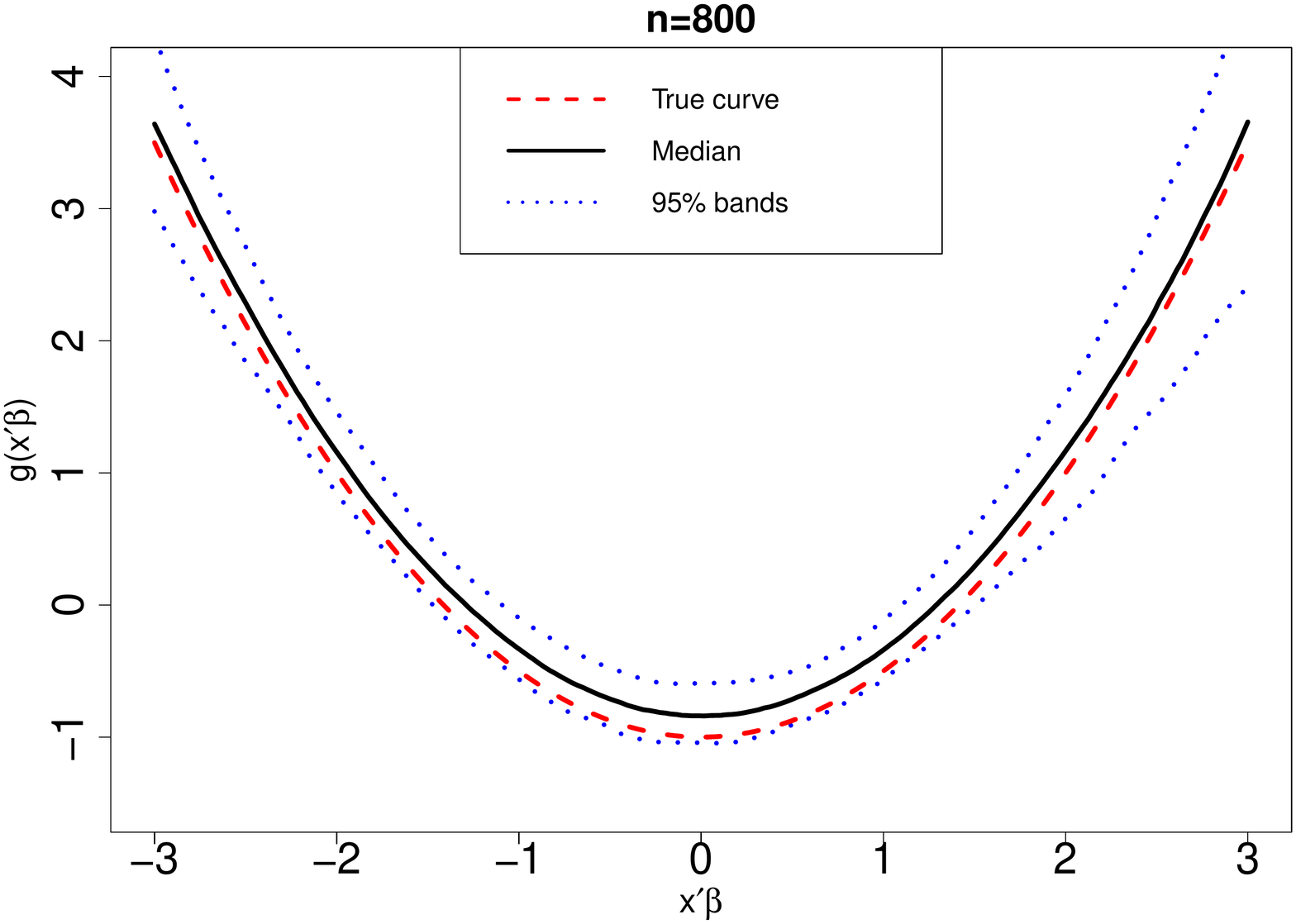}
			\includegraphics[width=0.32\textwidth,height=0.35\textheight]{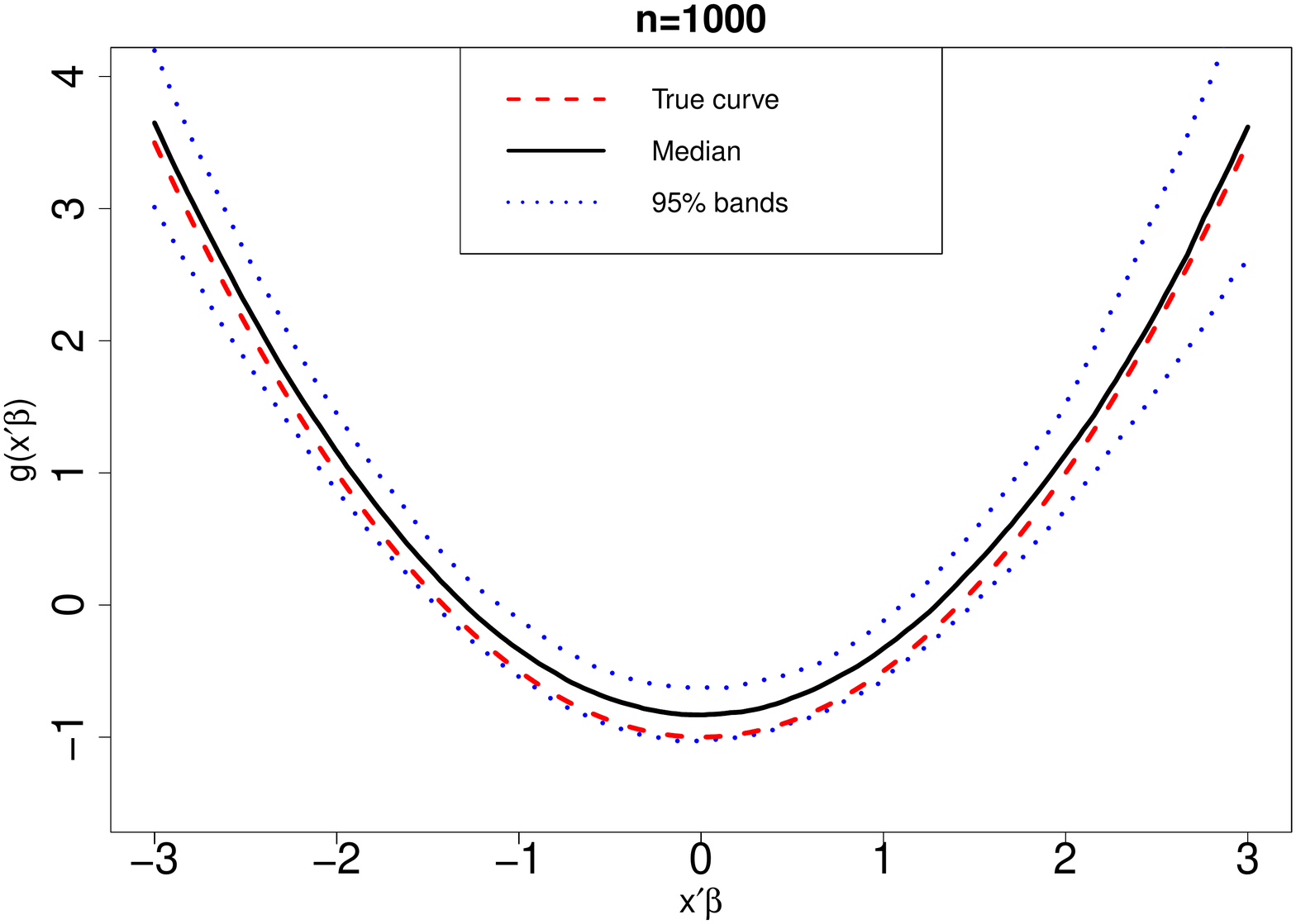}\\
			\includegraphics[width=0.32\textwidth,height=0.35\textheight]{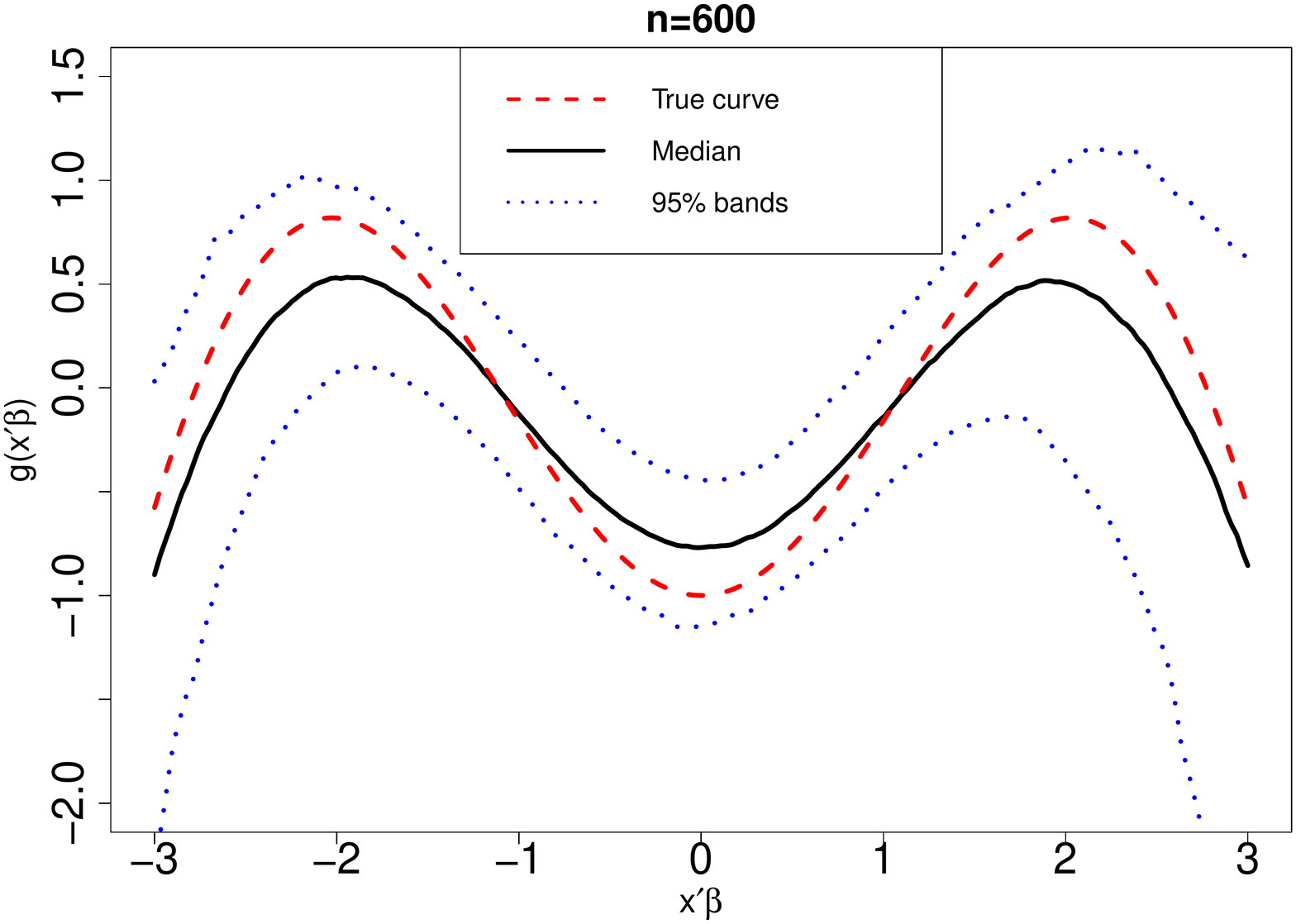}
			\includegraphics[width=0.32\textwidth,height=0.35\textheight]{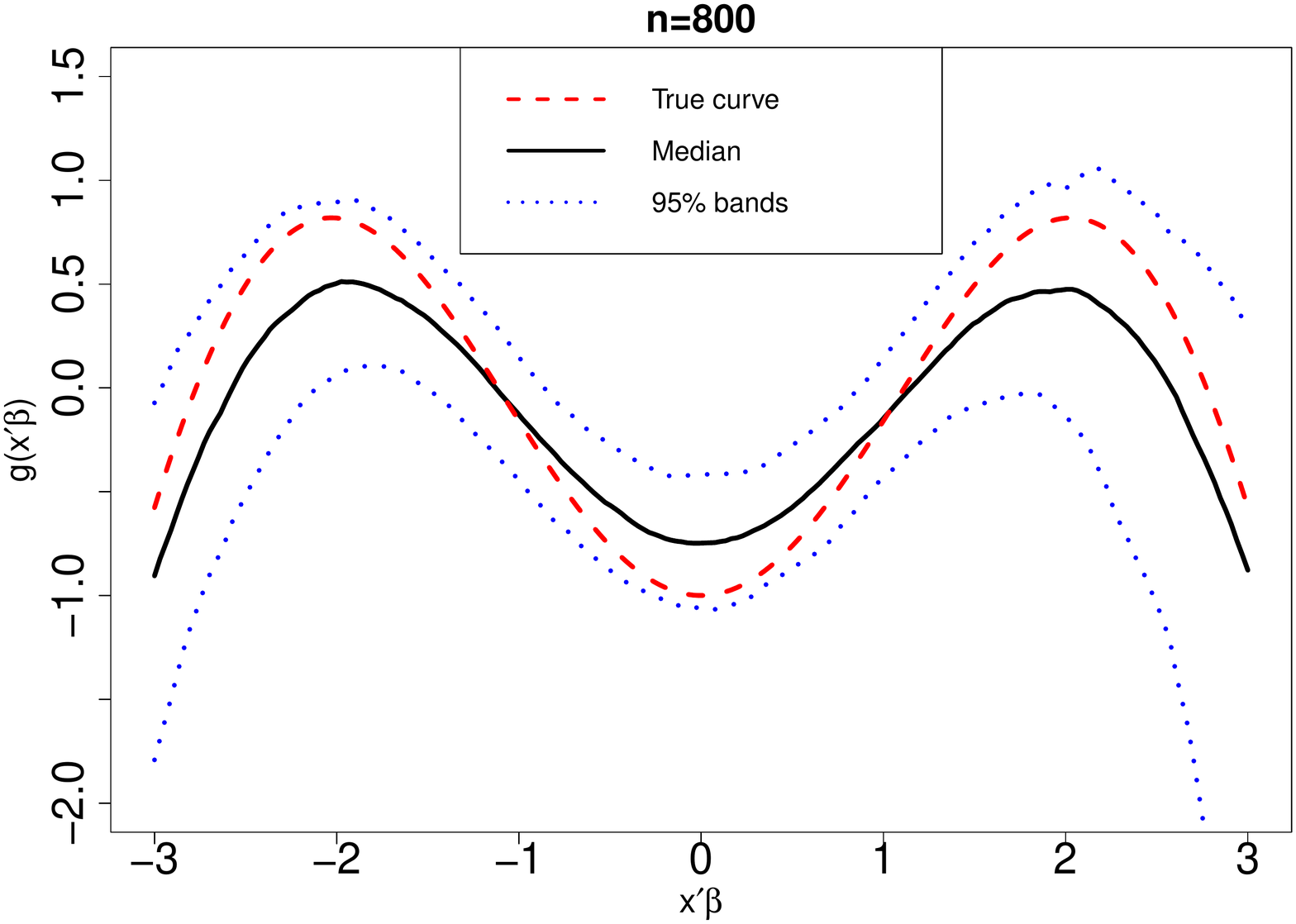}
			\includegraphics[width=0.32\textwidth,height=0.35\textheight]{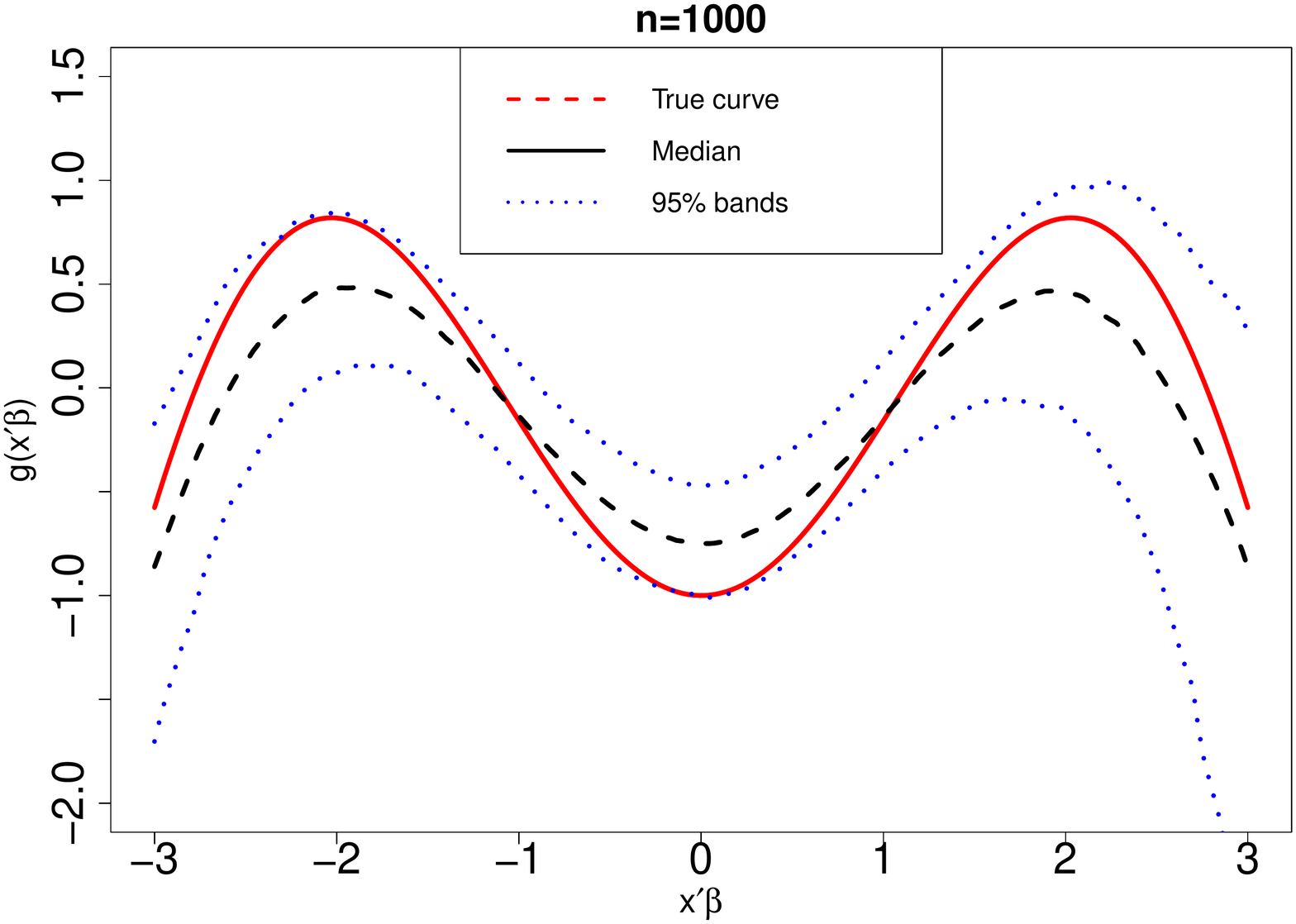}
			\caption{\footnotesize{Estimation of $g_1(\cdot)$ (top) and $g_2(\cdot)$ (bottom), where $g_1(\x'\beta)=0.5(\x'\bb)^2-1$ and $g_2(\x'\beta)=(\x'\bb)\sin(\x'\bb)-1$ in Simulation 5.}}
			\label{figure_dis1}
		\end{center}
	\end{figure}

	As we have repeatedly observed in previous simulations that Method I is superior to other
	three methods, we only consider Method I to estimate the unknown
	parameter $\bb$ and to proceed with the inference of $\bb$ and the
	estimation and inference of $g_1(\cdot)$ and $g_2(\cdot)$. The results
	are summarized in the first parts of Tables
	\ref{table1_disz} and \ref{table2_disz}. In the second
	parts of the aforementioned tables, we provided the results
	obtained from SIM. From Table \ref{table1_disz}, we can see
	that using our proposed Method I, estimation bias is acceptable 
	and the coverage probability obtained from bootstrap confidence
	interval is close to the nominal level of 0.95. On the other hand, as
	expected, due to the non-monotone treatment effect models $g_1(\cdot)$
	and $g_2(\cdot)$, SIM yields large bias. Also, SIM
	performs poorly in terms of inference of $\bb$. From Table 
	\ref{table2_disz}, we can clearly observe that our proposed Method I
	yields higher PCD values and also small bias in estimating value
	function, compared to SIM. In Figure \ref{figure_dis1}, we
	computed 95\% confidence band and median for $g_1(\cdot)$ and
	$g_2(\cdot)$.   
	
	\begin{table}[!ht]
		\begin{center}
			\caption{Bias, standard deviation (SD) and coverage probabilities (CP) for the estimation of $\beta_2$ and $\beta_3$ in Simulation 5 using proposed method and SIM method.}
			\vspace{-0.3in}
			\begin{tabular}{lcccc|cccc}
				\\\hline
				&\multicolumn{4}{c}{$\beta_2$}  & \multicolumn{4}{c}{$\beta_3$} \\ \cline{2-9}
				$n$     &  Bias &SD& SE & CP &  Bias &  SD&SE & CP \\ \hline
				&\multicolumn{8}{c}{Proposed Method} \\ \cline{1-9}
				$600$      &0.0001&0.0859&0.1068 &  0.964 &-0.0085&0.0983&0.1140& 0.954\\ 
				$800$      &-0.0039&0.0764& 0.0919 &  0.956 &-0.0087&0.0796& 0.0981 & 0.962\\ 
				$1000$    &-0.0032&0.0663& 0.0885 &  0.964&-0.0057&0.0707& 0.0911 &0.958\\ \hline
				&\multicolumn{8}{c}{SIM Method} \\ \cline{1-9}
				$600$      &0.6166&2.3757&0.4028 &  0.200 &-0.5619&1.5774&0.3716& 0.400\\
				$800$     &1.9593&1.3710&0.2192 &  0.100 &1.5210&4.1547&0.2212 &0.400\\
				$1000$    &2.2497&1.6344&10.8646 &1.000&-0.1332&1.0503&10.8043 &1.000\\ \hline
			\end{tabular}
			\label{table1_disz}
		\end{center}
	\end{table}
	
	\begin{table}[!ht]
		\begin{center}
			\caption{The percentage of
				correct decisions (PCDs) with standard errors, 
				and the estimated value function (VF) with empirical standard deviations
				in Simulation 5, using proposed method and SIM method. Here, PCD1 and PCD2 are related to $g_1(\cdot)$ and $g_2(\cdot)$, respectively. True value function is 0.8056.
			}
			\vspace{-0.3in}
			\begin{tabular}{lcccccc}
				\\\hline
				$n$       &  PCD1 &  SD(PCD1) & PCD2  &  SD(PCD2) &  $\evf$ & SD($\evf$)\\\hline
				&\multicolumn{6}{c}{Proposed Method} \\ \cline{1-7}
				$600$   & 0.9455  & 0.0295 & 0.8821 & 0.0558 &0.7884& 0.1521\\
				$800$   & 0.9480  &0.0249 &   0.9011 &0.0450  & 0.7961 & 0.1356 \\
				$1000$ &0.9501 &0.0224 &   0.9077  &0.0475& 0.8174 &  0.1153\\ \hline
				&\multicolumn{6}{c}{SIM Method} \\ \cline{1-7}
				$600$   & 0.5175  & 0.0263 & 0.5000 & 0.0263 &-0.2997& 0.0958\\
				$800$   & 0.5059  &0.0112 &   0.5064 &0.0181  & -0.2859 & 0.1098 \\
				$1000$ &0.5040 &0.0132 &   0.4954  &0.0106&-0.3364 &  0.0817\\ \hline
			\end{tabular}
			\label{table2_disz}
		\end{center}
	\end{table}
	\vspace{-0.2in}
	\subsection{Simulation 6}\label{simu6}
	In the last setting, we consider continuous treatment indicator. We
	generate covariate $\X_i$ as in Simulation 4, and consider a
	quadratic treatment effect
	$Z_ig_1(\bb\trans\X_i)+Z_i^2g_2(\bb\trans\X_i)$. Here $Z_i$ follows 
	the uniform distribution on $[0,1]$,
	$g_1(\bb\trans\X_i)=0.5(\bb\trans\X_i)^3$ and
	$g_2(\bb\trans\X_i)=2-(\bb\trans\X_i)^2$ with
	$\bb=(1,-1)\trans$. We let $f(\X_i)=0.5(X_{i1}+X_{i2})$ and generated the
	regression error $\epsilon_i$ as in
	Simulation 2.  
	
\begin{figure}[!ht]
		\begin{center}
			\includegraphics[width=0.32\textwidth,height=0.35\textheight]{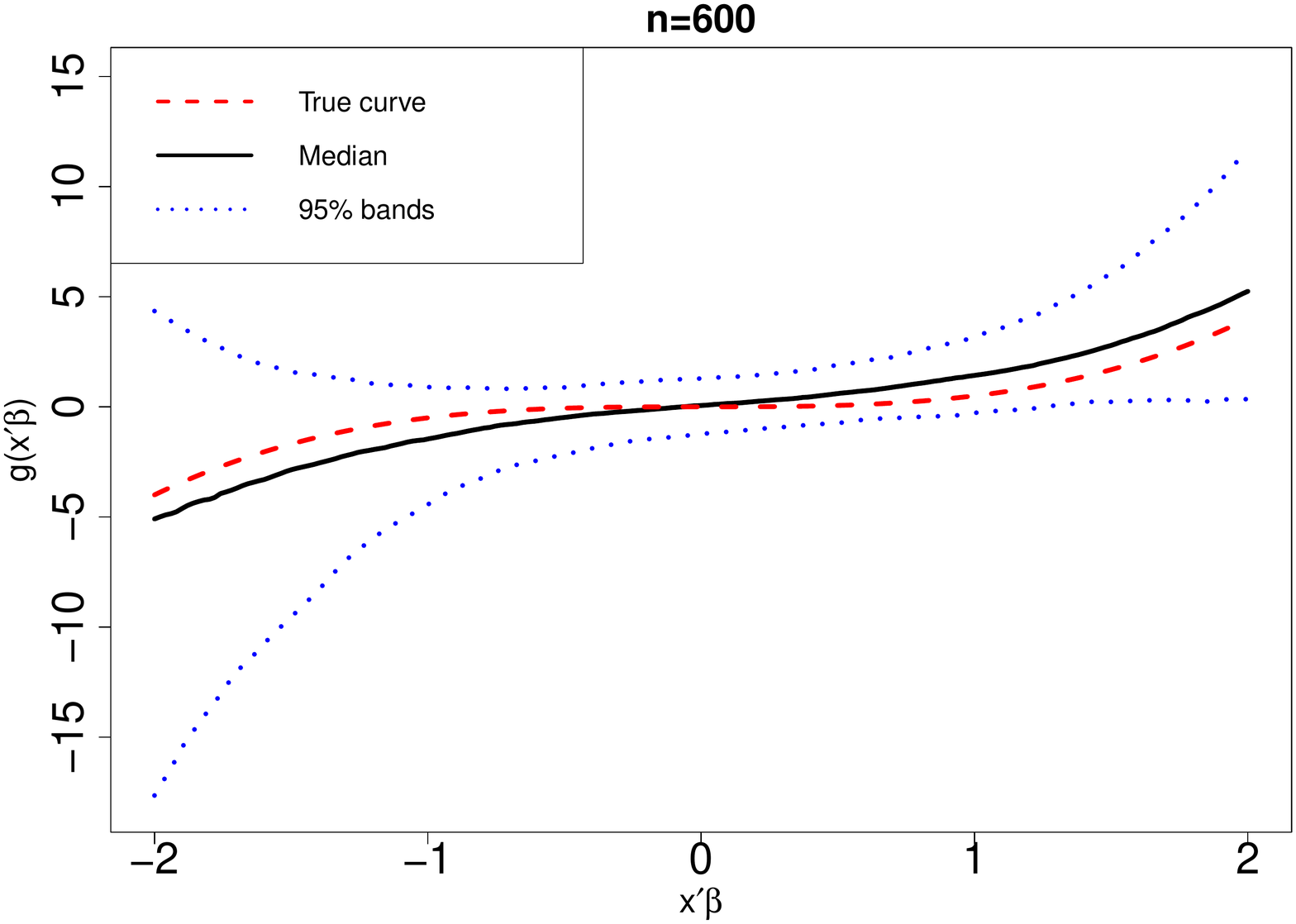}
			\includegraphics[width=0.32\textwidth,height=0.35\textheight]{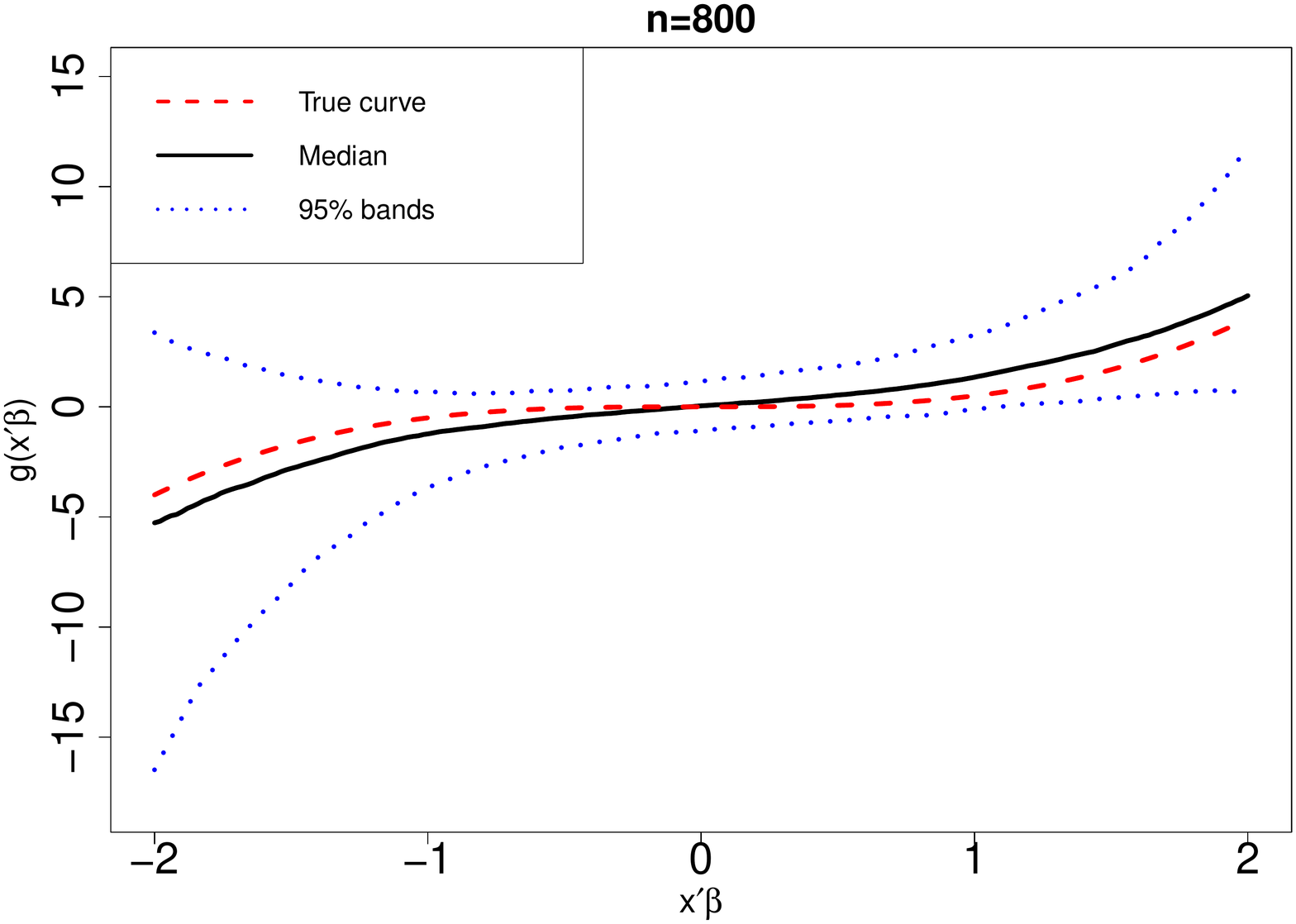}
			\includegraphics[width=0.32\textwidth,height=0.35\textheight]{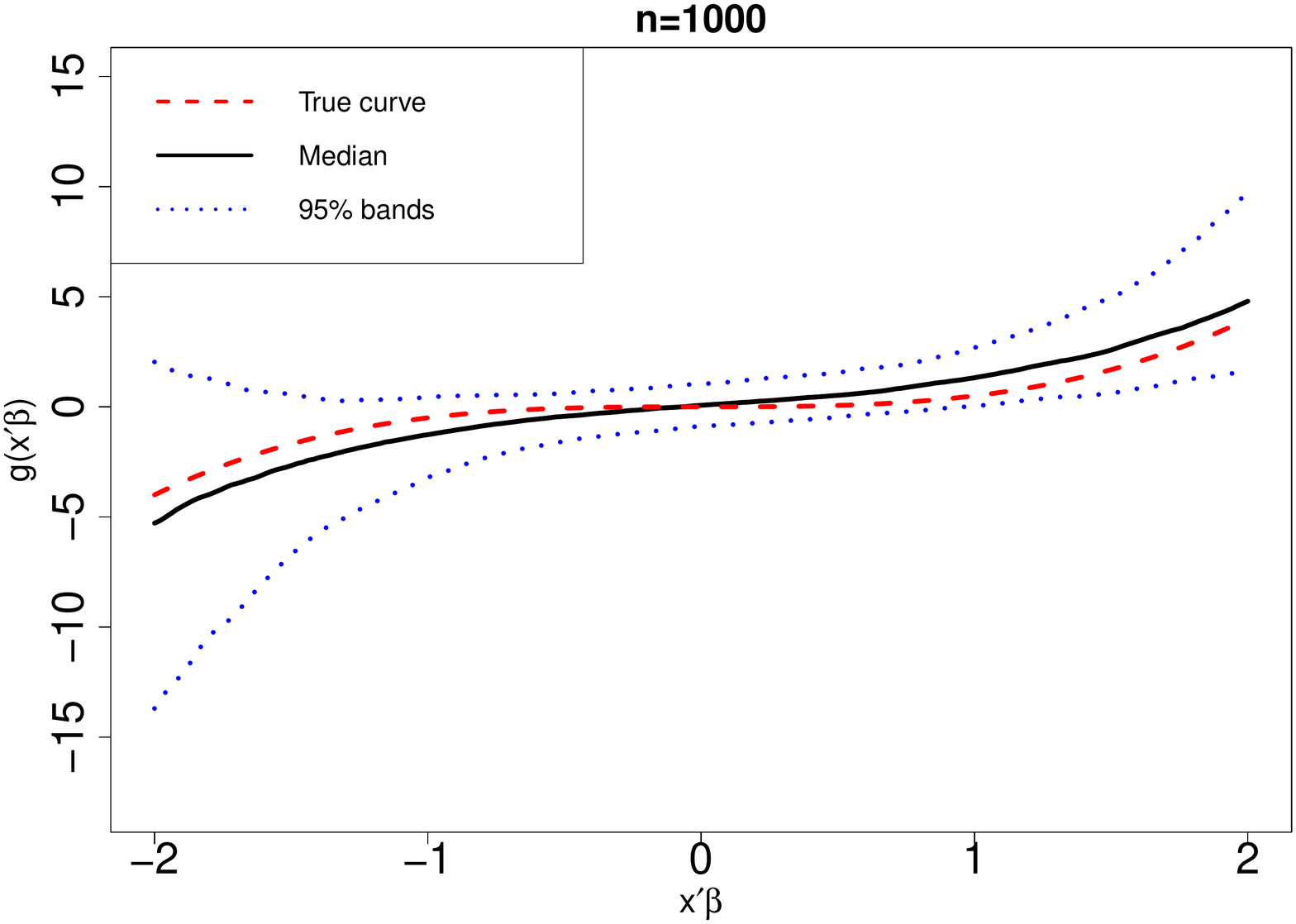}\\
			\includegraphics[width=0.32\textwidth,height=0.35\textheight]{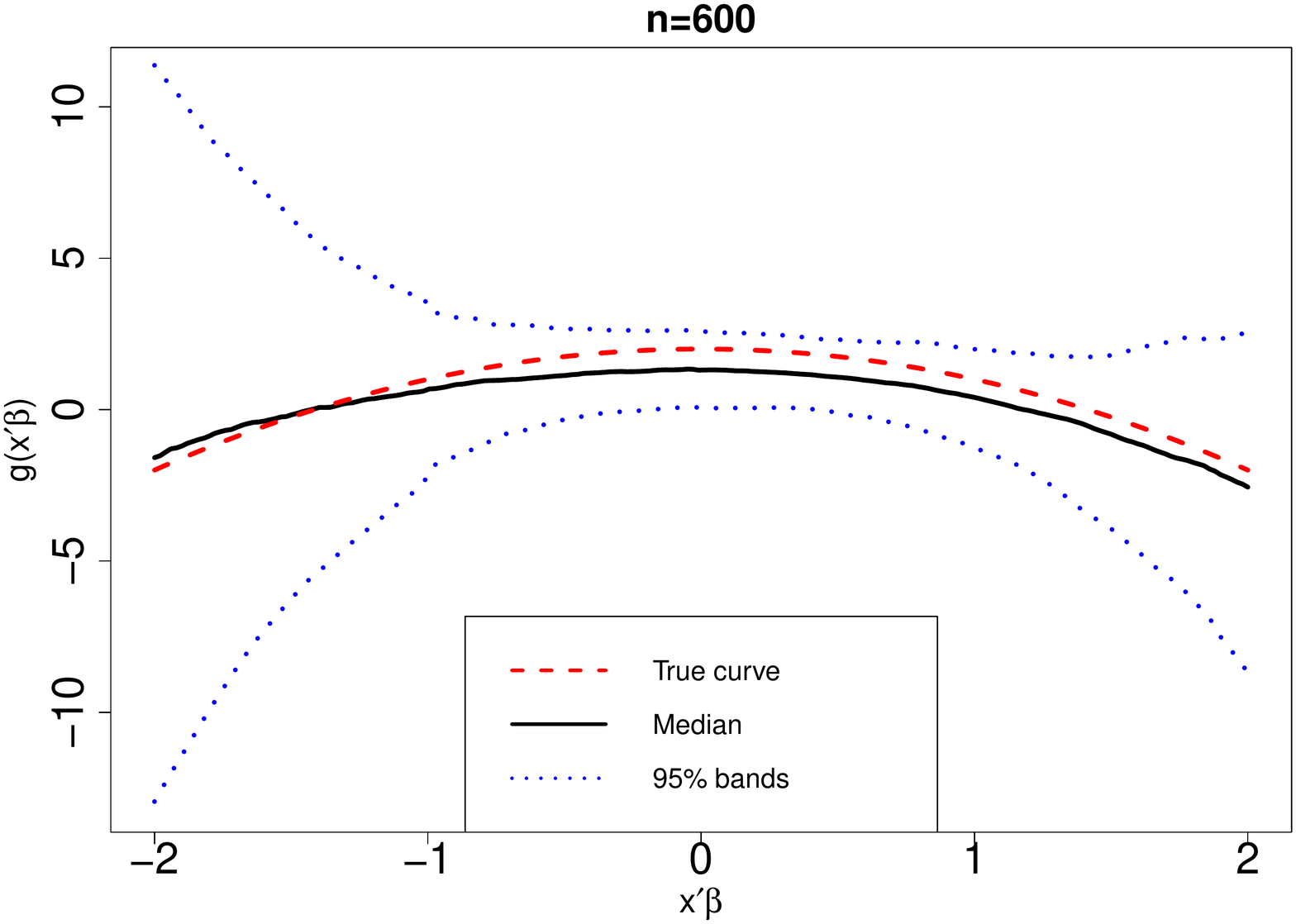}
			\includegraphics[width=0.32\textwidth,height=0.35\textheight]{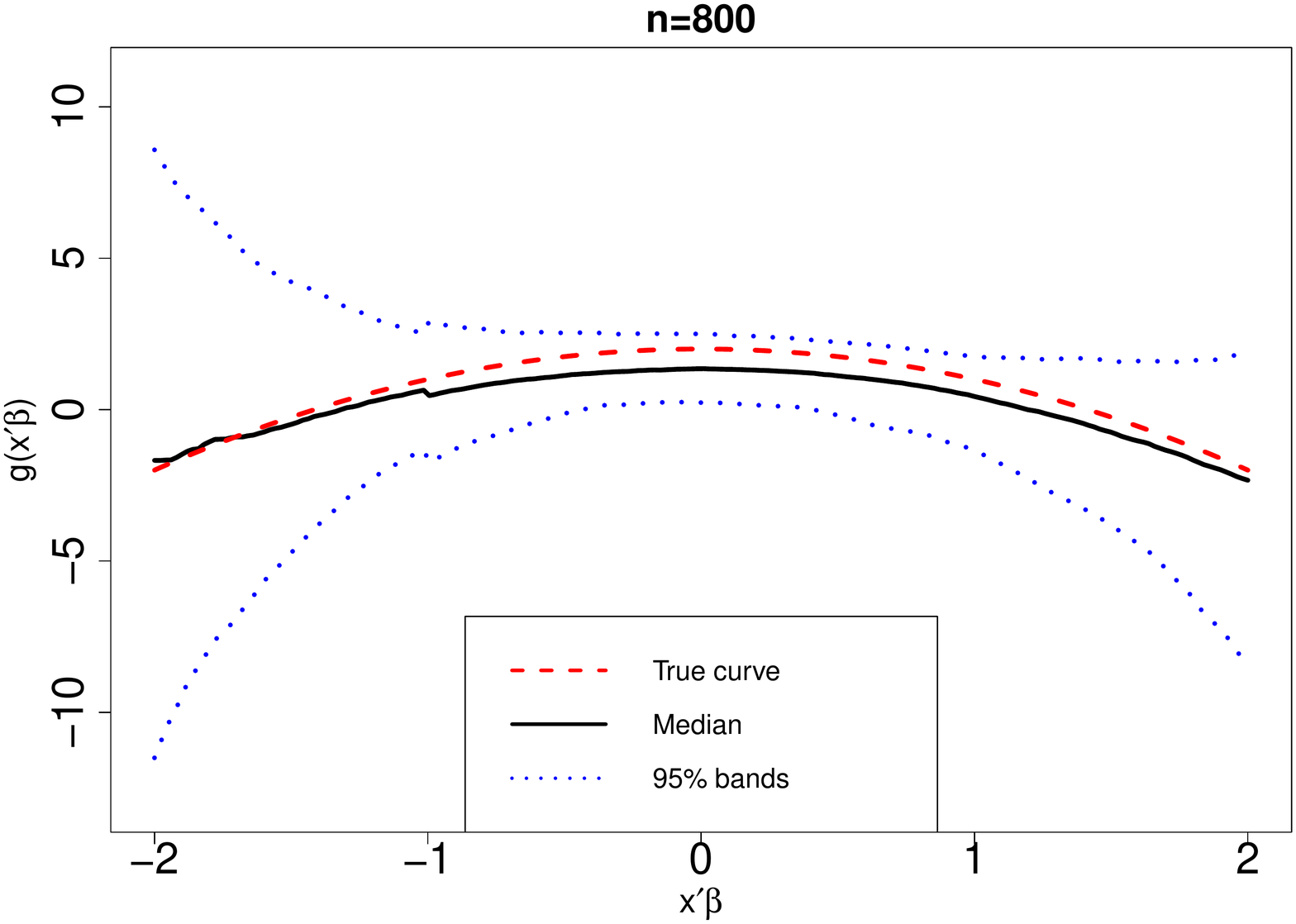}
			\includegraphics[width=0.32\textwidth,height=0.35\textheight]{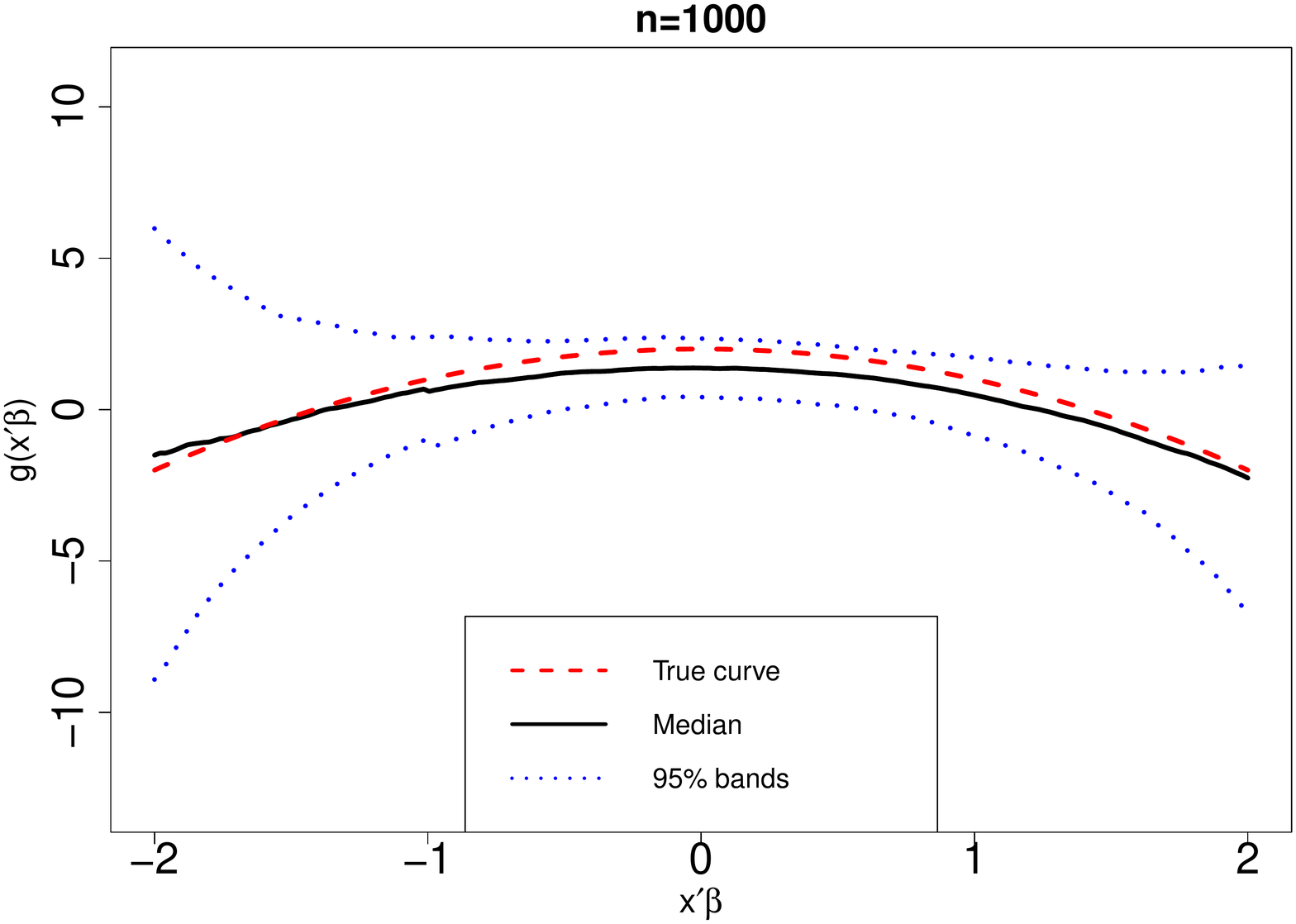}
			\caption{\footnotesize{Estimation of $g_1(\cdot)$ (top) and $g_2(\cdot)$ (bottom), where $g_1(\x'\beta)=0.5(\x'\bb)^3$ and $g_2(\x'\beta)=2-(\x'\bb)^2$ in Simulation 6.}}
			\label{figure_con1}
		\end{center}
	\end{figure}

	Similar to Simulation 5, we implement Method I to estimate the
	unknown parameter $\bb$ and proceed with the inference of $\bb$ and the
	estimation of value function and PCD for $g_1(\cdot)$ and $g_2(\cdot)$. The results
	are summarized in the upper parts of Tables \ref{table1_conz} and
	\ref{table2_conz}. From the upper part of Table \ref{table1_conz}, we
	can see that estimation bias is acceptable 
	and the coverage probability obtained from bootstrap confidence
	interval is close to the nominal level of 0.95 for our proposed
	method. Also, value function obtained by proposed method is close to
	true value. In Table \ref{table2_conz}, we summarized the mean and standard deviations
	for PCD for $g_1(\cdot)$ and $g_2(\cdot)$. In Figure \ref{figure_con1}, we also plotted 95\% confidence
	band and median for $g_1(\cdot)$ and $g_2(\cdot)$. We 
	also compared the performance of our method with the existing kernel
	assisted learning (KAL) method \citep{KAL2020}. From the lower part of
	Table \ref{table1_conz}, we clearly see that the our method performs
	better than KAL method. We suspect that due to their restrictive model
	set up, the value function estimation using KAL is not
	trustworthy. Indeed, \cite{KAL2020} consider the following set up : $E(Y\mid Z,\X)=f(\X)+Q\{Z-g(\wt\bb\trans\X)\}H(\X)$, where $Q(\cdot)$ is a unimodal function which is maximized at $0$ and $H(\X)$ is a non-negative function. Then $E(Y\mid Z,\X)$ is maximized at $Z=g(\wt\bb\trans\X)$, where $g : \mR \to \mA$ is a predefined link function which ensures that the suggested dose falls within safe dose range ($\mA$). On the other hand, we consider a general polynomial model, $E(Y\mid Z,\X)=f(\X)+\sum_{k=1}^{n}Z_kg_k(\bb\trans\X)$. In this particular simulation setting, we consider k=2. Thus, we try to re-write our model similar to their model so that we can easily implement their method in our simulation study. After re-writing, we get 
	
	\bse
	E(Y\mid Z,\X)=
	\left\{f(\X)-\frac{g_1^2(\bb\trans\X)}{4g_2(\bb\trans\X)}\right\} +
	\left[-\left\{Z-\frac{-g_1(\bb\trans\X)}{2g_2(\bb\trans\X)}\right\}^2\right] 
	\{-g_2(\bb\trans\X)\}
	\ese
	and our model satisfies the conditions of KAL when
	$\bb\trans\X>\sqrt{2}$. Thus, we use the subset of simulated data that
	fits into the KAL model requirement to
	to obtain the KAL estimate of  $\beta_2$ and the value function. 
	
	\begin{table}[!ht]
		\begin{center}
			\caption{Estimation and inference results of $\beta_2$ and estimation of Value function using our proposed method and KAL method \citep{KAL2020} in Simulation 6. True value function is 0.6652.}
			\vspace{-0.3in}
			\begin{tabular}{lcccccc}
				\\\hline
				$n$ & Bias($\beta_2$) & SD($\beta_2$)  & SE($\beta_2$)  & CP($\beta_2$)  & $\evf$ & SD($\evf$)\\\hline
				&\multicolumn{5}{c}{Proposed Method} \\ \cline{1-7}
				$600$ &-0.0376&0.1361&0.1910&0.966&0.3567&0.2771\\
				$800$ &-0.0031&0.1191&0.1589&0.954&0.3742&0.2702\\
				$1000$&-0.0207&0.1067&0.1272&0.95&0.4073&0.2572\\\hline
				&\multicolumn{5}{c}{KAL Method} \\ \cline{1-7}
				$600$ & 0.2944 &0.3045 &0.2719 & 0.868 & -161.77 & 214.52 \\
				$800$ & 0.1101 &0.2575 &0.2308 & 0.78 & -164.91 & 222.83 \\
				$1000$ & -0.0442 & 0.2441 &0.2174 & 0.79 & -167.22 & 221.02 \\\hline
			\end{tabular}
			\label{table1_conz}
		\end{center}
	\end{table}
	
	\begin{table}[!ht]
		\begin{center}
			\caption{The percentage of
				correct decisions (PCDs) with standard deviations obtained by proposed method
				in Simulation 6. Here, PCD1 and PCD2 are related to $g_1(\cdot)$ and $g_2(\cdot)$, respectively.
			}
			\vspace{-0.3in}
			\begin{tabular}{lcccc}
				\\\hline
				$n$       &  PCD1 &  SD(PCD1) & PCD2  &  SD(PCD2) \\\hline
				$600$   & 0.8617 &0.1364&0.7633&0.1180\\
				$800$   & 0.8351 &0.1218&0.7895&0.0961\\
				$1000$ &0.8567&0.1093& 0.8136&0.0812\\ \hline
			\end{tabular}
			\label{table2_conz}
		\end{center}
	\end{table}
	
	\vspace{-0.5in}
	\section{Real data example}\label{sec:example}
	In this section, 
	we demonstrate our proposed method using the Sequenced Treatment Alternatives to
	Relieve Depression (STAR*D) study. The STAR*D study was a sequential,
	multiple-assignment, randomized trial 
	(SMART, \cite{murphy2005experimental} and \cite{qian2013dynamic}) 
	for
	patients with non-psychotic major depressive disorder. This study aimed to
	determine what antidepressant medications should be given to patients
	to yield the optimal treatment effect. For illustration purpose, we
	focused on a subset of 319 participants who were given treatment 
	bupropion (BUP) or sertraline (SER). Among  these participants, 153
	patients were randomly assigned to the BUP treatment group
	and the rest of them were assigned to the SER treatment group. 
	The 16-item Quick Inventory of Depressive Symptomatology-Self-Report
	scores (QIDS-SR(16)) were recorded at treatment visits for each
	patient and considered as the outcome variable. We used
	$R=-\mbox{QIDS-SR(16)}$ as the reward
	to accommodate the assumption that the larger reward is more desirable.  
	
	In the original data set, there are a large number of covariates that
	describe participant features such as age, gender, socioeconomic
	status, and ethnicity. However, 
	many of them are not significantly related to the QIDS-SR(16).
	According to
	the study conducted by \cite{fan2014sequential},
	we included five 
	important covariates into our study. These five covariates are
	``fatigue or loss of energy'' in baseline protocol eligibility (DSMLE,
	$X_1$),  
	patient's age (Age, $X_2$), ``ringing in ears'' in patient rated
	inventory of side effects at Level 2 (EARNG-Level2, $X_3$), ``feeling
	of worthlessness or 
	guilt'' in baseline protocol eligibility (DSMFW, $X_4$) and ``hard to
	control worrying'' in psychiatric diagnostic,  screening questionnaire
	at baseline (WYCRL, $X_5$). 
	
\begin{figure}[!ht]
		\centering
		\begin{minipage}[b]{0.45\textwidth}
			\includegraphics[width=\textwidth,height=0.35\textheight]{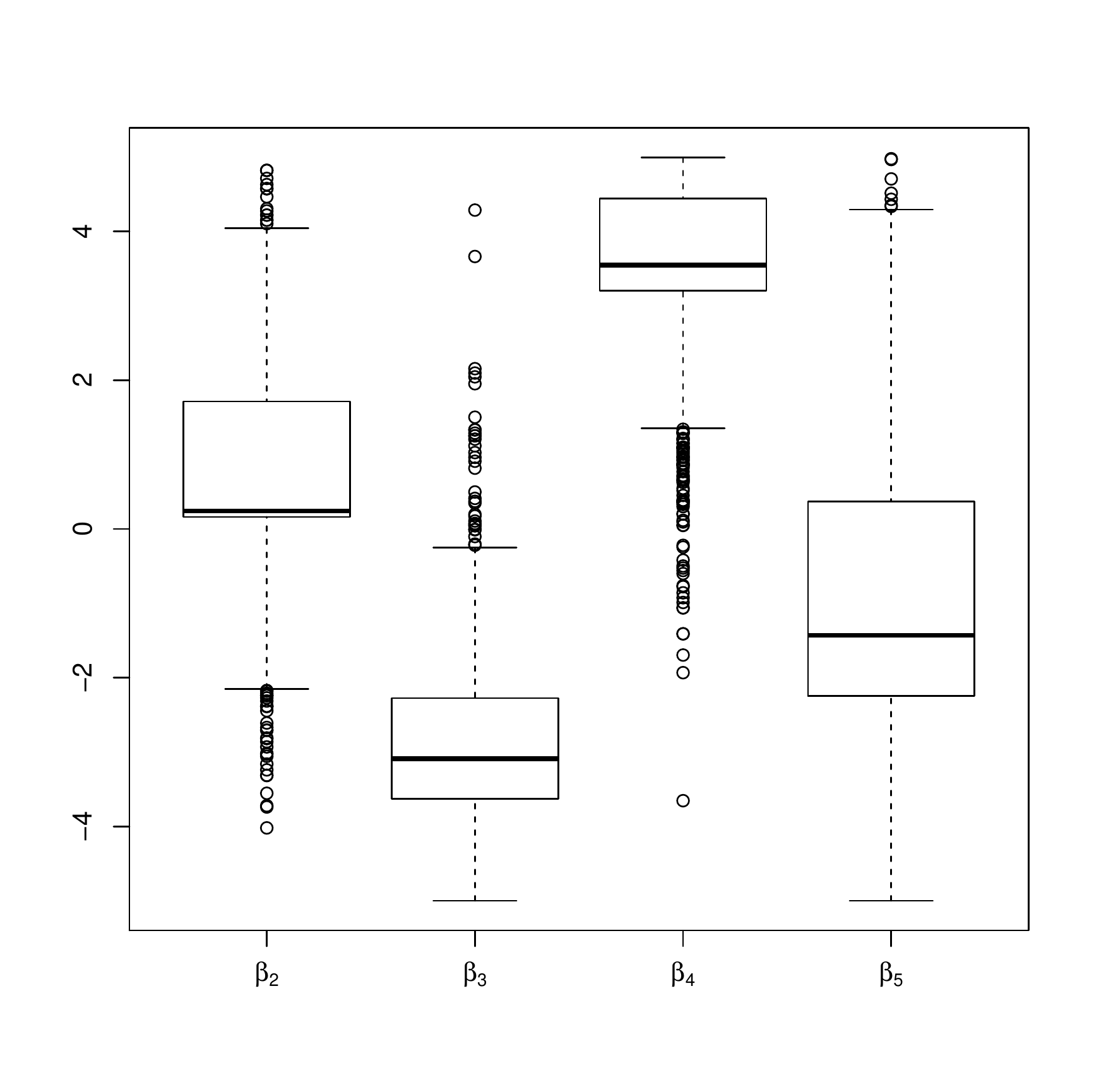}\par
			\vspace{-0.3in}
			\caption{\footnotesize{Box plots of the bootstrap estimators of the coefficients
					$\beta_2, \dots,\beta_4$.}}
			\label{fig:box-plot-beta}
		\end{minipage}
		\hfill
		\begin{minipage}[b]{0.45\textwidth}
			\includegraphics[width=\textwidth,height=0.35\textheight]{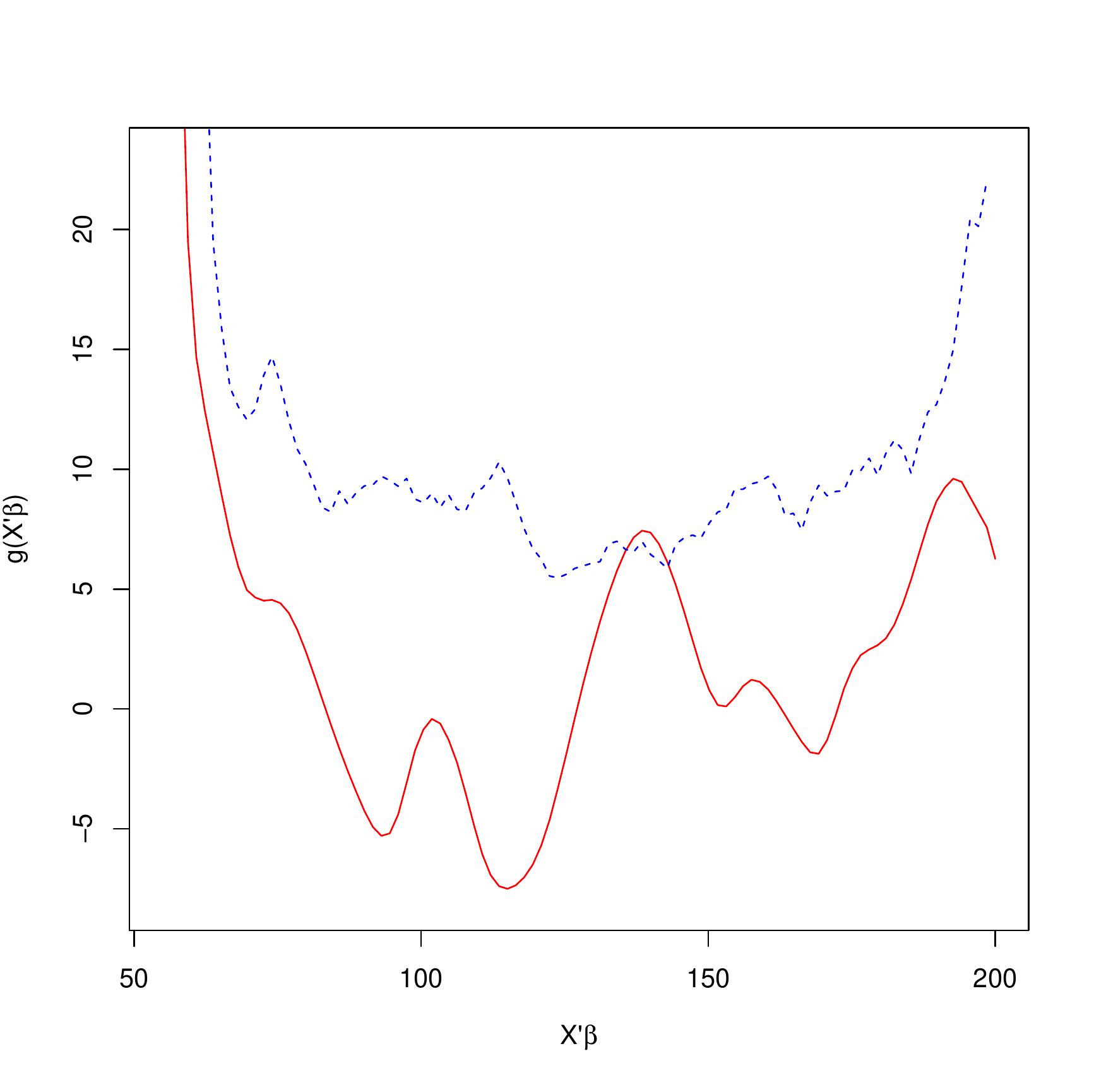}\par
			\vspace{-0.3in}
			\caption{\footnotesize{$\wh g(\cdot)$ and upper 95\% pointwise quantile 
					curve based on 1000 permuted data sets.}}
			\label{fig:gxfunction}
		\end{minipage}
	\end{figure}

	Let $\X=(X_1,X_2,X_3,X_4,X_5)\trans$ and let $Z_i$ be the treatment indicator
	where $Z_i=1$ means the participant was assigned to the BUP treatment,
	$Z_i=0$ represents 
	the SER group. We fitted the model $Y_i=f(\X_i)+Z_ig(\X_i\trans\bb)+\epsilon_i$. We are interested in finding the optimal treatment assignment to the
	patients.  Since treatment effects are described by $g(\X\trans\bb)$, we
	first obtained the estimators 
	for $\bb$ and the function $g(\cdot)$, using the proposed locally
	efficient estimator described in Method I.  We fixed the coefficient
	of $X_1$ to be 1 and estimated the remaining four
	coefficients. The estimator for $(\beta_2,\beta_3,\beta_4,\beta_5)\trans$
	was $(3.00985, -4.08772, 3.83337, 0.460488)\trans$.  
	To check if these coefficients are significant, we
	used bootstrap to obtain the estimation variance. We provide a box
	plot of the bootstrap estimators 
	of $\bb$ in Figure \ref{fig:box-plot-beta}. We also constructed the
	confidence intervals for $\beta_2$ to $\beta_5$ respectively. We found
	that the effects of $\beta_3$ and $\beta_4$ are significantly 
	different from zero. The 95\% confidence intervals for
	$\beta_2, \dots, \beta_5$ are, respectively, $(-2.001,  3.621)$, $(-4.737,
	-0.299)$, $(0.657, 4.920)$ 
	and $(-4.207, 3.224)$.
	
	We also estimated the function $g(\cdot)$ nonparametrically according
	to the method introduced in the Section \ref{sec:best} to obtain $\wh
	g$, with its plot
	in Figure \ref{fig:gxfunction}. 
	To evaluate the significance of the nonparametric function $\wh g(\cdot)$,
	we applied a permutation test.  Specifically, we randomly permute
	the treatment label $Z_i$'s and estimate 
	$g(\cdot)$ based on each permuted data. If the true
	$g(\cdot)$ function is zero, then the permutation should not alter the
	true function $g(\cdot)$ and the  function estimated from  
	the original real data, i.e. $\wh g$,  is not likely to be an extreme case among all the
	estimated $g(\cdot)$ functions based on the permuted data sets. 
	We plot the pointwise upper 95\% quantile curve based on 1000
	permuted data sets in Figure \ref{fig:gxfunction}. It is clear that
	there is one region where the estimated curve $\wh g$ is above the  
	95\% quantile curve. Therefore,  in this region, we have a
	significantly positive treatment effect using BUP.
	
	Finally, we compare the optimal treatment assignment with the real
	treatment assignment in this experiment. Under the estimated optimal
	ITR, we assign the participants according to their corresponding
	estimation of $g(\cdot)$, i.e. the  $i$th participant is assigned to the
	treatment BUP if $\wh g(\X_i\trans\wh\bb)>0$ and is
	assigned to SER if $\wh g(\X_i\trans\wh\bb)\le0$. Then we classified 
	the patients according to the optimal treatment assignments and the
	real treatment assignments. The results are displayed in Table
	\ref{table2by2}. We can see that a total of 179 participants were not assigned to
	their corresponding optimal treatment group. Therefore, the proposed
	method could have  been used
	for improving the patient satisfaction in this example.  
	
	\begin{table}[!ht]
		\begin{center}
			\caption{\label{table2by2} Real treatment assignments versus the optimal treatment assignments.}
			\begin{tabular}{ccc}
				\hline
				& \multicolumn{2}{c}{Optimal Assignment}\\ 
				Real Assignment       & BUP  & SER   \\\hline
				BUP     & 80 &   73  \\
				SER      & 106 &  60   \\\hline
			\end{tabular}
		\end{center}
	\end{table}
	\vspace{-0.4in}
	\section{Conclusion}\label{sec:discussion}
	
	We note that in the derivations of the proposed estimator, we did not
	use the fact that $Z^2=Z$ in the binary data. In fact, the estimation
	procedure, asymptotic theory and implementation details are directly applicable
	even when $Z$ is categorical or continuous. A categorical  $Z$
	corresponds to the case of comparing multiple treatments, while a
	continuous $Z$ could be used when a range of dosage levels are
	under examination. 
	
	In many clinical studies, the covariate is often of very high dimension. To
	develop optimal individualized treatment rules in this case, it will be important
	to develop simultaneous variable selection and estimation of
	individualized rules. It is also of great interest to extend the
	current approach to multi-stage randomized clinical studies.

	\bigskip
	\begin{center}
		{\large\bf SUPPLEMENTARY MATERIAL}
	\end{center}
	
	Supplementary Material contains the derivation of efficient score functions of our proposed  methods and the proofs of Theorem \ref{th:best} - \ref{th:simple2}.

	\bibliographystyle{agsm}
	\bibliography{tr}
	
\newpage
\begin{center}
	{\large\bf SUPPLEMENTARY MATERIAL}
\end{center}
\renewcommand{\theequation}{A.\arabic{equation}}
\renewcommand{\thesubsection}{A.\arabic{subsection}}
\setcounter{equation}{0}

\subsection{Derivation of the efficient score function
	$\S\eff(\x,z,y,\bb,g,f,\sigma)$ for binary $Z$ in Section \ref{sec:best}}\label{sec:nuisanceg}
Taking derivative with respect to $\eta, f, g$ and $\sigma$ of the
logarithm of (\ref{eq:pdfbin}), we obtain
\bse
\Lambda_\eta&=&\{\a(\X):\forall \a(\X)\in{\cal R}^{p-1} {\rm \, s.t. } E(\a)=\0\}\\
\Lambda_f&=&[\{Y-f(\X)-Zg(\bb\trans\X)\}\a(\X) :\forall
\a(\X)\in {\cal R}^{p-1}]=\{\epsilon\a(\X) :\forall \a(\X)\in{\cal R}^{p-1}\}\\
\Lambda_g&=&[\{Y-f(\X)-Zg(\bb\trans\X)\}Z\a(\bb\trans\X) :\forall \a(\bb\trans\X)\in{\cal R}^{p-1}]=\{\epsilon
Z\a(\bb\trans\X) :\forall \a(\bb\trans\X)\in{\cal R}^{p-1}\}\\
\Lambda_\sigma&=&\{(\epsilon^2-\sigma^2)\a: \forall 
\a\in{\cal R}^{p-1}\},
\ese
where $\epsilon=Y-f(\X)-Zg(\bb\trans\X)$,
and the nuisance tangent space is then
$\Lambda=\Lambda_\eta+\Lambda_f+\Lambda_g+\Lambda_\sigma$. It is easy to
verify that $\Lambda_\eta\perp(\Lambda_f+\Lambda_g+\Lambda_\sigma)$
and
$\Lambda_\sigma\perp(\Lambda_f+\Lambda_g)$. To further orthogonalize
$\Lambda_f$ and $\Lambda_g$, define
\bse
\Lambda_g'=[\{Y-f(\X)-Zg(\bb\trans\X)\}\{Z-E(Z\mid\X)\}\a(\bb\trans\X)]
=\{\epsilon \{Z-E(Z\mid\X)\}\a(\bb\trans\X)\}.
\ese
It is easy to verify that $\Lambda_f\perp\Lambda_g'$ and
$\Lambda_f+\Lambda_g=\Lambda_f\oplus\Lambda_g'$. 
Thus, we have 
$\Lambda=\Lambda_\eta\oplus\Lambda_f\oplus\Lambda_g'\oplus\Lambda_\sigma.$
Consequently, the nuisance tangent space orthogonal complement is
characterized through
\bse
\Lambda^\perp&=&\left(\h(\x,z,y)\in{\cal R}^{p-1}: E\{\h(\X,Z,Y)\mid\x\}=\0, 
E\{\epsilon\h(\X,Z,Y)\mid\x\}=\0, \right.\\
&&\left.E[\epsilon\{Z-E(Z\mid\X)\}\h(\X,Z,Y)\mid\bb\trans\x]=\0, 
E\{(\epsilon^2-\sigma^2)\h(\X,Z,Y)\}=\0\right).
\ese
We further calculate the score function to obtain
\bse
\S_\beta(\x,z,y,\bb,g,f,\sigma)=\sigma^{-2}\{y-f(\x)-zg(\bb\trans\x)\}zg_1(\bb\trans\x)\x_L=\sigma^{-2}\epsilon
zg_1(\bb\trans\x)\x_L,
\ese
where $\x_L$ is the lower $(p-1)$-dimensional sub-vector of $\x$. 
Let 
\bse
\u(\bb\trans\x)=\frac{E\left\{\var(Z\mid\X)\X_L\mid\bb\trans\x\right\}}
{E\left\{\var(Z\mid\X)\mid\bb\trans\x\right\}}.
\ese
We decompose the score function as
\bse
\S_\beta(\x,z,y,\bb,g,f,\sigma)&=&\sigma^{-2}\epsilon\{z- E(Z\mid\x)\}g_1(\bb\trans\x)\{\x_L-\u(\bb\trans\x)\}\\
&&+\sigma^{-2}\epsilon E(Z\mid\x)g_1(\bb\trans\x)\x_L
+\sigma^{-2}\epsilon \{z-E(Z\mid\x)\} g_1(\bb\trans\x)\u(\bb\trans\x).
\ese
Obviously,
\bse
&&\sigma^{-2}\epsilon E(Z\mid\X)g_1(\bb\trans\X)\X_L
\in\Lambda_f\subset\Lambda\\
&&\sigma^{-2}\epsilon \{Z-E(Z\mid\X)\} g_1(\bb\trans\X)\u(\bb\trans\X)
\in\Lambda_g'\subset\Lambda\\
&&\sigma^{-2}\epsilon\{Z- E(Z\mid\X)\}g_1(\bb\trans\X)\{\X_L-\u(\bb\trans\X)\}\in\Lambda^\perp,
\ese
hence the efficient score function is
\bse
\S\eff(\x,z,y,\bb,g,f,\sigma)&=&
\sigma^{-2}\epsilon\{z-
E(Z\mid\x)\}g_1(\bb\trans\x)\{\x_L-\u(\bb\trans\x)\}\\
&=&\sigma^{-2}\{y-f(\x)-zg(\bb\trans\x)\}
\{z-
E(Z\mid\x)\}g_1(\bb\trans\x)\{\x_L-\u(\bb\trans\x)\}.
\ese

\subsection{Derivation of the estimating equation (\ref{eq:g})}\label{sec:nuisancenog}

Following the local linear estimation idea, we replace
$g(\bb\trans\x)$ with $\alpha_c+\alpha_1\bb\trans(\x-\x_0)$.
Our goal is to estimate $\ba=(\alpha_c, \alpha_1)\trans$. 
The corresponding pdf of $(\X,Z,Y)$ then becomes
\bse
f_{\X,Z,Y}(\x,z,y)=\frac{1}{\sqrt{2\pi}\sigma}\exp\left(\frac{[y-f(\x)-z\{
	\alpha_c+\alpha_1\bb\trans(\x-\x_0)\}]^2}{-2\sigma^2}\right)
\eta(\x)f_{Z\mid\X}(z,\x).
\ese
Viewing $\ba$ together with $\bb$ as the parameter of interest, $\eta, f$ and $\sigma$ as
nuisance parameters, at any fixed $\bb$,
following the similar lines of derivation as in Section \ref{sec:nuisanceg}, we have
\bse
\Lambda_\eta&=&\{\a(\X):\forall \a(\X)\in{\cal R}^{p+1} {\rm \, s.t. } E(\a)=\0\}\\
\Lambda_f&=&([Y-f(\X)-Z\{\alpha_c+\alpha_1\bb\trans(\X-\x_0)\}]\a(\X) :\forall \a(\X) \in
{\cal R}^{p+1})=\{\epsilon\a(\X) :\forall \a(\X)\in {\cal R}^{p+1}\}\\
\Lambda_\sigma&=&\{\sigma^{-3}(\epsilon^2-\sigma^2)\a: \forall 
\a\in{\cal R}^{p+1}\},
\ese
where
$\epsilon=Y-f(\X)-Z\{\alpha_c+\alpha_1\bb\trans(\X-\x_0)\}$. 
Obviously,
$\Lambda=\Lambda_\eta\oplus\Lambda_f\oplus\Lambda_\sigma$, and
\bse
\Lambda^\perp&=&\left(\h(\x,z,y)\in{\cal R}^{p+1}: E\{\h(\X,Z,Y)\mid\x\}=\0, 
E\{\epsilon\h(\X,Z,Y)\mid\x\}=\0, \right.\\
&&\left.E\{(\epsilon^2-\sigma^2)\h(\X,Z,Y)\}=\0\right).
\ese
Further, the score functions with respect to the parameters of interest $\bb$,
$\ba$ are respectively
\bse
\S_\beta(\x,z,y,\bb,\ba,f,\sigma)&=&
\sigma^{-2}\epsilon z\alpha_1(\x_L-\x_{L0}),\\
\S_\alpha(\x,z,y,\bb,\ba,f,\sigma)&=&
\sigma^{-2}\epsilon z\{1, \bb\trans(\x-\x_0)\}\trans.
\ese
Decomposing the score functions as
\bse
\S_\beta(\x,z,y,\bb,\ba,f,\sigma)&=&\sigma^{-2}\epsilon\{z-E(Z\mid\x)\}\alpha_1(\x_L-\x_{L0})
+\sigma^{-2}\epsilon E(Z\mid\x)\alpha_1(\x_L-\x_{L0})\\
\S_\alpha(\x,z,y,\bb,\ba,f,\sigma)&=&\sigma^{-2}\epsilon \{z-E(Z\mid\x)\}\{1, \bb\trans(\x-\x_0)\}\trans
+\sigma^{-2}\epsilon E(Z\mid\x)\{1, \bb\trans(\x-\x_0)\}\trans,
\ese
we verify that
\bse
&&\sigma^{-2}\epsilon E(Z\mid\X)\{\alpha_1(\X_L-\x_{L0})\trans,
1, \bb\trans(\X-\x_0)\}\trans\in\Lambda_f\subset\Lambda,\\
&&\sigma^{-2}\epsilon\{Z-E(Z\mid\X)\}
\{\alpha_1(\X_L-\x_{L0})\trans,1, \bb\trans(\X-\x_0)\}\trans\in\Lambda^\perp.
\ese
Therefore, 
\bse
&&\S\eff(\x,z,y,\bb,\ba,f,\sigma)\\
&=&\sigma^{-2}\epsilon\{z-E(Z\mid\x)\}
\{\alpha_1(\x_L-\x_{L0})\trans,1, \bb\trans(\x-\x_0)\}\trans\\
&=&\sigma^{-2}[y-f(\x)-z\{\alpha_c+\alpha_1\bb\trans(\x-\x_0)\}]\{z-E(Z\mid\x)\}
\{\alpha_1(\x_L-\x_{L0})\trans,1, \bb\trans(\x-\x_0)\}\trans.
\ese
Extracting the last two components of the efficient score function
provides us the element to construct the estimating equations for
$\ba$. Since our goal is to estimate $g(\cdot)$ at $\bb\trans\x_j$, we
set $\x_0=\x_j$ for $j=1, \dots, n$,
and weigh the
evaluation of the estimating function at the $i$th observation using
$\w_{ij}$. This yields (\ref{eq:g}).

\subsection{List of regularity conditions}\label{sec:conditions}

\begin{enumerate}
	\item[C1.] The univariate kernel function $K(\cdot)$ is symmetric,
	has compact support and is
	Lipschitz continuous on its support. It satisfies
	\bse
	\int K(u) du= 1,  \ \int uK(u) du= 0,   \ 0\ne \int u^2  K(u)du<\infty.
	\ese
	Denote $C_2=\int u^2  K(u)du$.
	
	\item[C2.] The probability density function
	of $\bb\trans\x$  denoted
	by $f_1\left(\bb\trans\x\right)$ is bounded away from 0 and $\infty$.
	
	\item[C3.] Let $\r_1(\bb\trans\x) = E\{ \a_1(\X,Y)\mid \bb\trans\x\}f_1( \bb\trans\x)$,
	where $\a_1(\X,Z,Y)$ can be any $Y$, $Z$ and $\x$.
	The $(m-1)$-th derivatives of $\r_1(\bb\trans\x)$ and
	$f_1(\bb\trans\x)$  and $g(\bb\trans\x)$  are
	locally Lipschitz-continuous.
	
	\item[C4.] $E\left(\|\x\|^4\right)<\infty$,  $E(Y^4)<\infty$
	and $E\left\{\|g_1(\bb\trans\X)\|^4\right\} <\infty$.
	
	\item[C5.]
	The bandwidths satisfy
	$nh_g^4h_u^4\to 0$, $nh_g^8\to 0$, $nh_u^8\to 0$,
	$nh_gh_u\to\infty$, $nh_g^2\to\infty$, $nh_u^2\to\infty$.
	
	\item[C6.]
	In a domain of the parameter values $\cal B$, the limiting estimating
	equations for $\bb$ when $n\to\infty$ has a unique solution.
\end{enumerate}

\subsection{Proof of Theorem \ref{th:best}}\label{sec:bestproof}

We decompose the estimating equation as
\bse
\0&=&n^{-1/2}\sumi\{y_i-f^*(\x_i)-z_i\wh\alpha_c(\wh\bb,i)\}
\{z_i-E(Z\mid\x_i)\}\wh\alpha_1(\wh\bb,i)\{\x_{Li}-\wh\u(\wh\bb\trans\x_i)\}\\
&=&n^{-1/2}\sumi\{y_i-f^*(\x_i)-z_i\wh\alpha_c(\bb,i)\}
\{z_i-E(Z\mid\x_i)\}\wh\alpha_1(\bb,i)\{\x_{Li}-\wh\u(\bb\trans\x_i)\}\\
&&+\left[n^{-1}\sumi \frac{\partial \{y_i-f^*(\x_i)-z_i\wh\alpha_c(\bb,i)\}
	\{z_i-E(Z\mid\x_i)\}\wh\alpha_1(\bb,i)\{\x_{Li}-\wh\u(\bb\trans\x_i)\}}{\partial\bb_L\trans}+o_p(1)\right]\\
&&\times\sqrt{n}(\wh\bb_L-\bb_L).
\ese
It is easy to obtain that
\bse
&&n^{-1}\sumi \frac{\partial \{y_i-f^*(\x_i)-z_i\wh\alpha_c(\bb,i)\}
	\{z_i-E(Z\mid\x_i)\}\wh\alpha_1(\bb,i)\{\x_{Li}-\wh\u(\bb\trans\x_i)\}}{\partial\bb_L\trans}+o_p(1)\\
&=&E \left[\frac{\partial \{Y-f^*(\X)-Zg(\bb\trans\X)\}
	\{Z-E(Z\mid\X)\}g_1(\bb\trans\X)\{\X_L-\u(\bb\trans\X)\}}{\partial\bb_L\trans}\right]+o_p(1)\\
&=&E \left[-Z
\{Z-E(Z\mid\X)\}g_1^2(\bb\trans\X)\{\X_L-\u(\bb\trans\X)\}\X_L\trans\right]\\
&&+E \left[ \{\epsilon+f(\X)-f^*(\X)\}
\{Z-E(Z\mid\X)\}\frac{\partial
	g_1(\bb\trans\X)\{\X_L-\u(\bb\trans\X)\}}{\partial\bb_L\trans}\right]+o_p(1)\\
&=&E \left[-
\var(Z\mid\X)g_1^2(\bb\trans\X)\{\X_L-\u(\bb\trans\X)\}\X_L\trans\right]\\
&&+E \left[ \{\epsilon+f(\X)-f^*(\X)\}
\{Z-E(Z\mid\X)\}\frac{\partial
	g_1(\bb\trans\X)\{\X_L-\u(\bb\trans\X)\}}{\partial\bb_L\trans}\right]+o_p(1)\\
&=&-\A+o_p(1).
\ese
On the other hand, we make the following expansion
\be
&&n^{-1/2}\sumi\{y_i-f^*(\x_i)-z_i\wh\alpha_c(\bb,i)\}
\{z_i-E(Z\mid\x_i)\}\wh\alpha_1(\bb,i)\{\x_{Li}-\wh\u(\bb\trans\x_i)\}\nonumber\\
&=&n^{-1/2}\sumi\{z_ig(\bb\trans\x_i)-z_i\wh\alpha_c(\bb,i)\}
\{z_i-E(Z\mid\x_i)\}\{\wh\alpha_1(\bb,i)-g_1(\bb\trans\x_i)\}\{\u(\bb\trans\x_i)-\wh\u(\bb\trans\x_i)\}\nonumber\\
&&+n^{-1/2}\sumi\{y_i-f^*(\x_i)-z_ig(\bb\trans\x_i)\}
\{z_i-E(Z\mid\x_i)\}\{\wh\alpha_1(\bb,i)-g_1(\bb\trans\x_i)\}\{\u(\bb\trans\x_i)-\wh\u(\bb\trans\x_i)\}\nonumber\\
&&+n^{-1/2}\sumi\{z_ig(\bb\trans\x_i)-z_i\wh\alpha_c(\bb,i)\}
\{z_i-E(Z\mid\x_i)\}\{\wh\alpha_1(\bb,i)-g_1(\bb\trans\x_i)\}\{\x_{Li}-\u(\bb\trans\x_i)\}\nonumber\\
&&+n^{-1/2}\sumi\{z_ig(\bb\trans\x_i)-z_i\wh\alpha_c(\bb,i)\}
\{z_i-E(Z\mid\x_i)\}g_1(\bb\trans\x_i)\{\u(\bb\trans\x_i)-\wh\u(\bb\trans\x_i)\}\nonumber\\
&&+n^{-1/2}\sumi\{y_i-f^*(\x_i)-z_ig(\bb\trans\x_i)\}
\{z_i-E(Z\mid\x_i)\}g_1(\bb\trans\x_i)\{\u(\bb\trans\x_i)-\wh\u(\bb\trans\x_i)\}\nonumber\\
&&+n^{-1/2}\sumi\{y_i-f^*(\x_i)-z_ig(\bb\trans\x_i)\}
\{z_i-E(Z\mid\x_i)\}\{\wh\alpha_1(\bb,i)-g_1(\bb\trans\x_i)\}\{\x_{Li}-\u(\bb\trans\x_i)\}\nonumber\\
&&+n^{-1/2}\sumi\{z_ig(\bb\trans\x_i)-z_i\wh\alpha_c(\bb,i)\}
\{z_i-E(Z\mid\x_i)\}g_1(\bb\trans\x_i)\{\x_{Li}-\u(\bb\trans\x_i)\}\nonumber\\
&&+n^{-1/2}\sumi\{y_i-f^*(\x_i)-z_ig(\bb\trans\x_i)\}
\{z_i-E(Z\mid\x_i)\}g_1(\bb\trans\x_i)\{\x_{Li}-\u(\bb\trans\x_i)\}.\label{eq:expand}
\ee
and we inspect each of the above eight terms. It is easy to see that
under Condition C5, the first four terms will go to zero in
probability when $n\to\infty$. Plugging (\ref{eq:u}) in to the fifth
term, we have
\bse
&&n^{-1/2}\sumi\{y_i-f^*(\x_i)-z_ig(\bb\trans\x_i)\}
\{z_i-E(Z\mid\x_i)\}g_1(\bb\trans\x_i)\{\u(\bb\trans\x_i)-\wh\u(\bb\trans\x_i)\}\\
&=&n^{-1/2}\sumi\{y_i-f^*(\x_i)-z_ig(\bb\trans\x_i)\}
\{z_i-E(Z\mid\x_i)\}g_1(\bb\trans\x_i)\\
&&\times\left[\u(\bb\trans\x_i)-
\frac{n^{-1}\sumj K_h\{\bb\trans(\x_j-\x_i)\}\var(Z\mid\x_j)\x_{Lj}}
{n^{-1}\sumj K_h\{\bb\trans(\x_j-\x_i)\}\var(Z\mid\x_j)}\right]\\
&=&n^{-1/2}\sumi\{y_i-f^*(\x_i)-z_ig(\bb\trans\x_i)\}
\{z_i-E(Z\mid\x_i)\}g_1(\bb\trans\x_i)
\left[\u(\bb\trans\x_i)\right.\\
&&\left.-
\frac{E\{\var(Z\mid\X)\X_L\mid\bb\trans\x_i\} +\sumj w_{ji}\var(Z\mid\x_j)\x_{Lj}-E\{\var(Z\mid\X)\X_L\mid\bb\trans\x_i\} }
{E\{\var(Z\mid\X)\mid\bb\trans\x_i\}+  \sumj
	w_{ji}\var(Z\mid\x_j)-E\{\var(Z\mid\X)\mid\bb\trans\x_i\}}\right]\\
&=&n^{-1/2}\sumi\{y_i-f^*(\x_i)-z_ig(\bb\trans\x_i)\}
\{z_i-E(Z\mid\x_i)\}g_1(\bb\trans\x_i)\\
&&\times\left(-\frac{\sumj w_{ji}\var(Z\mid\x_j)\x_{Lj}-E\{\var(Z\mid\X)\X_L\mid\bb\trans\x_i\}}
{E\{\var(Z\mid\X)\mid\bb\trans\x_i\}}\right.\\
&&\left.+
\frac{E\{\var(Z\mid\X)\X_L\mid\bb\trans\x_i\}[\sumj w_{ji}\var(Z\mid\x_j)-E\{\var(Z\mid\X)\mid\bb\trans\x_i\}]}
{[E\{\var(Z\mid\X)\mid\bb\trans\x_i\}]^2}\right)+o_p(1)\\
&=&-n^{-1/2}\sumi\{y_i-f^*(\x_i)-z_ig(\bb\trans\x_i)\}
\{z_i-E(Z\mid\x_i)\}g_1(\bb\trans\x_i)\frac{1}{E\{\var(Z\mid\X)\mid\bb\trans\x_i\}}\\
&&\times\left(\frac{n^{-1}\sumj K_h\{\bb\trans(\x_j-\x_i)\}[
	\var(Z\mid\x_j)\x_{Lj}-E\{\var(Z\mid\X)\X_L\mid\bb\trans\x_i\}]}
{f_1(\bb\trans\x_i)+n^{-1}\sumj K_h\{\bb\trans(\x_j-\x_i)\} -f_1(\bb\trans\x_i)}\right.\\
&&\left.-
\u(\bb\trans\x_i)
\frac{n^{-1}\sumj 
	K_h\{\bb\trans(\x_j-\x_i)\}[
	\var(Z\mid\x_j)-E\{\var(Z\mid\X)\mid\bb\trans\x_i\}]}
{f_1(\bb\trans\x_i)+n^{-1}\sumj
	K_h\{\bb\trans(\x_j-\x_i)\}-f_1(\bb\trans\x_i)}
\right)+o_p(1)\\
&=&-n^{-1/2}\sumi\{y_i-f^*(\x_i)-z_ig(\bb\trans\x_i)\}
\{z_i-E(Z\mid\x_i)\}g_1(\bb\trans\x_i)\frac{1}{E\{\var(Z\mid\X)\mid\bb\trans\x_i\}}\\
&&\times\left(\frac{n^{-1}\sumj K_h\{\bb\trans(\x_j-\x_i)\}[
	\var(Z\mid\x_j)\x_{Lj}-E\{\var(Z\mid\X)\X_L\mid\bb\trans\x_i\}]}
{f_1(\bb\trans\x_i)}\right.\\
&&\left.-
\u(\bb\trans\x_i)
\frac{n^{-1}\sumj 
	K_h\{\bb\trans(\x_j-\x_i)\}[
	\var(Z\mid\x_j)-E\{\var(Z\mid\X)\mid\bb\trans\x_i\}]}
{f_1(\bb\trans\x_i)}
\right)+o_p(1)\\
&=&-n^{-3/2}\sumi\sumj\{y_i-f^*(\x_i)-z_ig(\bb\trans\x_i)\}
\{z_i-E(Z\mid\x_i)\}g_1(\bb\trans\x_i)\frac{K_h\{\bb\trans(\x_j-\x_i)\}}{E\{\var(Z\mid\X)\mid\bb\trans\x_i\}}\\
&&\times\left(\frac{\var(Z\mid\x_j)\x_{Lj}-E\{\var(Z\mid\X)\X_L\mid\bb\trans\x_i\}
	-\u(\bb\trans\x_i)[\var(Z\mid\x_j)-E\{\var(Z\mid\X)\mid\bb\trans\x_i\}]}
{f_1(\bb\trans\x_i)}\right)\\
&&+o_p(1).
\ese
Denote the last summand $\a(\bo_i, \bo_j)$, where $\bo_i=(\x_i,z_i,y_i)$
is the $i$th observation. Write
\bse
\b(\x_i)&=&E\left\{
K_h\{\bb\trans(\x_j-\x_i)\}\left(
\var(Z\mid\x_j)\x_{Lj}-E\{\var(Z\mid\X)\X_L\mid\bb\trans\x_i\}
\right.\right.\\
&&\left.\left. -\u(\bb\trans\x_i)[\var(Z\mid\x_j)-E\{\var(Z\mid\X)\mid\bb\trans\x_i\}]\right) \mid\x_i,z_i,y_i\right\},
\ese
then $\b(\X_i)=O_p(h^2)$ and
\bse
&&n^{-1/2}\sumi E\{\a(\bo_i, \O_j)\mid\bo_i\}\\
&=&-n^{-1/2}\sumi\{y_i-f^*(\x_i)-z_ig(\bb\trans\x_i)\}
\{z_i-E(Z\mid\x_i)\}\frac{g_1(\bb\trans\x_i)\b(\x_i)}{E\{\var(Z\mid\X)\mid\bb\trans\x_i\}
	f_1(\bb\trans\x_i)}\\
&=&-n^{-1/2}\sumi\epsilon_i
\frac{\{z_i-E(Z\mid\x_i)\} g_1(\bb\trans\x_i)\b(\x_i)}{E\{\var(Z\mid\X)\mid\bb\trans\x_i\}
	f_1(\bb\trans\x_i)}\\
&&-n^{-1/2}\sumi
\{z_i-E(Z\mid\x_i)\}\frac{\{g(\x_i)-f^*(\x_i)\}g_1(\bb\trans\x_i)\b(\x_i)}{E\{\var(Z\mid\X)\mid\bb\trans\x_i\}
	f_1(\bb\trans\x_i)}\\
&=&O_p(h^2)=o_p(1).
\ese
On the other hand
\bse
n^{-1/2}\sumj E\{\a(\O_i, \bo_j)\mid\bo_j\}=\0.
\ese
Hence, using the U-statistic property, the fifth term in
(\ref{eq:expand}) is of order $o_p(1)$.
We now examine the sixth and seventh term in (\ref{eq:expand}).
From Section \ref{sec:best}, we can easily obtain
\bse
v_{0i}&=&\sumk w_{ki}z_k\{z_k-E(Z\mid\x_k)\}
=E\{\var(Z\mid\X)\mid\bb\trans\x_i\}+o_p(1),\\
v_{1i}&=&\sumk w_{ki}z_k\{z_k-E(Z\mid\x_k)\}\bb\trans(\x_k-\x_i)
=h^2C_2 \frac{\partial[ E\{\var(Z\mid\X)\mid\bb\trans\x_i\}f_1(\bb\trans\x_i)]}{f_1(\bb\trans\x_i)\partial(\bb\trans\x_i)}+o_p(1),\\
v_{2i}&=&\sumk w_{ki}z_k\{z_k-E(Z\mid\x_k)\}\{\bb\trans(\x_k-\x_i)\}^2
=h^2C_2E\{\var(Z\mid\X)\mid\bb\trans\x_i\}+o_p(1).
\ese
Similarly, we can obtain
\bse
&&\sumj
w_{ji}\{y_j-f^*(\x_j)\}\{z_j-E(Z\mid\x_j)\}\bb\trans(\x_j-\x_i)\\
&=&h^2C_2\frac{\partial[E\{\var(Z\mid\X)\mid\bb\trans\x_i\}g(\bb\trans\x_i)f_1(\bb\trans\x_i)]}{f_1(\bb\trans\x_i)\partial(\bb\trans\x_i)}+o_p(1)\\
&=&h^2C_2\frac{\partial[E\{\var(Z\mid\X)\mid\bb\trans\x_i\}f_1(\bb\trans\x_i)]}{f_1(\bb\trans\x_i)\partial(\bb\trans\x_i)}g(\bb\trans\x_i)\\
&&+h^2C_2E\{\var(Z\mid\X)\mid\bb\trans\x_i\}g_1(\bb\trans\x_i)
+o_p(1),
\ese
and
\bse
\sumj
w_{ji}\{y_j-f^*(\x_j)\}\{z_j-E(Z\mid\x_j)\}
=E\{\var(Z\mid\X)\mid\bb\trans\x_i\}g(\bb\trans\x_i)+o_p(1).
\ese
Combining the above results, we obtain
\bse
\wh\alpha_c(\bb,i)&=&
\frac{\sumj w_{ji}\{y_j-f^*(\x_j)\}\{z_j-E(Z\mid\x_j)\}\{v_{2i}
	-v_{1i}\bb\trans(\x_j-\x_i)\}}{v_{0i}v_{2i}-v_{1i}^2}\\
&=&g(\bb\trans\x_i)+o_p(1)\\
\wh\alpha_1(\bb,i) 
&=&\frac{\sumj
	w_{ji}\{y_j-f^*(\x_j)\}\{z_j-E(Z\mid\x_j)\}\{v_{0i}\bb\trans(\x_j-\x_i)-v_{1i}\}
}{v_{0i}v_{2i}-v_{1i}^2}\\
&=&g_1(\bb\trans\x_i)+o_p(1).
\ese
Combining with the definition of $\u(\bb\trans\x_i)$, these results directly lead to
the conclusion that the sixth and seventh terms in (\ref{eq:expand})
are  of order $o_p(1)$ as well. Therefore, we have obtained
\bse
\0&=&n^{-1/2}\sumi\{y_i-f^*(\x_i)-z_ig(\bb\trans\x_i)\}
\{z_i-E(Z\mid\x_i)\}g_1(\bb\trans\x_i)\{\x_{Li}-\u(\bb\trans\x_i)\}\\
&&-\A\sqrt{n}(\wh\bb_L-\bb_L)+o_p(1).
\ese
It is easy to check that 
\bse
\B=\var \{Y_i-f^*(\X_i)-Z_ig(\bb\trans\X_i)\}
\{Z_i-E(Z\mid\X_i)\}g_1(\bb\trans\X_i)\{\X_{Li}-\u(\bb\trans\X_i)\},
\ese
and when $f^*(\cdot)=f(\cdot)$, this is the variance of the efficient
score function, 
hence we have proved Theorem \ref{th:best}.

\subsection{Proof of Theorem \ref{th:simple1}}\label{sec:simple1proof}
We decompose the estimating equation as
\bse
\0&=&n^{-1/2}\sumi\{y_i-f^*(\x_i)-z_ig^*(\wh\bb\trans\x_i)\}
\{z_i-E(Z\mid\x_i)\}h^*(\wh\bb\trans\x_i)\{\x_{Li}-\wh\u(\wh\bb\trans\x_i)\}\\
&=&
n^{-1/2}\sumi\{y_i-f^*(\x_i)-z_ig^*(\bb\trans\x_i)\}
\{z_i-E(Z\mid\x_i)\}h^*(\bb\trans\x_i)\{\x_{Li}-\wh\u(\bb\trans\x_i)\}\\
&&+n^{-1}\sumi\left[\frac{\partial \{y_i-f^*(\x_i)-z_ig^*(\bb\trans\x_i)\}
	\{z_i-E(Z\mid\x_i)\}h^*(\bb\trans\x_i)\{\x_{Li}-\wh\u(\bb\trans\x_i)\}}{\partial\bb_L}+o_p(1)\right]\\
&&\times\sqrt{n}(\wh\bb_L-\bb_L).
\ese
It is easy to obtain that
\bse
&&n^{-1}\sumi\frac{\partial \{y_i-f^*(\x_i)-z_ig^*(\bb\trans\x_i)\}
	\{z_i-E(Z\mid\x_i)\}h^*(\bb\trans\x_i)\{\x_{Li}-\wh\u(\bb\trans\x_i)\}}{\partial\bb_L}+o_p(1)\\
&=&E \left[\frac{\partial \{Y-f^*(\X)-Zg^*(\bb\trans\X)\}
	\{Z-E(Z\mid\X)\}h^*(\bb\trans\X)\{\X_L-\u(\bb\trans\X)\}}{\partial\bb_L\trans}\right]+o_p(1)\\
&=&E \left[-Z
\{Z-E(Z\mid\X)\}h^{*}(\bb\trans\X) g_1^{*}(\bb\trans\X)\{\X_L-\u(\bb\trans\X)\}\X_L\trans\right]\\
&&+E \left[ \{\epsilon+f(\X)-f^*(\X)\}
\{Z-E(Z\mid\X)\}\frac{\partial
	h^*(\bb\trans\X)\{\X_L-\u(\bb\trans\X)\}}{\partial\bb_L\trans}\right]\\
&&+E \left[ Z\{g(\bb\trans\X)-g^*(\bb\trans\X)\}
\{Z-E(Z\mid\X)\}\frac{\partial
	h^*(\bb\trans\X)\{\X_L-\u(\bb\trans\X)\}}{\partial\bb_L\trans}\right]
+o_p(1)\\
&=&E \left[-\var(Z\mid\X)h^{*}(\bb\trans\X) g_1^{*}(\bb\trans\X)\{\X_L-\u(\bb\trans\X)\}\X_L\trans\right]\\
&&+E \left[ \var(Z\mid\X)\{g(\bb\trans\X)-g^*(\bb\trans\X)\}\frac{\partial
	h^*(\bb\trans\X)\{\X_L-\u(\bb\trans\X)\}}{\partial\bb_L\trans}\right]
+o_p(1)\\
&=&-\A+o_p(1).
\ese
Using the form of  $\wh u(\cdot)$ in (\ref{eq:u}), we write 
\bse
&&n^{-1/2}\sumi\{y_i-f^*(\x_i)-z_ig^*(\bb\trans\x_i)\}
\{z_i-E(Z\mid\x_i)\}h^*(\bb\trans\x_i)\{\x_{Li}-\wh\u(\bb\trans\x_i)\}\\
&=&
n^{-1/2}\sumi\{y_i-f^*(\x_i)-z_ig^*(\bb\trans\x_i)\}
\{z_i-E(Z\mid\x_i)\}h^*(\bb\trans\x_i)\left\{\x_{Li}-\u(\bb\trans\x_i)\right\}\\
&&n^{-1/2}\sumi\left[
\frac{E\left\{\var(Z\mid\X)\X_L\mid\bb\trans\x_i\right\}}
{E\left\{\var(Z\mid\X)\mid\bb\trans\x_i\right\}}
-
\frac{\sumj w_{ji}\var(Z\mid\x_j)\x_{Lj}}
{\sumj w_{ji}\var(Z\mid\x_j)}
\right]\\
&=&
n^{-1/2}\sumi\{y_i-f^*(\x_i)-z_ig^*(\bb\trans\x_i)\}
\{z_i-E(Z\mid\x_i)\}h^*(\bb\trans\x_i)\left\{\x_{Li}-\u(\bb\trans\x_i)\right\}\\
&&-n^{-1/2}\sumi\frac{1}{E\left\{\var(Z\mid\X)\mid\bb\trans\x_i\right\}}
\left[\sumj w_{ji}\var(Z\mid\x_j)\x_{Lj}
- E\left\{\var(Z\mid\X)\X_L\mid\bb\trans\x_i\right\}\right]\\
&&+n^{-1/2}\sumi
\frac{\u(\bb\trans\x_i)}
{E\left\{\var(Z\mid\X)\mid\bb\trans\x_i\right\}}
\left[\sumj
w_{ji}\var(Z\mid\x_j)-E\left\{\var(Z\mid\X)\mid\bb\trans\x_i\right\}\right]+o_p(1)\\
&=&
n^{-1/2}\sumi\{y_i-f^*(\x_i)-z_ig^*(\bb\trans\x_i)\}
\{z_i-E(Z\mid\x_i)\}h^*(\bb\trans\x_i)\left\{\x_{Li}-\u(\bb\trans\x_i)\right\}\\
&&-n^{-3/2}\sumi\sumj
\frac{K_h\{\bb\trans(\x_j-\x_i)\}\left[\var(Z\mid\x_j)\x_{Lj}
	- E\left\{\var(Z\mid\X)\X_L\mid\bb\trans\x_i\right\}\right]}{E\left\{\var(Z\mid\X)\mid\bb\trans\x_i\right\} f_1(\bb\trans\x_i)}\\
&&+n^{-3/2}\sumi \sumj
\frac{\u(\bb\trans\x_i)
	K_h\{\bb\trans(\x_j-\x_i)\}
	\left[\var(Z\mid\x_j)-E\left\{\var(Z\mid\X)\mid\bb\trans\x_i\right\}\right]}{
	E\left\{\var(Z\mid\X)\mid\bb\trans\x_i\right\}
	f_1(\bb\trans\x_i)}+o_p(1).
\ese
Denote the summand in the last equation $\a(\bo_i, \bo_j)$.
Taking expectation of $\a(\bo_i, \bo_j)$ conditional on the $i$th
observation, we have
\bse
E\{\a(\bo_i, \O_j)\mid \bo_i\}
=\{y_i-f^*(\x_i)-z_ig^*(\bb\trans\x_i)\}
\{z_i-E(Z\mid\x_i)\}h^*(\bb\trans\x_i)\left\{\x_{Li}-\u(\bb\trans\x_i)\right\}.
\ese
Taking expectation of $\a(\bo_i, \bo_j)$ conditional on the $j$th
observation, we have
\bse
&&E\{\a(\O_i, \bo_j)\mid \bo_j\}\\
&=&-\frac{\var(Z\mid\x_j)\x_{Lj}
	-
	E\left\{\var(Z\mid\X)\X_L\mid\bb\trans\x_j\right\}}{E\left\{\var(Z\mid\X)\mid\bb\trans\x_j\right\}}
+\u(\bb\trans\x_j)\left[\frac{\var(Z\mid\x_j)}{E\left\{\var(Z\mid\X)\mid\bb\trans\x_j\right\}}-1\right]
\\
&=&\frac{
	\var(Z\mid\x_j)
	\{\u(\bb\trans\x_j)-
	\x_{Lj}\}}{E\left\{\var(Z\mid\X)\mid\bb\trans\x_j\right\}}.
\ese
Obviously, $E\{\a(\O_i, \O_j)\}=\0$. Thus, using the U-statistic
property, we obtain
\bse
&&n^{-1/2}\sumi\{y_i-f^*(\x_i)-z_ig^*(\bb\trans\x_i)\}
\{z_i-E(Z\mid\x_i)\}h^*(\bb\trans\x_i)\{\x_{Li}-\wh\u(\bb\trans\x_i)\}\\
&=&n^{-1/2}\sumi \{y_i-f^*(\x_i)-z_ig^*(\bb\trans\x_i)\}
\{z_i-E(Z\mid\x_i)\}h^*(\bb\trans\x_i)\left\{\x_{Li}-\u(\bb\trans\x_i)\right\}\\
&&+n^{-1/2}\sumi \frac{
	\var(Z\mid\x_i)
	\{\u(\bb\trans\x_i)-
	\x_{Li}\}}{E\left\{\var(Z\mid\X)\mid\bb\trans\x_i\right\}}+o_p(1)\\
&=&n^{-1/2}\sumi \epsilon_i
\{z_i-E(Z\mid\x_i)\}h^*(\bb\trans\x_i)\left\{\x_{Li}-\u(\bb\trans\x_i)\right\}\\
&&+n^{-1/2}\sumi \{f(\x_i)-f^*(\x_i)\}
\{z_i-E(Z\mid\x_i)\}h^*(\bb\trans\x_i)\left\{\x_{Li}-\u(\bb\trans\x_i)\right\}\\
&&+n^{-1/2}\sumi z_i\{g(\bb\trans\x_i)-g^*(\bb\trans\x_i)\}
\{z_i-E(Z\mid\x_i)\}h^*(\bb\trans\x_i)\left\{\x_{Li}-\u(\bb\trans\x_i)\right\}\\
&&+n^{-1/2}\sumi \frac{
	\var(Z\mid\x_i)
	\{\u(\bb\trans\x_i)-
	\x_{Li}\}}{E\left\{\var(Z\mid\X)\mid\bb\trans\x_i\right\}}+o_p(1).
\ese
From central limit theorem, the above expression converges to a normal
distribution with mean zero and variance given in (\ref{eq:B1}).

\subsection{Proof of Theorem \ref{th:simple2}}\label{sec:simple2proof}

We decompose the estimating equation as
\bse
\0&=&
n^{-1/2}\sumi\{y_i-f^*(\x_i)-z_i\wh\alpha_c(\wh\bb,i)\}
\{z_i-E(Z\mid\x_i)\}\wh\alpha_1(\wh\bb,i)\{\x_{Li}-\u^*(\wh\bb\trans\x_i)\}\\
&=&n^{-1/2}\sumi\{y_i-f^*(\x_i)-z_i\wh\alpha_c(\bb,i)\}
\{z_i-E(Z\mid\x_i)\}\wh\alpha_1(\bb,i)\{\x_{Li}-\u^*(\bb\trans\x_i)\}\\
&&+\left[n^{-1}\sumi\{y_i-f^*(\x_i)-z_i\wh\alpha_c(\bb,i)\}
\{z_i-E(Z\mid\x_i)\}\wh\alpha_1(\bb,i)\{\x_{Li}-\u^*(\bb\trans\x_i)\}+o_p(1)\right]\\
&&\times\sqrt{n}(\wh\bb_L-\bb_L).
\ese
It is easy to see that
\bse
&&n^{-1}\sumi\{y_i-f^*(\x_i)-z_i\wh\alpha_c(\bb,i)\}
\{z_i-E(Z\mid\x_i)\}\wh\alpha_1(\bb,i)\{\x_{Li}-\u^*(\bb\trans\x_i)\}+o_p(1)\\
&=&E\left[\frac{\partial
	\{Y-f^*(\X)-Zg(\bb\trans\X)\}
	\{Z-E(Z\mid\X)\}g_1(\bb\trans\X)\{\X_{L}-\u^*(\bb\trans\X)\}}{\partial\bb_L\trans}\right]+o_p(1)\\
&=&E\left[-Zg_1^2(\bb\trans\X)
\{Z-E(Z\mid\X)\}\{\X_{L}-\u^*(\bb\trans\X)\}\X_L\trans
\right]\\
&&+E\left[
\{Y-f^*(\X)-Zg(\bb\trans\X)\}\{Z-E(Z\mid\X)\}\frac{\partial
	g_1(\bb\trans\X)\{\X_{L}-\u^*(\bb\trans\X)\}}{\partial\bb_L\trans}\right]+o_p(1)\\
&=&E\left[-\var(Z\mid\X)
g_1^2(\bb\trans\X)\{\X_{L}-\u^*(\bb\trans\X)\}\X_L\trans
\right]+o_p(1)\\
&=&-\A+o_p(1).
\ese
In addition, 
\be
&&n^{-1/2}\sumi\{y_i-f^*(\x_i)-z_i\wh\alpha_c(\bb,i)\}
\{z_i-E(Z\mid\x_i)\}\wh\alpha_1(\bb,i)\{\x_{Li}-\u^*(\bb\trans\x_i)\}\nonumber\\
&=&n^{-1/2}\sumi\{z_ig(\bb\trans\x_i)-z_i\wh\alpha_c(\bb,i)\}
\{z_i-E(Z\mid\x_i)\}\{\wh\alpha_1(\bb,i)-g_1(\bb\trans\x_i)\}\{\x_{Li}-\u^*(\bb\trans\x_i)\}\nonumber\\
&&+n^{-1/2}\sumi\{y_i-f^*(\x_i)-z_ig(\bb\trans\x_i)\}
\{z_i-E(Z\mid\x_i)\}\{\wh\alpha_1(\bb,i)-g_1(\bb\trans\x_i)\}\{\x_{Li}-\u^*(\bb\trans\x_i)\}\nonumber\\
&&-n^{-1/2}\sumi\{z_i\wh\alpha_c(\bb,i) -z_ig(\bb\trans\x_i)\}
\{z_i-E(Z\mid\x_i)\}g_1(\bb\trans\x_i)\{\x_{Li}-\u^*(\bb\trans\x_i)\}\nonumber\\
&&+n^{-1/2}\sumi\{y_i-f^*(\x_i)-z_ig(\bb\trans\x_i)\}
\{z_i-E(Z\mid\x_i)\}g_1(\bb\trans\x_i)\{\x_{Li}-\u^*(\bb\trans\x_i)\}.\label{eq:expand2}
\ee
The first term in (\ref{eq:expand2}) is of order $o_p(1)$ under
condition C5. Following the same derivation of the sixth term in the
proof of Theorem \ref{th:best}, the second term in (\ref{eq:expand2})
is of order $o_p(1)$ as well. 
The third term in (\ref{eq:expand2}) can
be written as
\be\label{eq:expand3}
&&n^{-1/2}\sumi\left[
\frac{\sumj w_{ji}\{y_j-f^*(\x_j)\}\{z_j-E(Z\mid\x_j)\}\{v_{2i}
	-v_{1i}\bb\trans(\x_j-\x_i)\}}{v_{0i}v_{2i}-v_{1i}^2}
-g(\bb\trans\x_i)\right]\nonumber\\
&&\times z_i
\{z_i-E(Z\mid\x_i)\}g_1(\bb\trans\x_i)\{\x_{Li}-\u^*(\bb\trans\x_i)\}\nonumber\\
&=&n^{-1/2}\sumi\sumj w_{ji}\{y_j-f^*(\x_j)-g(\bb\trans\x_i) z_j\}\{z_j-E(Z\mid\x_j)\}
\nonumber\\
&&\times\frac{ z_i
	\{z_i-E(Z\mid\x_i)\}g_1(\bb\trans\x_i)\{\x_{Li}-\u^*(\bb\trans\x_i)\}}
{E\{\var(Z\mid\X)\mid\bb\trans\x_i\}}\nonumber\\
&&-n^{-1/2}\sumi\sumj
w_{ji}\{y_j-f^*(\x_j)-g(\bb\trans\x_i)z_j\}\{z_j-E(Z\mid\x_j)\}\bb\trans(\x_j-\x_i)\nonumber\\
&&\times\frac{\partial[
	E\{\var(Z\mid\X)\mid\bb\trans\x_i\}f_1(\bb\trans\x_i)]}{f_1(\bb\trans\x_i)[E\{\var(Z\mid\X)\mid\bb\trans\x_i\}]^2\partial(\bb\trans\x_i)}\nonumber\\
&&\times z_i
\{z_i-E(Z\mid\x_i)\}g_1(\bb\trans\x_i)\{\x_{Li}-\u^*(\bb\trans\x_i)\}
+o_p(1)\nonumber\\
&=&n^{-3/2}\sumi\sumj K_h(\bb\trans\x_i-\bb\trans\x_j)\{y_j-f^*(\x_j)-g(\bb\trans\x_i) z_j\}\{z_j-E(Z\mid\x_j)\}
\nonumber\\
&&\times\frac{ z_i
	\{z_i-E(Z\mid\x_i)\}g_1(\bb\trans\x_i)\{\x_{Li}-\u^*(\bb\trans\x_i)\}}
{f_1(\bb\trans\x_i) E\{\var(Z\mid\X)\mid\bb\trans\x_i\}}\nonumber\\
&&-n^{-3/2}\sumi\sumj
K_h(\bb\trans\x_i-\bb\trans\x_j)
\{y_j-f^*(\x_j)-g(\bb\trans\x_i)z_j\}\{z_j-E(Z\mid\x_j)\}\nonumber\\
&&\times\bb\trans(\x_j-\x_i)\frac{\partial[
	E\{\var(Z\mid\X)\mid\bb\trans\x_i\}f_1(\bb\trans\x_i)]}{f_1^2(\bb\trans\x_i)[E\{\var(Z\mid\X)\mid\bb\trans\x_i\}]^2\partial(\bb\trans\x_i)}\nonumber\\
&&\times z_i
\{z_i-E(Z\mid\x_i)\}g_1(\bb\trans\x_i)\{\x_{Li}-\u^*(\bb\trans\x_i)\}
+o_p(1).
\ee
We first handle the first term in (\ref{eq:expand3}).
Denote the summand $\a(\bo_i, \bo_j)$. 
Taking expectation of $\a(\bo_i, \bo_j)$ conditional on the $i$th 
observation, we have 
\bse
E\{\a(\bo_i, \O_j)\mid\bo_i\}=\0.
\ese
Taking expectation of $\a(\bo_i, \bo_j)$ conditional on the $j$th 
observation, we have 
\bse
&&E\{\a(\O_i, \bo_j)\mid\bo_j\}\\
&=&E\left[K_h(\bb\trans\x_i-\bb\trans\x_j)\{y_j-f^*(\x_j)-g(\bb\trans\x_i) z_j\}
\frac{
	g_1(\bb\trans\x_i)\{\u(\bb\trans\x_i)-\u^*(\bb\trans\x_i)\}}
{f_1(\bb\trans\x_i) }\mid\bo_j\right]\\
&&\times \{z_j-E(Z\mid\x_j)\}\\
&=&\{y_j-f^*(\x_j)-g(\bb\trans\x_j) z_j\}
g_1(\bb\trans\x_j) \{z_j-E(Z\mid\x_j)\}\{\u(\bb\trans\x_j)-\u^*(\bb\trans\x_j)\}.
\ese
Thus we obtain the final expansion
\bse
&&n^{-1/2}\sumi\{y_i-f^*(\x_i)-z_i\wh\alpha_c(\bb,i)\}
\{z_i-E(Z\mid\x_i)\}\wh\alpha_1(\bb,i)\{\x_{Li}-\u^*(\bb\trans\x_i)\}\nonumber\\
&=&
-n^{-1/2}\sumi\{y_i-f^*(\x_i)-g(\bb\trans\x_i) z_i\}
g_1(\bb\trans\x_i) \{z_i-E(Z\mid\x_i)\}\{\u(\bb\trans\x_i)-\u^*(\bb\trans\x_i)\}\\
&&+n^{-1/2}\sumi\{y_i-f^*(\x_i)-z_ig(\bb\trans\x_i)\}
\{z_i-E(Z\mid\x_i)\}g_1(\bb\trans\x_i)\{\x_{Li}-\u^*(\bb\trans\x_i)\}+o_p(1)\\
&=&n^{-1/2}\sumi\{y_i-f^*(\x_i)-z_ig(\bb\trans\x_i)\}
\{z_i-E(Z\mid\x_i)\}g_1(\bb\trans\x_i)\{\x_{Li}-\u(\bb\trans\x_i)\}+o_p(1).
\ese
We now handle the second term in (\ref{eq:expand3}).
Denote the summand $\b(\bo_i, \bo_j)$. 
Taking expectation of $\b(\bo_i, \bo_j)$ conditional on the $i$th 
observation, we have 
\bse
E\{\b(\bo_i, \O_j)\mid\bo_i\}=\0.
\ese
Taking expectation of $\b(\bo_i, \bo_j)$ conditional on the $j$th 
observation, we have 
\bse
&&E\{\b(\O_i, \bo_j)\mid\bo_j\}=\0
\ese
as well, hence the last term is of order $o_p(1)$.
Central limit theorem then directly leads to the result, with the
variance given in (\ref{eq:B2}).
\qed

\subsection{Derivation of the efficient score function
	$\S\eff(\x,z,y,\bb,g,f,\sigma)$ for continuous $Z$ in Section \ref{sec:cont}}\label{sec:conteff}
Taking derivative with respect to $\eta, f, g$ and $\sigma$ of the
logarithm of (\ref{eq:pdfcont}), we obtain
\bse
f_{\X,Z,Y}(\X,Z,Y)=\frac{1}{\sqrt{2\pi}\sigma}\exp\left[\frac{\{Y-f(\X)-\sum_{k=1}^KZ^kg_k(\bb\trans\X)\}^2}{-2\sigma^2}\right]
\eta_1(\X)f_{Z\mid\X}(Z,\X).
\ese
The nuisance tangent space with respect to the various nonparametric
components are respectively
\bse
\Lambda_\eta&=&\{\a(\X):\forall \a(\X)\in{\cal R}^{p-1} {\rm \, s.t. } E(\a)=\0\}\\
\Lambda_f&=&[\{Y-f(\X)-\sum_{k=1}^KZ^kg_k(\bb\trans\X)\}\a(\X) :\forall
\a(\X)\in {\cal R}^{p-1}]=\{\epsilon\a(\X) :\forall \a(\X)\in{\cal R}^{p-1}\}\\
\Lambda_{g_k}&=&[\{Y-f(\X)-\sum_{k=1}^KZ^kg_k(\bb\trans\X)\}Z^k\a(\bb\trans\X)
:\forall \a(\bb\trans\X)\in{\cal R}^{p-1}]\\
&=&\{\epsilon
Z^k\a(\bb\trans\X) :\forall \a(\bb\trans\X)\in{\cal R}^{p-1}\}\\
\Lambda_\sigma&=&\{(\epsilon^2-\sigma^2)\a: \forall 
\a\in{\cal R}^{p-1}\},
\ese
where $\epsilon=Y-f(\X)-\sum_{k=1}^KZ^kg_k(\bb\trans\X)$,
and the nuisance tangent space is then
$\Lambda=\Lambda_\eta+\Lambda_f+\Lambda_{g_1}+\dots+\Lambda_{g_K}+\Lambda_\sigma$. It is easy to
verify that $\Lambda_\eta\perp(\Lambda_f+\Lambda_{g_1}+\dots+\Lambda_{g_K}+\Lambda_\sigma)$
and
$\Lambda_\sigma\perp(\Lambda_f+\Lambda_{g_1}+\dots+\Lambda_{g_K})$. To further orthogonalize
let
\bse
\Lambda_{g_k}'=[\{Y-f(\X)-\sum_{k=1}^KZ^kg_k(\bb\trans\X)\}\{Z^k-E(Z^k\mid\X)\}\a(\bb\trans\X)]
=\{\epsilon \{Z^k-E(Z^k\mid\X)\}\a(\bb\trans\X)\}.
\ese
It is easy to verify that $\Lambda_f\perp\Lambda_{g_k}'$ for $k=1,
\dots K$,  and 
$\Lambda_f+\Lambda_{g_1}+\dots+\Lambda_{g_K}=\Lambda_f\oplus(\Lambda_{g_1}+\dots+\Lambda_{g_K}
)'$. We now further orthogonalize the $\Lambda_{g_k}'$'s. 

Define $\wt W_1=Z-E(Z\mid\X)$, $W_1=\wt W_1/\sd(\wt W_1\mid\bb\trans\X)$. For $j<k$, $k=2, \dots, K$,  let
\bse
C_{jk}(\bb\trans\X)&=&E(Z^kW_j\mid\bb\trans\X)\\
\wt W_k&=&Z^k-E(Z^k\mid\X)-\sum_{j=1}^{k-1}C_{jk}(\bb\trans\X)W_j\\
W_k&=&\wt W_k/\sd(\wt W_k\mid\bb\trans\X).
\ese
We can verify that $E(W_jW_k\mid\bb\trans\X)=I(j=k)$.
It is easy to see that the $W_1, \dots, W_k$ is an orthonormal
representation of $Z-E(Z\mid\X), \dots, Z^k-E(Z^k\mid\X)$.
Although the relation is complex, it is a linear transformation hence
using matrix notation, we can write the relation as 
\bse 
\A(\bb\trans\X)\left(\begin{array}{c}
	W_1\\
	\vdots\\
	W_k\end{array}\right) 
=\left(\begin{array}{c}
	Z-E(Z\mid\X)\\
	\vdots\\
	Z^k-E(Z^k\mid\X)
\end{array}\right). 
\ese

Define further
\bse
\Lambda_{g_k}''=[\{Y-f(\X)-\sum_{k=1}^KZ^kg_k(\bb\trans\X)\}W_k\a(\bb\trans\X)]
=\{\epsilon W_k\a(\bb\trans\X)\}.
\ese
Then, we have 
$\Lambda=\Lambda_\eta\oplus\Lambda_f\oplus\Lambda_{g_1}''\oplus\dots\oplus\Lambda_{g_K}''\oplus\Lambda_\sigma.$
Consequently, the nuisance tangent space orthogonal complement is
characterized through
\bse
\Lambda^\perp&=&\left[\h(\x,z,y)\in{\cal R}^{p-1}: E\{\h(\X,Z,Y)\mid\x\}=\0, 
E\{\epsilon\h(\X,Z,Y)\mid\x\}=\0, \right.\\
&&\left.E\{\epsilon W_k\h(\X,Z,Y)\mid\bb\trans\x\}=\0, \mbox{
	for }k=1, \dots, K,
E\{(\epsilon^2-\sigma^2)\h(\X,Z,Y)\}=\0\right].
\ese
We further calculate the score function to obtain
\bse
\S_\beta(\x,z,y,\bb,g,f,\sigma)=\sigma^{-2}\{y-f(\x)-\sum_{k=1}^Kz^kg_k(\bb\trans\x)\}\{\sum_{k=1}^Kz^kg_{k1}(\bb\trans\x)\}\x_L=\sigma^{-2}\epsilon 
\{\sum_{k=1}^Kz^kg_{k1}(\bb\trans\x)\}\x_L,
\ese
where $\x_L$ is the lower $(p-1)$-dimensional sub-vector of $\x$. 
We decompose the score function as
\bse
&&\S_\beta(\x,z,y,\bb,g,f,\sigma) \\
&=&\sigma^{-2}\epsilon\sum_{k=1}^K\{z^k- E(Z^k\mid\x)\}g_{k1}(\bb\trans\x)\x_L
+\sigma^{-2}\epsilon \sum_{k=1}^KE(Z^k\mid\x)g_{k1}(\bb\trans\x)\x_L\\
&=&\sigma^{-2}\epsilon\sum_{k=1}^K\sum_{j=1}^KA_{kj}(\bb\trans\x)W_jg_{k1}(\bb\trans\x)\x_L
+\sigma^{-2}\epsilon \sum_{k=1}^KE(Z^k\mid\x)g_{k1}(\bb\trans\x)\x_L\\
&=&\sigma^{-2}\epsilon\sum_{j=1}^K
W_j\left\{\sum_{k=1}^Kg_{k1}(\bb\trans\x)
A_{kj}(\bb\trans\x)\right\}\x_L
+\sigma^{-2}\epsilon \sum_{k=1}^KE(Z^k\mid\x)g_{k1}(\bb\trans\x)\x_L\\
&=&
\sigma^{-2}\epsilon\sum_{j=1}^K 
W_j\left\{\sum_{k=1}^Kg_{k1}(\bb\trans\x) 
A_{kj}(\bb\trans\x)\right\}\x_L\\
&&-
\sigma^{-2}\epsilon \sum_{r=1}^K W_r 
E\left[W_r\sum_{j=1}^K 
W_j\left\{\sum_{k=1}^Kg_{k1}(\bb\trans\x) 
A_{kj}(\bb\trans\x)\right\}\X_L\mid\bb\trans\x\right]\\
&&+
\sigma^{-2}\epsilon \sum_{r=1}^K W_r 
E\left[W_r\sum_{j=1}^K 
W_j\left\{\sum_{k=1}^Kg_{k1}(\bb\trans\x) 
A_{kj}(\bb\trans\x)\right\}\X_L\mid\bb\trans\x\right]\\
&&+\sigma^{-2}\epsilon 
\sum_{k=1}^KE(Z^k\mid\x)g_{k1}(\bb\trans\x)\x_L.
\ese
Obviously,
\bse
&&\sigma^{-2}\epsilon  W_r 
E\left[W_r\sum_{j=1}^K 
W_j\left\{\sum_{k=1}^Kg_{k1}(\bb\trans\x) 
A_{kj}(\bb\trans\x)\right\}\X_L\mid\bb\trans\x\right]
\in\Lambda_{g_r}''\subset\Lambda\\
&&\sigma^{-2}\epsilon\sum_{k=1}^K E(Z^k\mid\x)g_{k1}(\bb\trans\x)\x_L
\in\Lambda_f\subset\Lambda\\
&&\sigma^{-2}\epsilon\sum_{j=1}^K 
W_j\left\{\sum_{k=1}^Kg_{k1}(\bb\trans\x) 
A_{kj}(\bb\trans\x)\right\}\x_L\\
&&-
\sigma^{-2}\epsilon \sum_{r=1}^K W_r 
E\left[W_r\sum_{j=1}^K 
W_j\left\{\sum_{k=1}^Kg_{k1}(\bb\trans\x) 
A_{kj}(\bb\trans\x)\right\}\X_L\mid\bb\trans\x\right]
\in\Lambda^\perp,
\ese
hence the efficient score function is
\bse
\S\eff(\x,z,y,\bb,g,f,\sigma)&=&
\sigma^{-2}\epsilon
\x_L
\sum_{j=1}^K 
W_j\left\{\sum_{k=1}^Kg_{k1}(\bb\trans\x) 
A_{kj}(\bb\trans\x)\right\}\\
&&-\sigma^{-2}\epsilon \sum_{r=1}^K W_r 
E\left[W_r\sum_{j=1}^K 
W_j\left\{\sum_{k=1}^Kg_{k1}(\bb\trans\x) 
A_{kj}(\bb\trans\x)\right\}\X_L\mid\bb\trans\x\right]\\
&=&\sigma^{-2}\epsilon 
\left[
\sum_{k=1}^Kg_{k1}(\bb\trans\x) 
\left\{ Z^k-E(Z^k\mid\X)\right\}\x_L\right.\\
&&\left.-\sum_{r=1}^K W_r 
E\left[\X_L W_r\sum_{k=1}^Kg_{k1}(\bb\trans\x) 
\left\{ Z^k-E(Z^k\mid\X)\right\}
\mid\bb\trans\x\right]\right]\\
&=&\sigma^{-2}\epsilon 
\left\{\u
-\sum_{r=1}^K W_r 
E\left(W_r\U
\mid\bb\trans\x\right)\right\}.
\ese
\qed

\subsection{Derivation of the efficient score function
	$\S\eff(\x,\z,y,\bb,g,f,\sigma)$ for categorical $Z$ in Section \ref{sec:cat}}\label{sec:cateff}
Taking derivative with respect to $\eta, f, g$ and $\sigma$ of the
logarithm of (\ref{eq:pdfcat}), we obtain
\bse
f_{\X,\Z,Y}(\X,\Z,Y)=\frac{1}{\sqrt{2\pi}\sigma}\exp\left[\frac{\{Y-f(\X)-\sum_{k=1}^KZ_kg_k(\bb\trans\X)\}^2}{-2\sigma^2}\right]
\eta_1(\X)f_{\Z\mid\X}(\Z,\X).
\ese
The nuisance tangent space with respect to the various nonparametric
components are respectively
\bse
\Lambda_\eta&=&\{\a(\X):\forall \a(\X)\in{\cal R}^{p-1} {\rm \, s.t. } E(\a)=\0\}\\
\Lambda_f&=&[\{Y-f(\X)-\sum_{k=1}^KZ_kg_k(\bb\trans\X)\}\a(\X) :\forall
\a(\X)\in {\cal R}^{p-1}]=\{\epsilon\a(\X) :\forall \a(\X)\in{\cal R}^{p-1}\}\\
\Lambda_{g_k}&=&[\{Y-f(\X)-\sum_{k=1}^KZ_kg_k(\bb\trans\X)\}Z_k\a(\bb\trans\X)
:\forall \a(\bb\trans\X)\in{\cal R}^{p-1}]\\
&=&\{\epsilon
Z_k\a(\bb\trans\X) :\forall \a(\bb\trans\X)\in{\cal R}^{p-1}\}\\
\Lambda_\sigma&=&\{(\epsilon^2-\sigma^2)\a: \forall 
\a\in{\cal R}^{p-1}\},
\ese
where $\epsilon=Y-f(\X)-\sum_{k=1}^KZ_kg_k(\bb\trans\X)$,
and the nuisance tangent space is then
$\Lambda=\Lambda_\eta+\Lambda_f+\Lambda_{g_1}+\dots+\Lambda_{g_K}+\Lambda_\sigma$. It is easy to
verify that $\Lambda_\eta\perp(\Lambda_f+\Lambda_{g_1}+\dots+\Lambda_{g_K}+\Lambda_\sigma)$
and
$\Lambda_\sigma\perp(\Lambda_f+\Lambda_{g_1}+\dots+\Lambda_{g_K})$. To further orthogonalize
let
\bse
\Lambda_{g_k}'=[\{Y-f(\X)-\sum_{k=1}^KZ_kg_k(\bb\trans\X)\}\{Z_k-E(Z_k\mid\X)\}\a(\bb\trans\X)]
=\{\epsilon \{Z_k-E(Z_k\mid\X)\}\a(\bb\trans\X)\}.
\ese
It is easy to verify that $\Lambda_f\perp\Lambda_{g_k}'$ for $k=1,
\dots K$,  and 
$\Lambda_f+\Lambda_{g_1}+\dots+\Lambda_{g_K}=\Lambda_f\oplus(\Lambda_{g_1}+\dots+\Lambda_{g_K}
)'$. We now further orthogonalize the $\Lambda_{g_k}'$'s. 

Define $\wt W_1=Z_1-E(Z_1\mid\X)$, $W_1=\wt W_1/\sd(\wt W_1\mid\bb\trans\X)$. For $j<k$, $k=2, \dots, K$,  let
\bse
C_{jk}(\bb\trans\X)&=&E(Z_kW_j\mid\bb\trans\X)\\
\wt W_k&=&Z_k-E(Z_k\mid\X)-\sum_{j=1}^{k-1}C_{jk}(\bb\trans\X)W_j\\
W_k&=&\wt W_k/\sd(\wt W_k\mid\bb\trans\X).
\ese
We can verify that $E(W_jW_k\mid\bb\trans\X)=I(j=k)$.
It is easy to see that the $W_1, \dots, W_k$ is an orthonormal
representation of $Z_1-E(Z_1\mid\X), \dots, Z_k-E(Z_k\mid\X)$. 
Although the relation is complex, it is a linear transformation hence
using matrix notation, we can write the relation as 
\bse 
\A(\bb\trans\X)\left(\begin{array}{c}
	W_1\\
	\vdots\\
	W_k\end{array}\right) 
=\left(\begin{array}{c}
	Z_1-E(Z_1\mid\X)\\
	\vdots\\
	Z_k-E(Z_k\mid\X)
\end{array}\right). 
\ese

Define further
\bse
\Lambda_{g_k}''=[\{Y-f(\X)-\sum_{k=1}^KZ_kg_k(\bb\trans\X)\}W_k\a(\bb\trans\X)]
=\{\epsilon W_k\a(\bb\trans\X)\}.
\ese
Then, we have 
$\Lambda=\Lambda_\eta\oplus\Lambda_f\oplus\Lambda_{g_1}''\oplus\dots\oplus\Lambda_{g_K}''\oplus\Lambda_\sigma.$
Consequently, the nuisance tangent space orthogonal complement is
characterized through
\bse
\Lambda^\perp&=&\left[\h(\x,\z,y)\in{\cal R}^{p-1}: E\{\h(\X,\Z,Y)\mid\x\}=\0, 
E\{\epsilon\h(\X,\Z,Y)\mid\x\}=\0, \right.\\
&&\left.E\{\epsilon W_k\h(\X,\Z,Y)\mid\bb\trans\x\}=\0, \mbox{
	for }k=1, \dots, K,
E\{(\epsilon^2-\sigma^2)\h(\X,\Z,Y)\}=\0\right].
\ese
We further calculate the score function to obtain
\bse
\S_\beta(\x,\z,y,\bb,g,f,\sigma)=\sigma^{-2}\{y-f(\x)-\sum_{k=1}^Kz_kg_k(\bb\trans\x)\}\{\sum_{k=1}^Kz_kg_{k1}(\bb\trans\x)\}\x_L=\sigma^{-2}\epsilon 
\{\sum_{k=1}^Kz_kg_{k1}(\bb\trans\x)\}\x_L,
\ese
where $\x_L$ is the lower $(p-1)$-dimensional sub-vector of $\x$. 
We decompose the score function as
\bse
&&\S_\beta(\x,\z,y,\bb,g,f,\sigma) \\
&=&\sigma^{-2}\epsilon\sum_{k=1}^K\{z_k- E(Z_k\mid\x)\}g_{k1}(\bb\trans\x)\x_L
+\sigma^{-2}\epsilon \sum_{k=1}^KE(Z_k\mid\x)g_{k1}(\bb\trans\x)\x_L\\
&=&\sigma^{-2}\epsilon\sum_{k=1}^K\sum_{j=1}^KA_{kj}(\bb\trans\x)W_jg_{k1}(\bb\trans\x)\x_L
+\sigma^{-2}\epsilon \sum_{k=1}^KE(Z_k\mid\x)g_{k1}(\bb\trans\x)\x_L\\
&=&\sigma^{-2}\epsilon\sum_{j=1}^K
W_j\left\{\sum_{k=1}^Kg_{k1}(\bb\trans\x)
A_{kj}(\bb\trans\x)\right\}\x_L
+\sigma^{-2}\epsilon \sum_{k=1}^KE(Z_k\mid\x)g_{k1}(\bb\trans\x)\x_L\\
&=&
\sigma^{-2}\epsilon\sum_{j=1}^K 
W_j\left\{\sum_{k=1}^Kg_{k1}(\bb\trans\x) 
A_{kj}(\bb\trans\x)\right\}\x_L\\
&&-
\sigma^{-2}\epsilon \sum_{r=1}^K W_r 
E\left[W_r\sum_{j=1}^K 
W_j\left\{\sum_{k=1}^Kg_{k1}(\bb\trans\x) 
A_{kj}(\bb\trans\x)\right\}\X_L\mid\bb\trans\x\right]\\
&&+
\sigma^{-2}\epsilon \sum_{r=1}^K W_r 
E\left[W_r\sum_{j=1}^K 
W_j\left\{\sum_{k=1}^Kg_{k1}(\bb\trans\x) 
A_{kj}(\bb\trans\x)\right\}\X_L\mid\bb\trans\x\right]\\
&&+\sigma^{-2}\epsilon 
\sum_{k=1}^KE(Z_k\mid\x)g_{k1}(\bb\trans\x)\x_L.
\ese
Obviously,
\bse
&&\sigma^{-2}\epsilon  W_r 
E\left[W_r\sum_{j=1}^K 
W_j\left\{\sum_{k=1}^Kg_{k1}(\bb\trans\x) 
A_{kj}(\bb\trans\x)\right\}\X_L\mid\bb\trans\x\right]
\in\Lambda_{g_r}''\subset\Lambda\\
&&\sigma^{-2}\epsilon\sum_{k=1}^K E(Z_k\mid\x)g_{k1}(\bb\trans\x)\x_L
\in\Lambda_f\subset\Lambda\\
&&\sigma^{-2}\epsilon\sum_{j=1}^K 
W_j\left\{\sum_{k=1}^Kg_{k1}(\bb\trans\x) 
A_{kj}(\bb\trans\x)\right\}\x_L\\
&&-
\sigma^{-2}\epsilon \sum_{r=1}^K W_r 
E\left[W_r\sum_{j=1}^K 
W_j\left\{\sum_{k=1}^Kg_{k1}(\bb\trans\x) 
A_{kj}(\bb\trans\x)\right\}\X_L\mid\bb\trans\x\right]
\in\Lambda^\perp,
\ese
hence the efficient score function is
\bse
\S\eff(\x,\z,y,\bb,g,f,\sigma)&=&
\sigma^{-2}\epsilon
\x_L
\sum_{j=1}^K 
W_j\left\{\sum_{k=1}^Kg_{k1}(\bb\trans\x) 
A_{kj}(\bb\trans\x)\right\}\\
&&-\sigma^{-2}\epsilon \sum_{r=1}^K W_r 
E\left[W_r\sum_{j=1}^K 
W_j\left\{\sum_{k=1}^Kg_{k1}(\bb\trans\x) 
A_{kj}(\bb\trans\x)\right\}\X_L\mid\bb\trans\x\right]\\
&=&\sigma^{-2}\epsilon 
\left[
\sum_{k=1}^Kg_{k1}(\bb\trans\x) 
\left\{ Z_k-E(Z_k\mid\X)\right\}\x_L\right.\\
&&\left.-\sum_{r=1}^K W_r 
E\left[\X_L W_r\sum_{k=1}^Kg_{k1}(\bb\trans\x) 
\left\{ Z_k-E(Z_k\mid\X)\right\}
\mid\bb\trans\x\right]\right]\\
&=&\sigma^{-2}\epsilon 
\left\{\u
-\sum_{r=1}^K W_r 
E\left(W_r\U
\mid\bb\trans\x\right)\right\}.
\ese
\qed
	
\end{document}